\documentclass[11pt, oneside]{article}   	

\usepackage{jheppub}

\usepackage{amsmath}
\usepackage{amssymb}
\usepackage{amsfonts}
\usepackage{amsthm}
\usepackage{amsbsy}
\usepackage{array}
\usepackage[usenames,dvipsnames]{xcolor}
\usepackage{tikz}

\usepackage{mathtools}

\usepackage{subcaption}

\usepackage{dsdshorthand}

\usepackage[vcentermath]{youngtab}

\newcommand\wL{\mathbf{L}}

\renewcommand\vol{\mathop{\mathrm{vol}}}

\newcommand{\thalf}{\tfrac{1}{2}}

\makeatletter
\def\@fpheader{\ }
\makeatother

\makeatletter
\newcommand*{\shifttext}[2]{%
	\settowidth{\@tempdima}{#2}%
	\makebox[\@tempdima]{\hspace*{#1}#2}%
}
\makeatother

\DeclareMathOperator{\tr}{tr}

\renewcommand{\O}{{{\mathbb{O}}}}
\renewcommand{\cO}{{\mathcal{O}}}

\title{Missing local operators, zeros, and twist-4 trajectories}
\author{Johan Henriksson${}^{1,2}$, Petr Kravchuk${}^3$, Brett Oertel${}^4$}
\affiliation{
${}^1$Department of Physics, University of Pisa and INFN,
Largo Pontecorvo 3, 56127 Pisa, Italy\\
	${}^2$Universit\'e Paris--Saclay, CEA, Institut de Physique Th\'eorique, 91191, Gif-sur-Yvette, France
\\
${}^3$Department of Mathematics, King's College London, Strand, London, WC2R 2LS, UK\\
${}^4$Department of Physics, Yale University, 217 Prospect St, New Haven, CT 06511, USA
}
\date{}
\abstract{
The number of local operators in a CFT below a given twist grows with spin. Consistency with analyticity in spin then requires that at low spin, infinitely many Regge trajectories must decouple from local correlation functions, implying infinitely many vanishing conditions for OPE coefficients. In this paper we explain the mechanism behind this infinity of zeros. 
Specifically, the mechanism is related to the two-point function rather than the three-point function, explaining the  vanishing of OPE coefficients in every correlator from a single condition. 
We illustrate our result by studying twist-4 Regge trajectories in the Wilson--Fisher CFT at one loop.
}

\begin{document}

\maketitle

\newpage

\section{Introduction and summary}

It is well-known that certain data in conformal field theories admit analytic continuation in spin $J$. Most rigorously, this is established by the Lorentzian inversion formula~\cite{Caron-Huot:2017vep,Simmons-Duffin:2017nub,Kravchuk:2018htv}. It shows that for each four-point correlation function one can define a function $C(\De,J)$, which for non-negative even integer $J$ encodes the scaling dimensions and OPE coefficients of local primary\footnote{In what follows, we will often omit the word ``primary'' when talking about primary operators. All our operators are primaries, unless explicitly stated otherwise.} operators through the positions of poles in $\De$ and their residues (see below). At the same time, this function can be analytically continued in $J$ to a domain $\Re J\geq J_0$ (typically, $J_0\leq 2$) while $\De$ is kept in a domain away from the aforementioned poles. One can then ask, what is the analytic structure of $C(\De,J)$ for non-integer spin $J$?

The simplest scenario is that the poles which for integer $J$ correspond to local operators for $J$ non-integer move on complex-analytic Regge trajectories, and no other singularities are generated. It was shown in~\cite{Kravchuk:2018htv} that in this scenario the locations of these poles encode quantum numbers of non-local light-ray operators. These light-ray operators then interpolate between the (light-transforms~\cite{Kravchuk:2018htv} of) local operators at various values of spin. Indeed: a pole that is present at $J=2$ and corresponds to a spin-2 local operator will move to a new position as $J$ is analytically continued to $J=4$, and will correspond to a new spin-4 local operator. Taking this idea to its logical conclusion, one arrives at the optimistic conjecture that all local operators (or at least those with $J\geq J_0$) should organise into families, with operators within each family connected by complex-analytic Regge trajectories of light-ray operators.

However, if we take this idea seriously, we quickly run into puzzles related to the fact that, in a well-defined sense, the number of local operators grows with spin $J$. 

Specifically, let $N_{\tau_0}(J)$ be the number of local operators of spin $J$ and twist $\tau=\Delta-J\leq \tau_0$. Then one can show that $\lim_{J\to \oo}N_{\tau_0}(J)=\oo$ for a sufficiently large\footnote{For $\tau_0$ a few times the lowest twist in the theory.} $\tau_0$. For example, in a free scalar theory, one can build operators of the schematic form
\be
\cO_{J_1,J_2}=\f \ptl^{J_1}\f \ptl^{J_2}\f,
\ee
which all have twist $\tau = 3\De_\f=3$ in $d=4$. For a fixed total spin $J=J_1+J_2$, the integer $J_1$ can take any value between $0$ and $J$, and therefore the number of such operators grows linearly with $J$. Taking into account the bosonic symmetry between the three $\f$'s, one can check that the number of independent twist-3 operators at spin $J$ is $J/6+O(1)$. In an interacting theory, one can argue for $\lim_{J\to \oo}N_{\tau_0}(J)=\oo$ by repeatedly applying light-cone bootstrap arguments.

\begin{figure}[t!]
	\centering
	\includegraphics[width=110mm]{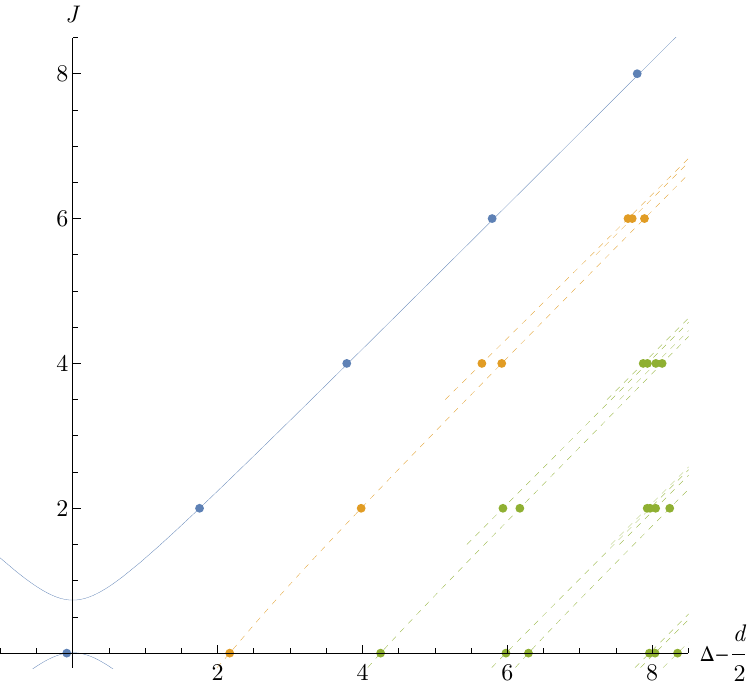}
	\caption{Schematic Chew--Frautschi plot in $\phi^4$ theory in $d=4-\e$, at small but finite $\epsilon$. The Regge trajectory of twist-2 operators is shown in solid blue, as described in~\cite{Caron-Huot:2020ouj,Caron-Huot:2022eqs}. The twist-4 operators are shown in orange. Schematic Regge trajectories for twist 4 and higher are shown by dashed lines. The anomalous dimensions of lower-twist operators have been exaggerated for greater visibility. For the data of local operators, see~\cite{Henriksson:2022rnm}.}\label{fig:CFplot}
\end{figure}

\begin{figure}[t!]
	\centering
	\includegraphics[width=120mm]{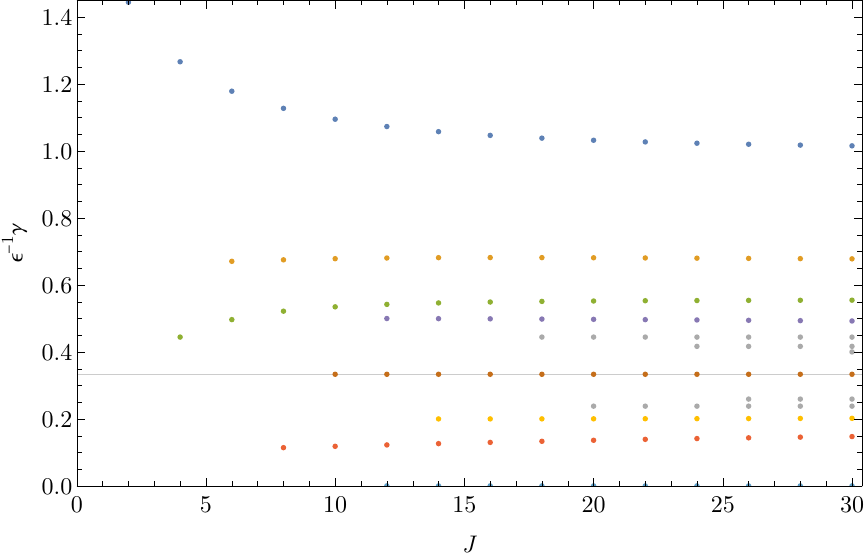}
	\caption{Anomalous dimensions of local primary twist-4 operators with spin up to 30. The number of local operators can be seen to grow with spin, and the operators themselves appear to organize into well-defined trajectories, encoded here by different colors. The top family also contains a scalar operator $\phi^4$ with $\e^{-1}\gamma=2$, which is outside the plotted range. The colours show the naive assignment of local operators to trajectories. As we will see in section~\ref{sec:twist4regges}, the situation is somewhat more subtle for the purple and the green ``trajectories'' shown here.}\label{fig:trajectories}
\end{figure}

The fact that $N_{\tau_0}(J)$ grows with $J$ without bound implies that in the limit of infinite $J$ there should exist infinitely many Regge trajectories of bounded twist. We illustrate this in figure~\ref{fig:CFplot} in $\f^4$ theory in $d=4-\e$. For example, if we focus on twist-4 operators, we can clearly see that their number grows with spin. This is shown in more detail in figure~\ref{fig:trajectories}, where we plot the one-loop anomalous dimensions of twist-4 local operators up to spin 30. 

This proliferation of local operators means that, at least na\"ively, we need to terminate some Regge trajectories in order to avoid having extra local operators at low spin. However, Regge trajectories cannot terminate in this way -- we expect them to be complex-analytic curves. But if they do not terminate, there should be infinitely many present at any given spin. Why is it then that we see only finitely many local operators of a given (approximate) twist at every spin?

A possible solution is that the Regge trajectories might not need to contain a local operator every time that they pass through an integer value of spin. While ultimately correct, this turns out to be not so easy to achieve. We can formalise this problem by considering a four-point function of local operators, say $\<\f\f\f\f\>$ where $\f$ is some scalar primary. Using the Euclidean inversion formula (see~\cite{Caron-Huot:2017vep,Karateev:2018oml} for recent accounts) one can associate to it the function $C(\De,J)$ mentioned above. It is meromorphic in $\De$ for any non-negative even integer spin $J$, with all poles (apart from some kinematic ones) related to the OPE data of local operators
\be
	C(\De,J)\sim \frac{r'_i(J)}{\De-\De'_i(J)}.
\ee
Here, $\De'_i(J)$ is the dimension of the $i$-th operator $\cO_{i,J}$ at spin $J$, and 
\be
	r'_i(J)=-\l_{\f\f\cO_{i,J}}^2,\label{eq:risminusff}
\ee
where $\lambda_{\f\f\cO_{i,J}}$ is the OPE coefficient. 
The Lorentzian inversion formula~\cite{Caron-Huot:2017vep} provides a canonical analytic continuation of $C(\De,J)$ in $J$. Under the hypothesis that all local operators lie on Regge trajectories, we expect to have at a complex spin $J$
\be
	C(\De,J)\sim \frac{r_i(J)}{\De-\De_i(J)},
\ee
where $\De_i(J)$ is the scaling dimension of the $i$-th Regge trajectory and $r_i(J)$ is related to the matrix elements of the associated light-ray operator~\cite{Kravchuk:2018htv}. Since we don't expect to necessarily be able to order local operators and Regge trajectories in the same way, we have used new symbols $r_i$ and $\De_i$ in place of $r_i'$ and $\De_i'$. Both $\De_i(J)$ and $r_i(J)$ are expected to be analytic functions of $J$.

The infinitude of Regge trajectories that we discussed above is then manifested in a large number of poles of $C(\De,J)$ with bounded $\tau=\De-J$ at large $J$. As we decrease $J$, these poles have to move in an analytic fashion, so we don't expect them to simply disappear. And yet, at even integer $J$ the only poles should be those related to local operators, and there are very few of these when $J$ is small (see figures~\ref{fig:CFplot} and~\ref{fig:trajectories}).

The simplest way in which this paradox can be resolved is if infinitely many residues $r_i(J)$ vanish at integer $J$, with more and more vanishing at smaller integers. This possibility has been recently tested in~\cite{Homrich:2022mmd}, where the authors studied the twist-3 single-trace Regge trajectories in $\cN=4$ SYM at one loop in the weak-coupling expansion. Extrapolating $r_i(J)$ from integer to complex values of $J$, they were able to observe the conjectured zeros. While~\cite{Homrich:2022mmd} gives encouraging evidence for this scenario, it does not explain the mechanism behind these zeros and relies on extrapolations of local operator data. Moreover, since each four-point function gives rise to a separate set of functions $r_i(J)$, there must be a mechanism that makes these functions have zeros for all four-point functions at once. 

In this paper we will explain the mechanism behind this vanishing of residues in a general CFT, thus confirming the simplest resolution of the paradox. In the process of doing so, we will clarify the properties of two-point functions of light-ray operators in general CFTs and their relation to the residues $r_i(J)$. Our main result is the formula~\eqref{eq:mainformulaIntro}, which shows that it is not $r_i(J)$ but rather the combination $r_i(J)/\sin (\pi J/2)$ that has a natural expression in terms of finite light-ray operators. The fact that needs explaining is therefore not that $r_i(J)$ vanishes infinitely often, but rather that it sometimes does not vanish. We explain the latter by the properties of the two-point functions of light-ray operators.

Although our arguments are quite general, to illustrate our discussion we will study the example of $\f^4$ theory in $d=4-\e$ dimensions, also known as the $\Z_2$ or Ising Wilson--Fisher theory~\cite{Wilson:1971dc,Wilson:1973jj}, at one loop. By perturbatively renormalising the light-ray operators directly at $J\in\C$ we will be able to confirm the predictions of our general analysis. Our perturbative study of higher-twist light-ray operators is also interesting it its own right. To the best of our knowledge this problem has not been considered before in the literature (although~\cite{Derkachov:2010zza} studied a closely-related problem).
Besides the intrinsic interest in understanding the $\f^4$ theory, we believe that it might also give a hint of what the theory of multi-twist operators might look like in general CFTs. As we will see, the one-loop dilatation operator becomes an integral kernel acting on a space of functions that define the light-ray operator. While at integer spin there is a finite-dimensional subspace corresponding to local operators, in general the problem is infinite-dimensional and we have to resort to numerical methods. 

As a byproduct of our work, we obtain a plenitude of one-loop results for twist-4 local operators, including anomalous dimensions with spins up to 700 and OPE coefficients with spin up to 30. These results allow us to explicitly verify various statements that are expected to hold both at large and at finite $J$. In particular, we find that while all twist-4 operators can be unambiguously assigned to double-twist Regge trajectories, these trajectories only begin to exhibit the behavior predicted by light-cone bootstrap~\cite{Komargodski:2012ek,Fitzpatrick:2012yx} at larger and larger spins.

The rest of this section contains a brief review of light-ray operators and a summary of our results. In section~\ref{sec:general} we prove the general formula~\eqref{eq:mainformulaIntro} for the residues $r_i(J)$ using the formalism of~\cite{Kravchuk:2018htv,Chang:2020qpj}. In section~\ref{sec:WF} we verify the results of section~\ref{sec:general} in the example of the Wilson--Fisher CFT. Section~\ref{sec:more} contains an extended discussion of twist-4 operators in Wilson--Fisher theory. We conclude in section~\ref{sec:discussion}. Appendices contain details and derivations omitted in the main text, and in particular a lot of data on OPE coefficients of twist-4 local operators.

\subsection{Light-ray operators}

Before stating our results, we briefly review some background on light-ray operators in CFTs and introduce our notation. For a more comprehensive discussion on light-ray operators, see~\cite{Kravchuk:2018htv,Kologlu:2019mfz,Chang:2020qpj,Caron-Huot:2022eqs}. 

In this paper we will only need to discuss traceless-symmetric local primary operators which we denote by $\cO^{\mu_1\cdots\mu_J}(x)$, where $J\geq 0$ is the spin and $x\in \R^{1,d-1}$.\footnote{We work exclusively in Lorentzian mostly plus signature.} It is convenient to use an index-free formalism in which we introduce a future-pointing null polarisation vector $z_\mu$, and define $\cO(x,z)=z_{\mu_1}\cdots z_{\mu_J}\cO^{\mu_1\cdots \mu_J}(x)$. The advantage of this formalism is that the spin is encoded by the equation
\be
	\cO(x,\l z)=\l^{J}\cO(x,z),\quad \forall \l>0,
\ee
and this definition of spin can be readily generalised to any $J\in \C$.

For any primary operator $\cO(x,z)$ (not necessarily local) with scaling dimension $\De$ and spin $J$ we define its light transform~\cite{Kravchuk:2018htv} as
\be\label{eq:Ldefn}
	\wL[\cO](x,z)=\int_{-\oo}^{+\oo} d\a (-\a)^{-\De-J} \cO(x-\a^{-1}z,z).
\ee
For $\a>0$ the integrand needs to be analytically continued into the next Poincare patch and, despite appearances, there is no ambiguity in the phase of the integrand. The key property of the light transform $\wL[\cO]$ is that it transforms as a primary operator of dimension $1-J$ and spin $1-\De$.

We are now ready to introduce light-ray operators and Regge trajectories. Suppose we have a family of canonically-normalised (i.e.\ their two-point functions take a standard form) local operators $\cO_J(x,z)$, labelled by spin $J$, and defined for even (odd) $J\in \cJ$, $\cJ\subseteq 2\Z$ ($\cJ\subseteq 2\Z+1$). We then say that this family forms an even-spin (odd-spin) Regge trajectory if there exists an analytic family of non-local primary operators $\O_J(x,z)$ labelled by\footnote{We expect that in general $\O_J$ will be a multi-valued function of $J$, analytic on an appropriate Riemann surface, see~\cite{Brower:2006ea, Gromov:2015wca,Caron-Huot:2022eqs,Klabbers:2023zdz} for examples. It is interesting to ask whether we should view $\O_J$ as being valued in a line bundle over that surface.} $J\in \C$ such that for every $J\in \cJ$
\be\label{eq:introLRlocalrelation}
	\O_J(x,z)=\cN_J\wL[\cO_J](x,z),
\ee
for some normalisation constants $\cN_J\neq 0$. We also require that $\cJ$ is complete: if for some $J$ it holds that $\O_J(x,z)=\wL[\cO](x,z)$ for a local operator $\cO$, then $J\in \cJ$.\footnote{This definition differs from the construction in~\cite{Kravchuk:2018htv} in that here we leave the normalisation of operators $\O_J$ arbitrary. We discuss this in detail in section~\ref{sec:general}.} The operators $\O_J$ for $J\in \C$ are referred to as light-ray operators. For~\eqref{eq:introLRlocalrelation} to respect conformal symmetry, $\O_J$ should have scaling dimension $1-J$ and spin $1-\De(J)$, where $\De(J)$ is an analytic continuation of the scaling dimensions of $\cO_J$. 

Note that we have to treat even and odd spins independently. For the most part, we will focus on even-spin operators in the rest of the paper, for example by only presenting the derivations for even spin trajectories (and identical operators where relevant). The generalisation to odd spins is straightforward.

Lastly, since there are many Regge trajectories (at the very least, many sheets of Regge trajectories), we will use an index $i$ to distinguish them. Specifically, we will write $\O_{i,J}$ for the light-ray operators, $\cJ_i$ for the subsets of $2\Z$ on which they reduce to light-transforms of local operators, and $\cN_{i,J}$ for the normalisation constants in 
\be
	\O_{i,J}=\cN_{i,J}\wL[\cO]\quad\text{for}\quad J\in \cJ_{i}.\label{eq:NiJdefinition}
\ee
We will usually not decorate the local operator $\cO$ in this formula with extra indices, since the identity $\O_{i,J}=\cN_{i,J}\wL[\cO]$ makes it clear which $\cO$ we are talking about. Importantly, we require that $\cO$ is canonically normalised.

\subsection{Summary of results}

Our main result is the following formula for the residues $r_i(J)$ in terms of the light-ray operators,
\be\label{eq:mainformulaIntro}
r_i(J)=\frac{1-e^{-i\pi J}}{i\pi}\p{\frac{\<0|\f\O_{i,J}\f|0\>}{\<0|\f\wL[\cO]\f|0\>_0}}^2\frac{\vol(\SO(1,1))\<\wL[\cO]\wL[\cO]\>_0}{\<\O_{i,J}\O_{i,J}\>}.
\ee
We derive it in section~\ref{sec:general}. Here, $r_i(J)$ is the residue of $C(\De,J)$ computed using the Lorentzian inversion formula for the four-point function $\<\f\f\f\f\>$ of a scalar primary operator $\f$. The ratios
\be
	\frac{\<0|\f\O_{i,J}\f|0\>}{\<0|\f\wL[\cO]\f|0\>_0}\quad\text{and}\quad\frac{\<\O_{i,J}\O_{i,J}\>}{\<\wL[\cO]\wL[\cO]\>_0}
\ee
express the 3-point and 2-point correlation functions
\be
	\<0|\f(x_1)\O_{i,J}(x,z)\f(x_2)|0\>\quad \text{and}\quad \<\O_{i,J}(x,z)\O_{i,J}(x',z')\>
\ee
as multiples of conventional conformally-invariant three-point and two-point tensor structures
\be
	\<0|\f(x_1)\wL[\cO](x,z)\f(x_2)|0\>_0\quad \text{and}\quad \<\wL[\cO](x,z)\wL[\cO](x',z')\>_0.
\ee
The fact that $r_i(J)$ depends on these conventional structures corresponds to the fact that there are conventions which go into writing down the Lorentzian inversion formula, and the dependence is precisely the same as in the ``natural'' form of the inversion formula derived in~\cite{Kravchuk:2018htv}.

An important feature of~\eqref{eq:mainformulaIntro} is that it involves the time-ordered two-point function $\<\O_{i,J}\O_{i,J}\>$. It is known that the Wightman two-point function of light-ray operators vanishes~\cite{Kravchuk:2018htv}. As we explain in section~\ref{sec:twoptgeneral}, the time-ordered two-point function is instead generically infinite. Fortunately, the infinity is universal and follows from a zero mode, which in~\eqref{eq:mainformulaIntro} is cancelled by the infinite volume factor $\vol(\SO(1,1))$.

However, when the light-ray operator $\O_{i,J}$ reduces to a light-transform of a local operator as in~\eqref{eq:NiJdefinition}, the two-point function $\<\O_{i,J}\O_{i,J}\>$ becomes the double light transform of $\<\cO\cO\>$, which was shown in~\cite{Kravchuk:2018htv} to be finite, see~\eqref{eq:standard2ptL}. This implies that the ratio
\be\label{eq:twoptvolratio}
	\<\O_{i,J}\O_{i,J}\>/\vol(\SO(1,1))
\ee
vanishes for $J\in \cJ_i$. The effect that this has in~\eqref{eq:mainformulaIntro} is to cancel the zero in the numerator coming from $1-e^{-i\pi J}$. As a result, we generically expect that $r_i(J)\neq 0$ for $J\in\cJ_i$.

On the other hand, for $J\in 2\Z_{\geq 0}$ but $J\not\in\cJ_i$, i.e.\ when the spin is an even integer but the light-ray operator is \textbf{not} related to a light-transform of a local operator, we do not know of a generic mechanism that would force~\eqref{eq:twoptvolratio} to vanish. Therefore, we expect that $r_i(J)=0$ due to the factor $1-e^{-i\pi J}$.

The above represents our resolution of the puzzle described in the introduction. It is perfectly self-consistent for a Regge trajectory to pass through an even integer spin $J$ without containing a local operator. In such cases there is a generic mechanism which makes the residue of the Regge trajectory in $C(\De,J)$ vanish, making sure that it doesn't contribute to the local conformal block expansion.

To verify our results, in section~\ref{sec:WF} we study even-spin twist-4 light-ray operators in $\f^4$ theory in $d=4-\e$ dimensions.\footnote{As a warm-up exercise, we also consider twist-2 operators. We do not consider twist-3 operators because only one twist-3 Regge trajectory has non-vanishing anomalous dimension at one loop~\cite{Kehrein:1992fn}. We do not (for the most part) consider the odd-spin twist-4 operators because their one-loop anomalous dimensions are independent of $J$~\cite{Derkachov:1995zr} (we have verified that our numerics reproduce this result).} We construct a generic twist-4 operator as
\be\label{eq:OpsidefnIntro}
\O_{\psi}(x,z) = \int_{-\infty}^{-\infty}d\alpha_1\cdots d\alpha_4(-\a_1)^{-\De_\f}\cdots (-\a_4)^{-\De_\f} \psi(\alpha_1,\cdots,\alpha_4):\!\f(x-\tfrac{z}{\alpha_1})\cdots\f(x-\tfrac{z}{\alpha_4})\!:.
\ee
Here, $\psi$ is a wavefunction which selects a specific twist-4 operator, and the spin $J$ is encoded in its homogeneity degree. Using the results of~\cite{Derkachov:2010zza} we derive the one-loop dilatation operator for $\O_\psi$, which becomes an integral operator acting on the wavefunctions $\psi$. Numerically diagonalising the one-loop dilatation operator, we obtain the quantum numbers and the wavefunctions $\psi$ for many twist-4 Regge trajectories.

Crucially, we are able to directly diagonalise the dilatation operator for generic complex spin $J$. This allows us to verify our predictions without relying on any interpolations or extrapolations as done in~\cite{Homrich:2022mmd}. For example, figure~\ref{fig:reggetrajectories_main} shows the anomalous dimensions of twist-4 Regge trajectories, and we can see many instances where a Regge trajectory is passing through an even integer $J$ with no local operator present. Interestingly, in figure~\ref{fig:reggetrajectories_main} we find an example of an ``avoided level-crossing'' of a pair of trajectories. Our methods allow us to study the full complex Regge trajectories in a neighborhood of this point and to confirm that the two trajectories are just two branches of a single Riemann surface, see figure~\ref{fig:3dshape}.

We are also able to obtain explicit wavefunctions $\psi$, with some examples shown in figure~\ref{fig:wavefunctions}. Using them, we compute in section~\ref{sec:twist4twomatrix} the individual quantities that enter in~\eqref{eq:mainformulaIntro}, as well the residues $r_i(J)$ themselves. We do this for the $\f^2\x\f^2$ and $\f\x\f$ OPEs. We find that the matrix elements in~\eqref{eq:mainformulaIntro} do not exhibit any interesting behavior (they have neither zeroes nor poles), while the two-point function behaves exactly as expected. As a result, we find residues $r_i(J)$ which agree with the known values for local operators  via~\eqref{eq:risminusff}, and vanish for those even integer values where there are no local operators, confirming the general picture outlined above. The key numerical results are presented in figures~\ref{fig:grid124},~\ref{fig:grid124ff},~\ref{fig:grid56} and~\ref{fig:residue56}.

As figure~\ref{fig:reggetrajectories_main} demonstrates, we find that all local twist-4 operators (with non-zero one-loop anomalous dimensions) lie on Regge trajectories. Furthermore, we show that all these trajectories can be given an interpretation as double-twist operators of pairs of lower-twist operators~\cite{Alday:2007mf,Fitzpatrick:2012yx,Komargodski:2012ek}.\footnote{In the perturbative setting this was observed already in~\cite{Derkachov:1995zr,Kehrein:1995ia}.} This interpretation as double-twist trajectories is supported by results from a direct computation of anomalous dimensions of local twist-4 operators with spin up to 700, and OPE coefficients involving local twist-4 operators with spin up to 30, discussed in detail in section~\ref{sec:more}. In other words, we do not see any evidence of Regge trajectories which contain local operators and yet do not have a double-twist interpretation.\footnote{One could have imagined that genuinely ``triple-twist,'' ``quadruple-twist,'' or higher trajectories might be required to account for all local operators. We do not see any evidence of this in the one-loop perturbative data.}

Finally, we find in section~\ref{sec:trace} that it is possible to compute traces of the powers of our dilatation operator. Interestingly, we find that these traces cannot be completely accounted for by the Regge trajectories that we have been able to identify numerically. This raises the possibility that there are additional twist-4 Regge trajectories which do not contain any local operators. We leave the investigation of this question to future work.

\section{Light-ray operators in general CFTs}
\label{sec:general}

In this section we derive the formula~\eqref{eq:mainformulaIntro} which expresses the residues $r_i(J)$ of the function $C(\De,J)$ in terms of the light-ray operators, in a general conformal field theory. Our derivation is based on the connection between light-ray operators and the Lorentzian inversion formula that was established in~\cite{Kravchuk:2018htv}. For simplicity, we will consider the even-spin light-ray operators which appear in an OPE of a scalar operator $\f$ with itself. Using the methods of~\cite{Kravchuk:2018htv,Chang:2020qpj} our results can be straightforwardly extended to even- and odd-spin light-ray operators in a completely general OPE at the cost of complicating the notation.

\subsection{Conformal block expansion}
\label{sec:CB}
Consider a four-point function of a scalar operator $\f$. As is standard, it can be expressed in terms of a function of cross-ratios,
\be
	\<\f(x_1)\f(x_2)\f(x_3)\f(x_4)\>=\frac{1}{x_{12}^{2\De_{\f}}x_{34}^{2\De_{\f}}}g(z,\barz),
\ee
where $x_{ij}=x_i-x_j$ and\footnote{We hope that our use of $z$ for both the polarization vectors and the cross-ratios will not cause confusion.} $z\barz=u$, $(1-z)(1-\barz)=v$ and
\be
	u=\frac{x_{12}^2x_{34}^2}{x_{13}^2x_{24}^2},\qquad 	v=\frac{x_{14}^2x_{23}^2}{x_{13}^2x_{24}^2}.
\ee
The function $g(z,\barz)$ has an expansion in terms of conformal blocks
\be
	g(z,\barz)=\sum_{\cO} \l_{\f\f\cO}^2 g_{\De_{\cO},J_{\cO}}(z,\barz),
\ee
where the sum is over a basis of local Hermitian operators $\cO$ that appear in the $\f\x\f$ OPE, $\De_\cO$ is the scaling dimension of $\cO$ and $J_\cO$ is its spin, and $\l_{\f\f\cO}$ is the OPE coefficient of $\cO$ in the $\f\x\f$ OPE. 

There are two equivalent ways in which we can describe our normalisation of the OPE coefficient $\l_{\f\f\cO}$, which is consistent with~\cite{Caron-Huot:2017vep, Kravchuk:2018htv}.
The first way is by specifying the normalisation of the conformal block $g_{\De,J}(z,\barz)$, which we take to be
\be\label{eq:glimit}
	g_{\De,J}(z,\barz)\sim z^{\frac{\De-J}{2}}\barz^{\frac{\De+J}{2}} \qquad(0\ll z\ll \barz \ll 1).
\ee
The second way is as follows: if the operator $\cO$ is normalised so that 
\be\label{eq:standard2pt}
	\<\cO(x_1,z_1)\cO(x_2,z_2)\>=\frac{(-2z_1\.I(x_{12})\.z_2)^{J_\cO}}{x_{12}^{2\De_\cO}},
\ee
where $I^\mu_\nu(x)=\de^\mu_\nu-2x^{-2}x^\mu x_\nu$, then
\be\label{eq:standardstructure}
	\<\f(x_1)\f(x_2)\cO(x_3,z)\>=\lambda_{\f\f\cO}\frac{(2z\.x_{23}x_{13}^2-2z\.x_{13}x_{23}^2)^{J_\cO}}{x_{12}^{2\De_\f-\De_\cO+J_\cO}
		x_{13}^{\De_\cO+J_\cO}
		x_{23}^{\De_\cO+J_\cO}
	}.
\ee
These structures follow the conventions of~\cite{Kravchuk:2018htv} and have somewhat unusual factors of $2^J$ compared to most of the CFT literature. These factors lead to the conformal blocks being normalised as in~\eqref{eq:glimit}.

\subsection{Lorentzian inversion formula}

The conformal block expansion can be inverted using the Lorentzian inversion formula, which also provides an analytic continuation of the CFT data in spin. In the most explicit form, for identical operators and even-spin trajectories, it reads~\cite{Caron-Huot:2017vep}
\be\label{eq:LIF}
	C(\De,J)=\frac{\kappa_{\De+J}}{2}\int dzd\barz\tfrac{|z-\barz|^{d-2}}{(z\barz)^d}g_{J+d-1,\De-d+1}(z,\barz)\mathrm{dDisc}[g(z,\barz)],
\ee
where $\kappa_\b$ is given by
\be
	\kappa_\b=\frac{\G(\tfrac{\b}{2})^4}{2\pi^2\G(\b-1)\G(\b)},
\ee
and the integral is taken over $z,\barz\in (0,1)$. The double-discontinuity $\mathrm{dDisc}[g(z,\barz)]$ is defined in detail in~\cite{Caron-Huot:2017vep} and can be viewed as the double-commutator $\<0|[\f,\f][\f,\f]|0\>$. We will not need its precise form.

Let $J_0$ be the Regge intercept of our theory.\footnote{In non-perturbative unitary CFTs we have $J_0\leq 1$~\cite{Caron-Huot:2017vep}. At a fixed order in perturbation theory $J_0$ can be larger. The chaos bound~\cite{Maldacena:2015waa} implies $J_0\leq 2$ at the leading order in large-$N$ theories.
} Then for even integer $J>J_0$ the function $C(\De,J)$ agrees with the analogous function obtained from the Euclidean inversion formula, and is a meromorphic function of $\De$ whose only poles are either dictated by conformal kinematics (see~\cite{Caron-Huot:2017vep} for a detailed classification) or arise from local operators,
\be\label{eq:physicalpoles}
	C(\De,J)\sim -\frac{\lambda_{\f\f\cO}^2}{\De-\De_\cO}.\qquad (J=J_\cO\in 2\Z_{>J_0})
\ee

A salient feature of the Lorentzian inversion formula~\eqref{eq:LIF} is that it defines an analytic continuation of $C(\De,J)$ to complex $J$ with $\Re J> J_0$. Strictly speaking, this analytic continuation is defined for $\De$ in the neighborhood of the principal series $\frac{d}{2}+i\R$, which is away from the physical poles~\eqref{eq:physicalpoles}. It is generally expected that, when maximally analytically continued in $\De$, it remains a meromorphic function of $\De$, with poles lying on Regge trajectories
\be
	C(\De,J)\sim \frac{r_i(J)}{\De-\De_i(J)}.
\ee
Here, $i$ is an index labelling different Regge trajectories. In~\cite{Kravchuk:2018htv} it was shown that $\De_i(J)$ can be related to quantum numbers of light-ray operators, and $r_i(J)$ to their matrix elements.

\subsection{$r_i(J)$ and light-ray operators}
\label{sec:LRcanonical}

Specifically,~\cite{Kravchuk:2018htv} constructs for each Regge trajectory a family of light-ray operators $\O_{i,J}^{(0)}$ such that 
\begin{enumerate}
	\item the scaling dimension of $\O_{i,J}^{(0)}$ is $\De_L=1-J$,
	\item the spin of $\O_{i,J}^{(0)}$ is $J_L=1-\De_i(J)$,
	\item for $J\in 2\Z_{>J_0}$ we have $\O_{i,J}^{(0)}=\lambda_{\f\f\cO}\wL[\cO]$ for a local operator $\cO$ normalised as in section~\ref{sec:CB}. The coefficient $\lambda_{\f\f\cO}$ is to be understood as equal to $0$ if there is no local operator at $J$.\footnote{We will not rely on this vanishing condition. In a sense, our goal is to derive it.}
\end{enumerate}
We put the superscript ${}^{(0)}$ on $\O_{i,J}^{(0)}$ to indicate that this light-ray operator comes with a very specific normalisation which will be explained below. The construction of $\O_{i,J}^{(0)}$ depends on a pair of local operators ($\f$ and $\f$ in the present case), from the OPE of which it is extracted. The standard assumption, which we adopt, is that the same (up to normalisation and vanishing of OPE coefficients due to symmetries) set of light-ray operators appears in every OPE.\footnote{We do not have any evidence to the contrary. In perturbation theory we can construct light-ray operators explicitly (as we do in this paper, for example); this construction is independent of the OPE, and agrees with the construction of~\cite{Kravchuk:2018htv} in explicit examples.}

The last property in the list above allows us to write a convenient formula for $r_i(J)$ in terms of $\O_{i,J}^{(0)}$. First, let $\<\f\f\cO\>_0$ denote the standard time-ordered three-point function obtained from~\eqref{eq:standardstructure} by setting $\lambda_{\f\f\cO}=1$, so that $\<\f\f\cO\>=\lambda_{\f\f\cO}\<\f\f\cO\>_0$. Using the standard $i\e$-prescriptions, this also defines the Wightman functions such as $\<0|\f\cO\f|0\>_0$.

Using this notation, consider for $J\in 2\Z_{>J_0}$ the matrix element
\be
	\<0|\f(x_1)\O_{i,J}^{(0)}(x,z)\f(x_2)|0\>&=\lambda_{\f\f\cO}\<0|\f(x_1)\wL[\cO](x,z)\f(x_2)|0\>\nn\\
	&=\lambda_{\f\f\cO}^2\<0|\f(x_1)\wL[\cO](x,z)\f(x_2)|0\>_0.
\ee
This implies that for $J\in 2\Z_{>J_0}$ (cf.~\eqref{eq:risminusff})
\be\label{eq:rFromCanonicalO}
	r_i(J)=-\frac{\<0|\f\O_{i,J}^{(0)}\f|0\>}{\<0|\f\wL[\cO]\f|0\>_0},
\ee
where we omit the coordinate parameters for brevity. We can try to interpret this formula as defining an analytic continuation of $r_i(J)$ in $J$. This is indeed possible, as long as we decide on the analytic continuation of the standard three-point function $\<0|\f\wL[\cO]\f|0\>_0$ in the denominator. For an appropriate choice of this analytic continuation, the analytic continuation of $r_i(J)$ will agree with the one defined by the Lorentzian inversion formula.

This choice is necessary since the Lorentzian inversion formula itself involves a choice -- we are, in principle, free to multiply the right-hand side of~\eqref{eq:LIF} by, say, $e^{i\pi J}$. Of course, this may break some properties such as reality or the asymptotic behavior at large imaginary $J$, but these are extra conditions which me may or may not want to impose. In fact, in~\cite{Kravchuk:2018htv} it was shown that the Lorentzian inversion formula can be naturally written in terms of analytic continuations of $\<0|\f\wL[\cO]\f|0\>_0$ and $\<\wL[\cO]\wL[\cO]\>_0$, the latter being the double light-transform of the canonical time-ordered two-point function. For an appropriate choice of these analytic continuations, which we describe below, the Lorentzian inversion formula takes the form~\eqref{eq:LIF}.

The expression~\eqref{eq:rFromCanonicalO} depends on the same choices, albeit in an implicit way: while $\<0|\f\wL[\cO]\f|0\>_0$ enters directly, the dependence on $\<\wL[\cO]\wL[\cO]\>_0$ is hidden in the definition of $\O_{i,J}^{(0)}$, which we describe in the next subsection.\footnote{The Lorentzian inversion formula for a general four-point function $\<\cO_1\cO_2\cO_3\cO_4\>$ depends on analytic continuations of $\<0|\cO_1\wL[\cO]\cO_2|0\>_0$, $\<0|\cO_3\wL[\cO]\cO_4|0\>_0$, and $\<\wL[\cO]\wL[\cO]\>_0$. The continuation of $\<0|\cO_3\wL[\cO]\cO_4|0\>_0$ will enter in~\eqref{eq:rFromCanonicalO}, while $\<0|\cO_1\wL[\cO]\cO_2|0\>_0$ and $\<\wL[\cO]\wL[\cO]\>_0$ will appear in the construction of $\O_{i,J}^{(0)}$.} With this understanding of conventions, it was shown in~\cite{Kravchuk:2018htv} that~\eqref{eq:rFromCanonicalO} is equivalent to the Lorentzian inversion formula. Supplemented with the definition of $\O_{i,J}^{(0)}$,~\eqref{eq:rFromCanonicalO} will be a key ingredient in the proof of~\eqref{eq:mainformulaIntro} in section~\ref{sec:theproofsubsection}.

Before proceeding, we define our conventions for the analytic continuations of $\<0|\f\wL[\cO]\f|0\>_0$ and $\<\wL[\cO]\wL[\cO]\>_0$. These are only needed for the explicit examples that we study --  our main results are convention-independent. We define them as light-transforms of analytic continuations of standard structures
\be
	\<0|\f\cO\f|0\>_0,\qquad \<\cO\cO\>_0.
\ee
The standard two-point function is defined for $J\in \C$ as
\be\label{eq:standard2ptNonIntegerJ}
\<\cO(x_1,z_1)\cO(x_2,z_2)\>_0=\frac{(-2z_1\.I(x_{12})\.z_2)^{J}}{x_{12}^{2\De}},
\ee
where $J$ and $\De$ are the scaling dimensions of the ``operator'' $\cO$.\footnote{Note that in this context $\cO$ is not a real operator and really just a notational device to keep track of the quantum numbers of the tensor structure.} It can be shown that the expression raised to the power $J$ in this formula is positive-definite~\cite{Kravchuk:2018htv} and so no phase ambiguity arises for non-integer $J$. The expression in the denominator is defined for $x_{12}^2<0$ using the standard time-ordered prescription $x_{12}^2\to x_{12}^2+i0$. Note that this expression agrees with~\eqref{eq:standard2pt} for integer $J$.

The standard Wightman three-point structure for $J\in \C$ is defined as
\be\label{eq:standardstructureNonIntegerJ}
\<0|\f(x_1)\cO(x_3,z)\f(x_2)|0\>_0=\frac{(2z\.x_{23}x_{13}^2-2z\.x_{13}x_{23}^2)^{J}}{x_{12}^{2\De_\f-\De+J}
	x_{13}^{\De_\cO+J_\cO}
	x_{23}^{\De_\cO+J_\cO}
}.
\ee
Here we do need to specify the branch of the numerator. We define $(2z\.x_{23}x_{13}^2-2z\.x_{13}x_{23}^2)^{J}$ to be a positive real number when $2z\.x_{23}x_{13}^2-2z\.x_{13}x_{23}^2$ is, while for all other configurations the Wightman function is defined using Wightman analyticity, which gives a non-ambiguous definition (see appendix A of~\cite{Kravchuk:2018htv}). Note that for integer $J$ this agrees with the definition~\eqref{eq:standardstructure}.

The agreement (at $J\in 2\Z$) of~\eqref{eq:standard2ptNonIntegerJ} and~\eqref{eq:standardstructureNonIntegerJ} with~\eqref{eq:standard2pt} and~\eqref{eq:standardstructureNonIntegerJ} respectively implies that the light-transforms of~\eqref{eq:standard2ptNonIntegerJ} and~\eqref{eq:standardstructureNonIntegerJ} do indeed define analytic continuations of the kind that we need. These light-transforms can be computed (the calculation is identical to~\cite{Kravchuk:2018htv} but here $J\in \C$) and take the form\footnote{Notice that it is not hard to imagine how the prefactors in these formulas combine to give the coefficient~$\kappa_{\De+J}/2$ in~\eqref{eq:LIF}.}
\be
	&\<0|\f(x_1)\wL[\cO](x_3,z)\f(x_2)|0\>_0\nn\\
	&=-2\pi i\frac{\G(\De+J-1)}{\G(\tfrac{\De+J}{2})^2} \frac{(2z\.x_{23}x_{13}^2-2z\.x_{13}x_{23}^2)^{1-\De}}{(x_{12}^2)^{\frac{2\De_\f-(1-J)+(1-\De)}{2}}(-x_{13}^2)^{\frac{(1-J)+(1-\De)}{2}}(x_{23}^2)^{\frac{(1-J)+(1-\De)}{2}}},\label{eq:standard3ptL}\\
	&\<\wL[\cO](x_1,z_1)\wL[\cO](x_2,z_2)\>_0=\frac{-2\pi i}{\De+J-1}\frac{(-2z_1\.I(x_{12})\.z_2)^{1-\De}}{x_{12}^{2(1-J)}}\label{eq:standard2ptL}.
\ee
Here, the expression for the three-point structure is valid for $x_3$ being in the past of $x_1$, while both are spacelike from $x_2$, in which case all the coordinate dependent factors are manifestly positive. It can be analytically continued to other configurations using the standard prescription for Wightman correlators. The expression for the two-point function is valid when $x_1$ is spacelike from $x_2$, and vanishes otherwise. As we will discuss in section~\ref{sec:twoptgeneral}, it is somewhat non-trivial that the double light-transform of the two-point function is finite. We will also need the generalisation of~\eqref{eq:standard3ptL} to non-equal scalar scaling dimensions,
\be
	&\<0|\phi_1(x_1)\wL[\cO](x_3,z)\phi_2(x_2)|0\>_0\nn\\\label{eq:standard3ptLdistinct}
	&=\frac{-2\pi i\G(\De+J-1)}{\G\left(\frac{\De+J+\Delta_{12}}{2}\right)\G\left(\tfrac{\De+J-\Delta_{12}}{2}\right)} \frac{(2z\.x_{23}x_{13}^2-2z\.x_{13}x_{23}^2)^{1-\De}}{(x_{12}^2)^{\frac{\De_1+\De_2+J-\De}{2}}(-x_{13}^2)^{\frac{\De_{12}+2-J-\De}{2}}(x_{23}^2)^{\frac{-\De_{12}-\De_1+2-J-\De}{2}}},
\ee
where $\De_{12}=\De_1-\De_2$.

Note that our convention for the tensor structures at $J\in \C$ is designed to match some standard expressions for local operator tensor structures. This comes at the expense of having various prefactors in the above expressions, which might vanish or can have poles (in fact the $x$-dependent factors might also have implicit poles for special values of $\De, J$). In some applications one might want to make different choices for these structures, perhaps unrelated to the standard local operator structures but better-behaved for all $\De,J$.

\subsection{Construction of $\O^{(0)}_{i,J}$}

The light-ray operators $\O^{(0)}_{i,J}$ were constructed in~\cite{Kravchuk:2018htv} as
\be\label{eq:OcanonicalDef}
	\O^{(0)}_{i,J}(x,z)=\mathrm{res}_{\De=\De_i(J)} \int'_{x\approx 2\atop x\approx 1^{-}} d^dx_1 d^dx_2 K_{\De,J}(x_1,x_2;x,z)(\f(x_1)\f(x_2)+\f(x_2)\f(x_1)),
\ee
where the integrals are taken over a neighborhood of the future null cone of $x$ (this restriction is indicated by the prime on the integration sign), subject to the specified causal constraints,\footnote{We follow the notation of~\cite{Kravchuk:2018htv} in which $a>b$ stands for ``$a$ is in the absolute future of $b$'', $a\approx b$ stands for ``$a$ is spacelike from $b$'', and $a^-$ stands for ``the image of $a$ in the preceding Poincare patch''.} and $K_{\De,J}$ is a conformally-invariant kernel. Plugging this definition into~\eqref{eq:rFromCanonicalO} one is able to recover the Lorentzian inversion formula~\cite{Kravchuk:2018htv}.

The construction of the kernel $K_{\De,J}$ in~\cite{Kravchuk:2018htv} relied significantly on Euclidean kinematics and is rather inconvenient. In~\cite{Chang:2020qpj} an improved equivalent definition for $K_{\De,J}$ was derived, which we now review. First, we require that the kernel $K_{\De,J}(x_1,x_2;x,z)$ has the analyticity and conformal transformation properties of 
\be\label{eq:Ktype}
	\<0|\tl\f(x_1)\cO^L(x,z)\tl\f(x_2)|0\>,
\ee
where $\tl\f$ is a scalar of dimension $d-\De_\f$ and $\cO^L$ is an operator with dimension $1-J$ and spin $1-\De$. This fixes $K_{\De,J}(x_1,x_2;x,z)$ up to normalisation, which in turn is fixed by the equation (note that both sides are only non-zero if $x\approx x'$)
\be\label{eq:Kdefinition}
	&\int_{2>x'>1^{-}\atop {x\approx 2,1^-}} \frac{d^dx_1d^dx_2}{\vol(\SO(1,1))^2}K_{\De,J}(x_1,x_2;x,z)\<0|\f(x_2)\wL[\cO](x',z')\f(x_1)|0\>_0\nn\\
	&=\frac{1}{2\pi i}\<\wL[\cO](x,z)\wL[\cO](x',z')\>_0,
\ee
where $\<0|\f\wL[\cO]\f|0\>_0$ and $\<\wL[\cO]\wL[\cO]\>_0$ denote the analytic continuations of the standard structures. Here, $\cO$ formally has spin $J$ and dimension $\De$. The integration over $x_1,x_2$ is performed over the domain on the Lorentzian cylinder defined by the causal relations $2>x'>1^{-}$ and $x\approx 2,1^-$.
There are two zero modes in this integral, related to boosts between $z$ and $z'$ and dilatations between $x$ and $x'$, which we mod out by dividing by the respective group volumes; the left-hand side should be properly understood as having these zero modes fixed by a Fadeev--Popov procedure. 
This definition is convenient because it is purely Lorentzian and exhibits how $\O_{i,J}^{(0)}$ depends on the choices of the analytic continuations of the light-transforms of the standard three-point and two-point structures.

\subsection{The formula for $r_i(J)$}
\label{sec:theproofsubsection}
We are now ready to derive the formula~\eqref{eq:mainformulaIntro} for the residues $r_i(J)$. The key idea is to use the formula~\eqref{eq:rFromCanonicalO} and express $\O^{(0)}_{i,J}$ in terms of the arbitrarily-normalised light-ray operators $\O_{i,J}$. To do this, we first compute the time-ordered two-point function $\<\O^{(0)}_{i,J}\O^{(0)}_{i,J}\>$.

This can be done using the definition~\eqref{eq:OcanonicalDef},
\be
	&\<\O^{(0)}_{i,J}(x,z)\O^{(0)}_{i,J}(x',z')\>\nn\\
	&=2\.\mathrm{res}_{\De=\De_i(J)} \int'_{x\approx 2\atop x\approx 1^{-}} d^dx_1 d^dx_2 K_{\De,J}(x_1,x_2;x,z)\<\f(x_1)\f(x_2)\O^{(0)}_{i,J}(x',z')\>.\label{eq:residueintegral}
\ee
We claim that this is equivalent to
\be
	\<\O^{(0)}_{i,J}(x,z)\O^{(0)}_{i,J}(x',z')\>=-2\int_{x\approx 2\atop x\approx 1^{-}} \frac{d^dx_1 d^dx_2}{\vol(\SO(1,1))} K_{\De_i(J),J}(x_1,x_2;x,z)\<\f(x_1)\f(x_2)\O^{(0)}_{i,J}(x',z')\>,\label{eq:FPintegral}
\ee
where we have removed the prime over the integral, indicating that the integration region is now extended to all $x_1,x_2$ which satisfy the causal constraints, and also set $\De=\De_i(J)$ in the integrand. To compensate for these changes, we have divided by $\vol(\SO(1,1))$. To illustrate the idea behind this replacement, consider the integral
\be
	\int'_{x>0} dx\, x^{-1-\De+\De_i(J)},
\ee
where the prime over the integral indicates that it is performed over a neighborhood of $x=0$, i.e.
\be
	\int'_{x>0} dx x^{-1-\De+\De_i(J)}=\int_0^\e dx x^{-1-\De+\De_i(J)}=-\frac{\e^{-\De+\De_i(J)}}{\De-\De_i(J)},
\ee
where the last equality is valid for $\De<\De_i(J)$. Note that we expect the integral~\eqref{eq:residueintegral} to converge for $\De$ to the left of the physical poles and close to the principal series $\Re\De=d/2$, which is why we chose the signs in this way. Taking the residue we find
\be
	\mathrm{res}_{\De=\De_i(J)}\int'_{x>0} dx\, x^{-1+\De-\De_i(J)}=-1.
\ee
If we instead set $\De=\De_i(J)$ in the integrand, then the integrand $dx\, x^{-1}$ will possess a scaling symmetry $x\to \l x$. This symmetry will render the integral $\int_0^\oo dx\, x^{-1}$ infinite, but a finite answer can be obtained if we mod out by it,
\be
	\int_0^\oo \frac{dx}{\vol(\SO(1,1))} x^{-1}=1=-\mathrm{res}_{\De=\De_i(J)}\int'_{x>0} dx x^{-1+\De-\De_i(J)}.
\ee
This example is a faithful model for what happens in the integrals~\eqref{eq:residueintegral} and~\eqref{eq:FPintegral}. There is a boost-like symmetry which is a zero mode of the integral~\eqref{eq:FPintegral}. Integration over this mode in~\eqref{eq:residueintegral} (it isn't a \textit{zero} mode for~\eqref{eq:residueintegral} because $\De$ is held generic) is what leads to the pole in $\De$, and taking the residue of this pole is equivalent to computing~\eqref{eq:FPintegral}.

In~\eqref{eq:FPintegral} we are integrating a time-ordered three-point function $\<\f(x_1)\f(x_2)\O^{(0)}_{i,J}(x',z')\>$. Depending on the values of $x_1,x_2$, it is related to Wightman three-point functions with different orderings of the operators.  Light-ray operators annihilate the vacuum state~\cite{Kravchuk:2018htv}, so when the ordering is such that $\O^{(0)}_{i,J}(x',z')$ acts on the future or the past vacuum, the Wightman three-point function vanishes. In order to avoid this, it is necessary that one of the points $x_1,x_2$ is in the absolute future of $x'$ and the other is in the absolute past of $x'^+$ (recall that the light-ray operator $\O^{(0)}_{i,J}(x',z')$ is supported on the null-cone which stretches between $x'$ and $x'^+$). This means that we can write
\be
	\<\f(x_1)\f(x_2)\O^{(0)}_{i,J}(x',z')\>=&\<0|\f(x_1)\O^{(0)}_{i,J}(x',z')\f(x_2)|0\>\theta(1>x'>2^{-})\nn\\
	&+\<0|\f(x_2)\O^{(0)}_{i,J}(x',z')\f(x_1)|0\>\theta(2>x'>1^{-}).
\ee
Using this, we find
\be\label{eq:twopttwoterms}
	&\<\O^{(0)}_{i,J}(x,z)\O^{(0)}_{i,J}(x',z')\>\nn\\
	&=-2\int_{x\approx 2,1^-\atop 1>x'> 2^{-}} \frac{d^dx_1 d^dx_2}{\vol(\SO(1,1))} K_{\De_i(J),J}(x_1,x_2;x,z)\<0|\f(x_1)\O^{(0)}_{i,J}(x',z')\f(x_2)|0\>\nn\\
	&\quad-2\int_{x\approx 2,1^-\atop 2>x'> 1^{-}} \frac{d^dx_1 d^dx_2}{\vol(\SO(1,1))} K_{\De_i(J),J}(x_1,x_2;x,z)\<0|\f(x_2)\O^{(0)}_{i,J}(x',z')\f(x_1)|0\>.
\ee
The second term has a form which is very similar to~\eqref{eq:Kdefinition}. We will now show that the first term is in fact proportional to the second term.

First of all, consider the integral in first term in~\eqref{eq:twopttwoterms}. We claim that 
\be\label{eq:integral1claim}
	\int_{x\approx 1^{-}} {d^dx_1} K_{\De_i(J),J}(x_1,x_2;x,z)\<0|\f(x_1)\O^{(0)}_{i,J}(x',z')\f(x_2)|0\>=0,
\ee
which we will prove later in this section. Using $\int_{x\approx 1^{-}}=\int_{x\approx 1^-\atop 1>x'} +\int_{x\approx 1^-\atop 1^-<x'}$, which follows from a simple consideration of the causal relations, we conclude
\be
	&\int_{x\approx 1^-\atop 1>x'} d^dx_1 K_{\De_i(J),J}(x_1,x_2;x,z)\<0|\f(x_1)\O^{(0)}_{i,J}(x',z')\f(x_2)|0\>\nn\\
	&=-\int_{x\approx 1^-\atop 1^-<x'} d^dx_1 K_{\De_i(J),J}(x_1,x_2;x,z)\<0|\f(x_1)\O^{(0)}_{i,J}(x',z')\f(x_2)|0\>.
\ee
Similarly, we claim that
\be\label{eq:integral2claim}
\int_{x\approx 2} {d^dx_2} K_{\De_i(J),J}(x_1,x_2;x,z)\<0|\f(x_1)\O^{(0)}_{i,J}(x',z')\f(x_2)|0\>=0,
\ee
and therefore
\be
	&\int_{x\approx 2\atop x'>2^-} d^dx_2 K_{\De_i(J),J}(x_1,x_2;x,z)\<0|\f(x_1)\O^{(0)}_{i,J}(x',z')\f(x_2)|0\>\nn\\
	&=-\int_{x\approx 2\atop 2>x'} d^dx_2 K_{\De_i(J),J}(x_1,x_2;x,z)\<0|\f(x_1)\O^{(0)}_{i,J}(x',z')\f(x_2)|0\>.
\ee
Together, these identities imply
\be
&\int_{x\approx 2,1^-\atop 1>x'> 2^{-}} \frac{d^dx_1 d^dx_2}{\vol(\SO(1,1))} K_{\De_i(J),J}(x_1,x_2;x,z)\<0|\f(x_1)\O^{(0)}_{i,J}(x',z')\f(x_2)|0\>\nn\\
&=\int_{x\approx 2,1^-\atop 2>x'> 1^{-}} \frac{d^dx_1 d^dx_2}{\vol(\SO(1,1))} K_{\De_i(J),J}(x_1,x_2;x,z)\<0|\f(x_1)\O^{(0)}_{i,J}(x',z')\f(x_2)|0\>,
\ee
i.e. the only difference between the first and the second term in~\eqref{eq:twopttwoterms} is the order of the operators in the three-point function. So,~\eqref{eq:twopttwoterms} becomes
\be\label{eq:twoptprefinal}
&\<\O^{(0)}_{i,J}(x,z)\O^{(0)}_{i,J}(x',z')\>=-2\int_{x\approx 2,1^-\atop 2>x'> 1^{-}} \frac{d^dx_1 d^dx_2}{\vol(\SO(1,1))} K_{\De_i(J),J}(x_1,x_2;x,z)\nn\\
&\qquad\qquad\x(\<0|\f(x_1)\O^{(0)}_{i,J}(x',z')\f(x_2)|0\>+\<0|\f(x_2)\O^{(0)}_{i,J}(x',z')\f(x_1)|0\>).
\ee

Before proceeding, let us prove~\eqref{eq:integral1claim}. Since the integral is performed over $1^{-}\approx x$ or $1\approx x^+$, we will use coordinates where $x^+$ is at the spatial infinity. In these coordinates,
\be
	\int_{x\approx 1^-}d^dx_1 = \int d^dx_1,
\ee
where the integral in the right-hand side is the ordinary integral over Minkowski space.
Consider now the integrand,
\be
	&K_{\De_i(J),J}(x_1,x_2;x,z)\<0|\f(x_1)\O^{(0)}_{i,J}(x',z')\f(x_2)|0\>\nn\\
	&\sim\<0|\tl\f(x_1)\cO^L(x,z)\tl\f(x_2)|0\>\<0|\f(x_1)\O^{(0)}_{i,J}(x',z')\f(x_2)|0\>,
\ee
where we have replaced $K$ by~\eqref{eq:Ktype} which represents its transformation and analyticity properties. Importantly, in both factors the operator at $x_1$ is acting on the future vacuum, which implies that the integrand is analytic in $x_1$ when $x_1$ is continued to positive Euclidean times, i.e.\ when $\Im x_1^0<0$. We can therefore deform the integration contour to show that the integral is 0. This is legitimate because at large Lorentzian times $x_1^0$ the integrand decays as $(x_1^0)^{-2(d-\De_\f)-2\De_\f}=(x_1^0)^{-2d}$. The proof of~\eqref{eq:integral2claim} is analogous.

Finally, we would like to relate the term 
\be
\<0|\f(x_1)\O^{(0)}_{i,J}(x',z')\f(x_2)|0\>
\ee
to 
\be
\<0|\f(x_2)\O^{(0)}_{i,J}(x',z')\f(x_1)|0\>.
\ee 
These two expressions are in fact one and the same Wightman three-point function, evaluated at different choices of coordinates. Since conformal symmetry fixes this Wightman function up to an OPE coefficient, the relationship between these two expressions can be found by peforming an analytic continuation of explicit three-point functions. We will instead use a more conceptual argument which doesn't rely on explicit expressions for the three-point functions, and is therefore more amenable to generalisations.

Without loss of generality, we can assume $x'\approx x$, otherwise the integration domain in~\eqref{eq:twoptprefinal} is empty. We switch to the coordinates in which $x$ is at spacelike infinity, in which case $x_2$, $x'$ and $x_1^-$ can be viewed as standard coordinates in the Minkowski space. We use translation invariance to set $x'=0$, choose $A=x'=0$ and $B=x=\oo$, and consider the CRT transformation $J_{AB}$ (see appendix~\ref{app:CRT}). We have
\be
	\<0|\f(x_1)\O^{(0)}_{i,J}(0,z')\f(x_2)|0\>&=[\<0|J_{AB}\f(x_1)\O^{(0)}_{i,J}(0,z')\f(x_2)J_{AB}^{-1}|0\>]^\dagger\nn\\
	&=[\<0|\f(R_{AB}(x_1))^\dagger \O^{(0)}_{i,J}(0,z')^\dagger \f(R_{AB}(x_2))^\dagger |0\>]^\dagger\nn\\
	&=\<0|\f(R_{AB}(x_2))\O^{(0)}_{i,J}(0,z') \f(R_{AB}(x_1))|0\>,
\ee
where $R_{AB}$ describes how $J_{AB}$ acts on the spacetime points, and we have used that $\O^{(0)}_{i,J}(0,z')$ is an even-spin light-ray operator. For details see appendix~\ref{app:CRT}.
In appendix~\ref{app:ipiD} we show that\footnote{This is true for the causal configuration here, but not in all causal configurations.}
\be
	\f(R_{AB}(x_1))|0\>&=e^{-i\pi D}\f(x_1)|0\>,\\
	\<0|\f(R_{AB}(x_2))&=\<0|\f(x_2)e^{i\pi D},
\ee
where $D$ is the (anti-Hermitian) dilatation operator in Minkowski space, and so
\be
\<0|\f(x_1)\O^{(0)}_{i,J}(0,z')\f(x_2)|0\>&=\<0|\f(x_2)e^{i\pi D}\O^{(0)}_{i,J}(0,z')e^{-i\pi D}\f(x_1)|0\>,\nn\\
&=e^{i\pi(1-J)}\<0|\f(x_2)\O^{(0)}_{i,J}(0,z')\f(x_1)|0\>,
\ee
where we used the fact that $\O^{(0)}_{i,J}(0,z')$ has scaling dimension $1-J$. Restoring translation invariance, we get
\be
	\<0|\f(x_1)\O^{(0)}_{i,J}(x',z')\f(x_2)|0\>&=e^{i\pi(1-J)}\<0|\f(x_2)\O^{(0)}_{i,J}(x',z')\f(x_1)|0\>.
\ee
A key point is that this result can be obtained by using CRT symmetry, and therefore generalises straightforwardly to more general OPEs and to odd-spin trajectories.

Plugging this into~\eqref{eq:twoptprefinal} we find
\be
&\<\O^{(0)}_{i,J}(x,z)\O^{(0)}_{i,J}(x',z')\>\\\nn
&=2(e^{-i\pi J}-1)\int_{x\approx 2,1^-\atop 2>x'> 1^{-}} \frac{d^dx_1 d^dx_2}{\vol(\SO(1,1))} K_{\De_i(J),J}(x_1,x_2;x,z)\<0|\f(x_2)\O^{(0)}_{i,J}(x',z')\f(x_1)|0\>.
\ee
Using~\eqref{eq:rFromCanonicalO} we find
\be
&\<\O^{(0)}_{i,J}(x,z)\O^{(0)}_{i,J}(x',z')\>\\\nn
&=2(1-e^{-i\pi J})r_i(J)\int_{x\approx 2,1^-\atop 2>x'> 1^{-}} \frac{d^dx_1 d^dx_2}{\vol(\SO(1,1))}  K_{\De,J}(x_1,x_2;x,z)\<0|\f(x_2)\wL[\cO](x',z')\f(x_1)|0\>_0.
\ee
The definition of the light-ray kernel~\eqref{eq:Kdefinition} simplifies this to
\be\label{eq:twoptCanonicalFinal}
\<\O^{(0)}_{i,J}(x,z)\O^{(0)}_{i,J}(x',z')\>=r_i(J)\frac{1-e^{-i\pi J}}{i\pi}\vol(\SO(1,1))\<\wL[\cO](x,z)\wL[\cO](x',z')\>_0.
\ee
Using this and~\eqref{eq:rFromCanonicalO} we verify the equation
\be
	r_i(J)=\frac{1-e^{-i\pi J}}{i\pi}\p{\frac{\<0|\f\O_{i,J}^{(0)}\f|0\>}{\<0|\f\wL[\cO]\f|0\>_0}}^2\frac{\vol(\SO(1,1))\<\wL[\cO]\wL[\cO]\>_0}{\<\O^{(0)}_{i,J}\O^{(0)}_{i,J}\>}.
\ee
Since this equation is independent of the normalisation of $\O_{i,J}^{(0)}$, we are free to remove the superscript ${}^{(0)}$,
\be\label{eq:rJfinalFromDerivation}
r_i(J)=\frac{1-e^{-i\pi J}}{i\pi}\p{\frac{\<0|\f\O_{i,J}\f|0\>}{\<0|\f\wL[\cO]\f|0\>_0}}^2\frac{\vol(\SO(1,1))\<\wL[\cO]\wL[\cO]\>_0}{\<\O_{i,J}\O_{i,J}\>},
\ee
which is precisely the advertised result. Note that $\frac{1-e^{-i\pi J}}{i\pi}=\frac2\pi e^{-\pi J i/2}\sin(\frac{\pi J}2)$, which vanishes linearly for even integer $J$.

\subsection{Two-point functions of light-ray operators}
\label{sec:twoptgeneral}

In this section we study two-point functions of light-ray operators. In the previous section we have derived equation~\eqref{eq:twoptCanonicalFinal}, which shows that two-point functions of light-ray operators are generally not finite and instead are proportional to $\vol(\SO(1,1))$, the finite quantity being
\be
	\frac{\<\O_{i,J}(x,z)\O_{i,J}(x',z')\>}{\vol(\SO(1,1))}.
\ee
This is to be expected, at least for a class of light-ray operators which can be defined in perturbation theory as integrals of fundamental fields along the null line spanned by $z$, starting at $x$. All explicit examples of light-ray operators that we study in this paper will be from this class.

To see this, note that the stability group of $x,x',z,z'$ (which preserves $z,z'$ up to rescaling) is $\SO(1,1)^2\x \SO(d-2)$, which has two non-compact generators. The first non-compact generator is a dilatation-like transformation that preserves $x$ and $x'$. Infinitesimally near $x$ it is a rescaling, and near $x'$ it is the inverse rescaling. The second non-compact generator is a boost-like transformation preserving $x$ and $x'$ and rescaling $z$ and $z'$ by reciprocal amounts. It is easy to see that one linear combination of these generators acts trivially on the null lines spanned by $z$ and $z'$, while another acts non-trivially.
The latter leads to a zero-mode in the integrals which compute the two-point function, leading to
\be
\<\O_{i,J}(x,z)\O_{i,J}(x',z')\>\propto \vol(\SO(1,1)).
\ee
We will see this explicitly in the examples below.

A case requiring special treatment is when $\O_{i,J}$ reduces to a light-transform of a local operator, $\O_{i,J}=\cN_{i,J}\wL[\cO]$. The above argument still holds, but the coefficient multiplying the infinite volume vanishes. Formally, we can write for generic $J$
\be\label{eq:CiJdefinition}
	\<\O_{i,J}(x,z)\O_{i,J}(x',z')\>=C_i(J)\vol(\SO(1,1))\<\wL[\cO](x,z)\wL[\cO](x',z')\>_0
\ee
for some function $C_i(J)$. This function then vanishes for $J\in \cJ_i$, 
\be
	C_i(J)=0,\quad J\in \cJ_i,
\ee
since in that case we have
\be\label{eq:OOasLOLO}
	\<\O_{i,J}(x,z)\O_{i,J}(x',z')\>=\cN_{i,J}^2\<\wL[\cO](x,z)\wL[\cO](x',z')\>=0\x \vol(\SO(1,1))+\text{finite}.
\ee
One way to see this is by simply computing the double light-transform of the two-point function $\<\cO\cO\>$ explicitly, which was done in~\cite{Kravchuk:2018htv}. The result of this computation is finite and is discussed in section~\ref{sec:LRcanonical} above. Another way is to note to compute the coefficient of $\vol(\SO(1,1))$ we need to mod out by $\SO(1,1)$ in the integral defining the double light-transform. Since each light-transform is a one-dimensional integral, this effectively eliminates one of the integrations, say in $\wL[\cO](x,z)$. This shows that $C_i(J)$ is proportional to a sum of terms of the form $\<\cO(x_0,z)\wL[\cO](x,z)\>$, where $x_0$ are representative points from orbits under the $\SO(1,1)$ action. However, all such terms vanish since for a fixed $x_0$ the time-ordered two-point function reduces to some Wightman-two point function and $\wL[\cO](x,z)|0\>=\<0|\wL[\cO](x,z)=0$.

We can also show that $C_i(J)$ has only simple zeroes when $J\in\cJ_i$, by deriving a simple formula for $\ptl_J C_i(J)$ at these points. Note that we can rewrite~\eqref{eq:rJfinalFromDerivation} as
\be
\frac{\<\O_{i,J}\O_{i,J}\>}{\vol(\SO(1,1))}=\frac{1-e^{-i\pi J}}{i\pi r_i(J)}\p{\frac{\<0|\f\O_{i,J}\f|0\>}{\<0|\f\wL[\cO]\f|0\>_0}}^2{\<\wL[\cO]\wL[\cO]\>_0}.
\ee
Computing $J$ derivative at $J\in\cJ_i\subseteq 2\Z$ and using $\O_{i,J}=\cN_{i,J}\wL[\cO]$
\be
\frac{\ptl_J \<\O_{i,J}\O_{i,J}\>}{\vol(\SO(1,1))}=\frac{1}{r_i(J)}\cN^2_{i,J}\p{\frac{\<0|\f\wL[\cO]\f|0\>}{\<0|\f\wL[\cO]\f|0\>_0}}^2{\<\wL[\cO]\wL[\cO]\>_0}.
\ee
Using the fact that $\<0|\f\wL[\cO]\f|0\>=\l_{\f\f\cO}\<0|\f\wL[\cO]\f|0\>_0$, $r_i(J)=-\lambda_{\f\f\cO}^2$, equation~\eqref{eq:OOasLOLO}, and the fact that $\cO$ is canonically normalised, i.e. $\<\wL[\cO]\wL[\cO]\>=\<\wL[\cO]\wL[\cO]\>_0$, we get
\be
\frac{\ptl_J \<\O_{i,J}\O_{i,J}\>}{\vol(\SO(1,1))}=-\<\O_{i,J}\O_{i,J}\>,\qquad J\in\cJ_{i}.
\ee
As discussed above, the right-hand side is finite. Using~\eqref{eq:CiJdefinition} we can rewrite this as
\be
	\ptl_J C_i(J)=-\cN_{i,J}^2,\qquad J\in\cJ_{i}.\label{eq:CiJderivative}
\ee
Since we assume that $\cN_{i,J}\neq 0$, this shows that $C_i(J)$ has a simple $0$ at such points. For the simplest choices of conventions, it is often the case that $\cN_{i,J}$ is real, and so $\ptl_J C_i(J)<0$. If $C_i(J)$ was real, this would imply that $C_i(J)$ should also have zeroes in between the points of $\cJ_i$. In examples we see that $C_i(J)$ is complex and such zeroes are avoided.

\section{Wilson--Fisher theory}
\label{sec:WF}

In this section we study explicit examples of light-ray operators in Wilson--Fisher theory and verify the general predictions derived in the previous section. We will first consider the case of twist-2 operators, where all calculations can be done analytically. Then we will switch to the much more non-trivial case of twist-4 operators, where we will have to resort to numerical methods.

The Wilson--Fisher theory is defined by the Lagrangian
\be
	\cL=-\tfrac{1}{2}(\ptl\f)^2-\tfrac{1}{4!}g\f^4
\ee
in $d=4-\e$ dimensions and has a $\Z_2$ symmetry $\f\to -\f$. The $\b$-function for $g$ is non-trivial and has a fixed point at $g=O(\e)$, the properties of which can be systematically studied in $\e$-expansion~\cite{Wilson:1971dc,Wilson:1973jj,Kleinert:2001hn}. As $\e\to 1$ this fixed point is believed to coincide with the 3d Ising CFT and at $\e\to 2$ with the 2d Ising CFT.

We will need the tree-level (free-theory) two-point function of the field $\f$, which is
\be
	\<\f(x_1)\f(x_2)\>=\frac{1}{(d-2)2\pi^{d/2}\G(d/2)}\frac{1}{x_{12}^{d-2}},
\ee
reducing to
\be
	\<\f(x_1)\f(x_2)\>=\frac{1}{4\pi^{2}}\frac{1}{x_{12}^{2}}
\ee
for $\e=0$ ($d=4$). Note that in our conventions, $\phi$ is the fundamental field in the Lagrangian, and therefore its two-point function is not canonically normalised.

\subsection{Local operators in perturbation theory}
\label{sec:localops}

Local operators in free theory ($\l=0$) can be built out of normal-ordered products of derivatives of $\f(x)$, e.g.,
\be
	:\!\f(x)\ptl_{\mu_1}\cdots \ptl_{\mu_J}\f(x)\!:.
\ee
This local operator is not a primary, and appropriate total derivatives need to be added to construct a primary. Restricting to two fields $\phi$, there is at each even spin $J$ a unique primary operator which takes the form
\be
	\cO_J^{\tau=2}(x,z)&=\frac{2\sqrt{2}\pi^2}{\sqrt{(2J)!}}\sum_{k=0}^{J}(-1)^k\binom{J}{k}^2:\!(z\.\ptl)^k\f(x)(z\.\ptl)^{J-k}\f(x)\!:\label{eq:OJdefn0}\\
	&=\frac{2\sqrt{2}\pi^{3/2} 4^J\G(J+\thalf)}{\sqrt{(2J)!} \G(J+1)}\p{:\!\f(x)(z\.\ptl)^J\f(x)\!:+(z\.\ptl)(\cdots)}\label{eq:OJdefn},
\ee
where the coefficients are valid to the leading order in $\e$ and are chosen so that the operator is a canonically-normalised primary, i.e. its two-point function has the form~\eqref{eq:standard2pt}. In $d=4$ ($\e=0$) these operators have scaling dimension $\De_{\cO_J^{\tau=2}}=2+J$ and twist $\tau=\De_{\cO_J^{\tau=2}}-J=2$. For this reason, we will refer to them as twist-2 operators.

Using more $\f$ insertions one can construct primary operators of higher twist. For $\Z_2$-even operators the next twist is $4$, and the non-trivial traceless-symmetric tensor operators in the free theory take the form
\be\label{eq:localconstruction}
	\cO(x,z)=\sum_{k_1,\cdots, k_4\geq 0\atop k_1+\cdots +k_4=J} c_{k_1,k_2,k_3,k_4}:\!(z\.\ptl)^{k_1}\f(x)(z\.\ptl)^{k_2}\f(x)(z\.\ptl)^{k_3}\f(x)(z\.\ptl)^{k_4}\f(x)\!:\,,
\ee
where the coefficients $c_{k_1,k_2,k_3,k_4}$ are chosen so as to form a primary operator. 
Notice that at one-loop level, which is what we consider in this paper, one can ignore the mixing with terms constructed using the Laplacian, $\square \phi(x)$.

\begin{figure}
  \centering
\includegraphics[width=120mm]{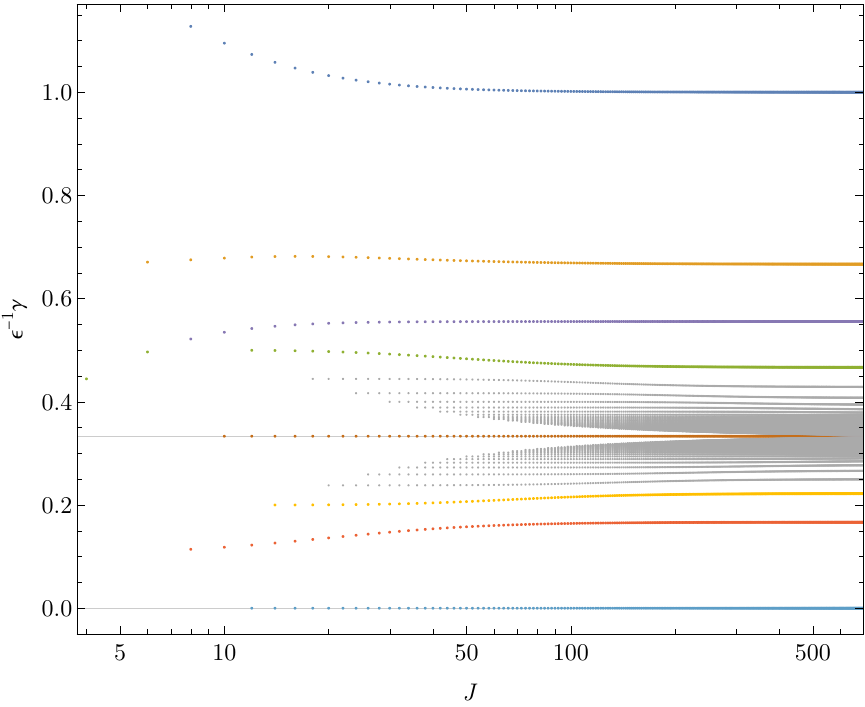}
  \caption{Perturbative spectrum of local twist-4 operators of even spin up to 700. Note that the first three operators in the top family are outside the scale.}\label{fig:twist-4-infty}
\end{figure}

The local one-loop dilatation operator of Wilson--Fisher theory has been determined completely~\cite{Kehrein:1992fn}.\footnote{See also~\cite{Kehrein:1994ff,Kehrein:1995ia,Hogervorst:2015akt,Liendo:2017wsn,Henriksson:2022rnm} for subsequent work using the one-loop dilatation operator.} 
At twist 4, one finds eigenvalues that are roots of polynomials of a degree which grows with spin $J$.
Comprehensive results from such studies were presented already in~\cite{Kehrein:1994ff}, and it is not difficult to determine the spectrum of local twist-4 operators up to some moderately large spin, like spin 30 of figure~\ref{fig:trajectories}. One bottleneck is that at each level the space of operators contains not only the primary operators, but also descendants from primaries of lower spin. 
As we will see below (section~\ref{sec:largespinlocal}), the use of lightray operators will reduce the space to include primaries only, and the computation of anomalous dimensions can be pushed much higher in spin (equivalent methods have been discussed in~\cite{Derkachov:2010zza}\footnote{We thank Jeremy Mann for bringing this work to our attention.}). For example, in figure~\ref{fig:twist-4-infty} we show the anomalous dimensions for even-spin twist-4 operators of spins up to 700.

\subsection{Light-ray operators in perturbation theory}
\label{sec:LRpertth}

In the free theory, a class\footnote{There are Regge trajectories for which the light-ray operators take a different form, see~\cite{Kravchuk:2018htv, Caron-Huot:2022eqs} for a discussion.} of light-ray operators can be constructed as null integrals of polynomials in $\f$,
\be
	\O_\psi(x,z)=\int_{-\oo}^{+\oo}d\a_1\cdots d\a_n& (-\a_1)^{-\De_\f} \cdots (-\a_n)^{-\De_\f}\psi(\a_1,\cdots,\a_n)\nn\\
	&\x:\!\f(x-z/\a_1)\cdots \f(x-z/\a_n)\!:.
\ee
Here, $\psi$ is a wave-function which generalises the coefficients $c_{k_1,\cdots,k_n}$ which enter the expression~\eqref{eq:localconstruction} for local operators. In order for this operator to have well-defined matrix elements, $\psi$ needs to be a well-defined distribution. We describe its other properties below. The $\a_i$ integrals have the same interpretation as in the definition of the light transform in~\eqref{eq:Ldefn}. The conformal transformation properties of this operator are more transparent when written in embedding space~\cite{Kravchuk:2018htv},
\be\label{eq:Opsidefn}
	\O_{\psi}(X,Z) = \int_{-\infty}^{-\infty}d\alpha_1\cdots d\alpha_n \psi(\alpha_1,\cdots,\alpha_n):\!\f(Z-\alpha_1X)\cdots\f(Z-\alpha_nX)\!:.
\ee
If we require this to be a primary with scaling dimension $\De_L$ and spin $J_L$, the embedding space rules~\cite{Costa:2011mg} require that for all $\l,\mu>0$ and $\nu\in\R$
\be
	\O_\psi(\l X,\mu Z+\nu X)=\l^{-\De_L}\mu^{J_L}\O_\psi(X,Z).
\ee
This requires, for $\lambda > 0$,
\be
	\psi (\lambda \alpha_1, \cdots, \lambda \alpha_n) &= \lambda^{\eta}\psi(\alpha_1,\cdots,\alpha_n),\label{eq:psihom}\\
	\psi(\alpha_1 + \lambda,\cdots, \alpha_n + \lambda) &= \psi(\alpha_1,\cdots,\alpha_n),\label{eq:psishift}
\ee
for some $\eta\in\C$, where
\be
		\De_L &= n+\eta ,\\
		J_L &= n(1-\Delta_{\phi}) +\eta.\label{eq:JLfrometa}
\ee
In terms of the quantum numbers $\De=1-J_L$ and $J=1-\De_L$ this is
\be
		\De &= n\Delta_{\phi} + 1  - n -\eta,  \\
		J &= 1 - n - \eta.\label{eq:Jfrometa}
\ee
It follows that all these operators have fixed tree-level twist $\tau = \De-J=n\Delta_{\phi}=n+O(\e)$. We therefore expect the $n=2,4$ light-ray operators to be related to the twist-2 and twist-4 local operators discussed in the previous section. 

In addition to~\eqref{eq:psihom} and~\eqref{eq:psishift}, $\psi$ has to be permutation-invariant,
\be
		\psi(\alpha_{\sigma(1)},\ldots,\alpha_{\sigma(n)}) = \psi(\alpha_1,\ldots,\alpha_n),\qquad \forall \s\in S_n.
\ee
Finally, the spin parity (or ``signature'') constrains $\psi$ to satisfy
\be
	\psi(-\a_1,\ldots,-\a_n)=\pm \psi(\a_1,\ldots,\a_n),
\ee
where the ``$+$'' sign is for even-spin trajectories and the ``$-$'' sign is for the odd-spin trajectories. See appendix~\ref{app:CRT} for details.

\subsection{Two-point functions and matrix elements of twist-2 operators}

In this section we study the relationship between the twist-2 local operators $\cO_J^{\tau=2}$ defined in section~\ref{sec:localops} and the light-ray operators defined in the section~\ref{sec:LRpertth}. We will work in the free theory at $d=4$ since the effects we are interested in can already be seen at this order. 

For twist-2 light-ray operators, an important consequence of the properties of the wavefunction $\psi$  discussed in the previous section is that it is uniquely fixed up to normalisation,
\be
\psi_J(\a_1,\a_2)=\frac{|\a_1-\a_2|^{-1-J}}{\G(-\tfrac{J+2}{2})},\qquad J\in\C,
\ee
and only exists for even spin parity. The $\G$-function in the denominator is necessary to get a well-defined distribution at $J\in \{0,2,4,\ldots\}$. Indeed, we have
\be
\psi_{J_0}(\a_1,\a_2)=\frac{(-1)^{J_0/2+1}\G(\tfrac{J_0}{2}+2)}{\G(J_0+1)}\de^{(J_0)}(\a_1-\a_2),\qquad J_0\in\cJ=\{0,2,4,\cdots\},
\ee
where $\delta^{(J_0)}(x)$ denotes the $J_0{}^{\text{th}}$ derivative of the Dirac delta function.
In terms of the light-ray operator~\eqref{eq:Opsidefn} this becomes
\be
\O_{\psi_{J_0}}(X,Z)=\frac{(-1)^{J_0/2+1}\G(\tfrac{J_0}{2}+2)}{\G(J_0+1)}\int_{-\oo}^{+\oo}d\a_1  O(Z-\a_1 X,-X),
\ee
where
\be
O(X,Z)=:\!\f(X)(Z\.\ptl_X)^{J_0}\f(X)\!:+(Z\.\ptl_X)\p{\cdots}.
\ee
Here, any expression can go into the parenthesis as it turns into a total $\a_1$-derivative upon insertion into the integral above. Comparing to~\eqref{eq:OJdefn} we find
\be
\O_{\psi_{J_0}}(X,Z)&=\cN_J \int_{-\oo}^{+\oo}d\a_1  \cO_{J_0}(Z-\a_1 X,-X),\\
&=\cN_{J}\wL[\cO_{J_0}](X,Z),
\ee
where
\be
\cN_J=\frac{(-1)^{J/2+1}\sqrt{(2J)!}\G(\tfrac{J}{2}+2)}{2\sqrt 2\pi^{3/2} 4^{J}\G(J+\thalf)}.
\ee
In what follows, we will use the notation $\O_{J}^{\tau=2}\equiv \O_{\psi_J}$, so that we can write
\be
	\O^{\tau=2}_{J}(x,z)=\cN_{J}\wL[\cO_J^{\tau=2}](x,z),\quad J\in \cJ = \{0,2,4,\cdots\},
\ee
which is precisely the setup that we studied in section~\ref{sec:general}.

\subsubsection{Light-ray operator two-point functions}
\label{sec:twist2two-pt}
Let us now compute the two-point function $\<\O^{\tau=2}_J(X,Z)\O^{\tau=2}_J(X',Z')\>$. We can use conformal symmetry to put $X$ at spatial infinity and $X'$ at $0$,\footnote{We write embedding-space vectors as $(X^+,X^-,X^\mu)$ and the embedding-space metric is $X^2=-X^+X^-+X^\mu X_\mu$.
\label{footnote:ESconvention}}
\be
	X=(0,1,0),\quad X'=(1,0,0),
\ee
with polarizations
\be
	Z=(0,0,z),\quad Z'=(0,0,z').
\ee
We can also use the homogeneity in $z,z'$ to set $-2z\.z'=1$. The time-ordered two-point function of $\f$ can be written as
\be
	\<\f(X_1)\f(X_2)\>=\frac{1}{4\pi^2}\frac{1}{-2X_1\.X_2+i\e}.
\ee
Finally, us define $X_\a=Z-\a X$ and $X'_\a=Z'-\a X'$. It then follows that $-2X_\a\.X'_{\a'}=-2z\.z'+\a\a'=1+\a\a'$.

With this preparation, we can now write
\be
	&\<\O^{\tau=2}_J(X,Z)\O^{\tau=2}_J(X',Z')\>\nn\\
	&=2\int d\a_1 d\a_2 d\a'_1 d\a'_2 \psi_J(\a_1,\a_2)\psi_J(\a'_1,\a'_2)\<\f(X_{\a_1})\f(X'_{\a'_1})\>\<\f(X_{\a_2})\f(X'_{\a'_2})\>\nn\\
	&=\frac{1}{8\pi^4}\int d\a_1 d\a_2 d\a'_1 d\a'_2 \frac{\psi_J(\a_1,\a_2)\psi_J(\a'_1,\a'_2)}{(1+\a_1\a_1'+i\e)(1+\a_2\a_2'+i\e)}.\label{eq:integralforapp}
\ee
We compute this integral in appendix~\ref{app:integral}. The result is
\be
	\<\O^{\tau=2}_J(X,Z)\O^{\tau=2}_J(X',Z')\>=\frac{i\G(J+1)\G(2+\tfrac{J}{2})e^{i\pi J/2}}{4^{J+1}\pi^{5/2}\G(-\tfrac{J+2}{2})\G(J+\tfrac{3}{2})}\vol(\SO(1,1)).
\ee
For our choice of $X,X',Z,Z'$ the standard analytic continuation~\eqref{eq:standard2ptL} takes the value
\be
	\<\wL[\cO](X,Z)\wL[\cO](X',Z')\>_0=\frac{-2\pi i}{\De+J-1}=\frac{-2\pi i}{2J+1}.
\ee
We therefore find that the two-point takes the form~\eqref{eq:CiJdefinition}
\be
\<\O^{\tau=2}_{J}(x,z)\O^{\tau=2}_{J}(x',z')\>=C(J)\vol(\SO(1,1))\<\wL[\cO](x,z)\wL[\cO](x',z')\>_0
\ee
with 
\be
	C(J)=-\frac{\G(J+1)\G(2+\tfrac{J}{2})e^{i\pi J/2}}{4^{J+1}\pi^{7/2}\G(-\tfrac{J+2}{2})\G(J+\tfrac{1}{2})}.
\ee

We can explicitly see that $C(J)$, as expected, has simple zeros at $J\in \cJ=\{0,2,4,\ldots\}$. Furthermore, we can explicitly verify that equation~\eqref{eq:CiJderivative} is satisfied,
\be
	\ptl_J C(J)=-\cN_J^2,\quad J\in \cJ.
\ee
Note that $C(J)$ has poles at negative integer values of $J$. We expect that these poles are related to the subtleties in defining $\psi_J$ as a distribution near $\a_i=\pm \oo$. Furthermore, $C(J)$ has zeros at negative half-integral $J$. We similarly expect that these zeroes are related to subtleties in defining~\eqref{eq:standard2ptL} at these values of $J$ (i.e.\ the position-dependent factors in~\eqref{eq:standard2ptL} might have poles in $J$). It would be interesting to explore the global structure of this $C(J)$ in more detail.

\subsubsection{Matrix elements}
\label{sec:twist2matrixelts}

We begin by computing the three-point functions $\<\f\f\cO_J^{\tau=2}\>$ involving the local twist-two operators $\cO^{\tau=2}_J(x,z)$. To do this, it is convenient to place one of the $\f$'s at $\oo$ and the other at $0$, so we focus on
\be
	\<\f(0)\f(\oo)\cO^{\tau=2}_J(x,z)\>.
\ee
The advantage of this is that the contractions of $\f(x)$ with $\f(\oo)$ are $x$-independent and only the $k=0$ and $k=J$ terms in~\eqref{eq:OJdefn} survive. Therefore, in $d=4$,
\be	
	\<\f(0)\f(\oo)\cO^{\tau=2}_J(x,z)\>=\frac{1}{8\pi^4}\frac{2\sqrt 2 \pi^2}{\sqrt{(2J)!}}(z\.\ptl_x)^J x^{-2}=\frac{\sqrt2J!}{4\pi^2\sqrt{(2J)!}}(-2z\.x)^Jx^{-2(J+1)}.
\ee
Comparing to~\eqref{eq:standardstructure} we find\footnote{If we define $\vf=2\pi \f$ so that the two-point function of $\vf$ is canonically normalised, we get $\lambda_{\vf\vf\cO^{\tau=2}_J}^2=\frac{2(J!)^2}{(2J)!}$, in agreement with well-known results (see e.g.~\cite{Dolan:2000ut}).}
\be\label{eq:fffOJ}
	\lambda_{\f\f\cO^{\tau=2}_J}=\frac{\sqrt2J!}{4\pi^2\sqrt{(2J)!}}.
\ee

The computation of the matrix element $\<0|\f(X_1)\O_J^{\tau=2}(X_3,Z)\f(X_2)|0\>$ is most easily performed in the embedding space, where we set $X_2$ to be the spatial infinity and $X_3$ to be at the origin of the Minkowski space
\be
	X_2=(0,1,0),\quad X_3=(1,0,0),\quad Z=(0,0,z).
\ee
In this case,
\be
	-2X_2\.(Z-\a X_3)=-\a.
\ee
Furthermore, if we place $X_1$ in the absolute future of the origin, we have
\be
	\<0|\f(X_1)\O^{\tau=2}_J(X_3,Z)\f(X_2)|0\>=\<\f(X_1)\f(X_2)\O^{\tau=2}_J(X_3,Z)\>
\ee
and we can use time-ordered propagators. This gives
\be\label{eq:twist2matrixeltintermediate}
	\<\f(X_1)\f(X_2)&\O^{\tau=2}_J(X_3,Z)\>=\frac{1}{8\pi^4}\int d\a_1d\a_2 \frac{\psi_J(\a_1,\a_2)}{(-\a_2+i\e)(-2z\.x_1-\a_1x_1^2+i\e)}\nn\\
	&=-\frac{1}{8\pi^4\G(-\frac{J+2}{2})}\int d\a_1d\a_2 \frac{|\a_2|^{-J-1}}{(\a_2+\a_1-i\e)(-2z\.x_1-\a_1x_1^2+i\e)}\nn\\
	&=-\frac{i}{4\pi^3(1-e^{i\pi J})\G(-\frac{J+2}{2})}\int d\a_1\frac{(\a_1-i\e)^{-J-1}}{(-x_1^2)(2z\.x_1x_1^{-2}+\a_1+i\e)}\nn\\
	&=\frac{e^{i\pi J}}{2\pi^2(1-e^{i\pi J})\G(-\frac{J+2}{2})}(-x_1^2)^{J}(-2z\.x_1)^{-J-1}\nn\\
	&=\frac{ie^{i\pi J/2}\G(2+\frac{J}{2})}{4\pi^3}(-x_1^2)^{J}(-2z\.x_1)^{-J-1}.
\ee
Comparing to~\eqref{eq:standard3ptL} we find
\be
	\<0|\f\O^{\tau=2}_J\f|0\>=-\frac{e^{i\pi J/2}\G(2+\frac{J}{2})\G(J+1)^2}{8\pi^4\G(2J+1)}\<0|\f\wL[\cO]\f|0\>_0.
\ee

Using~\eqref{eq:rJfinalFromDerivation} we can now compute the residues $r(J)$,
\be
	r(J)=\frac{1-e^{-i\pi J}}{i\pi}\frac{e^{i\pi J}\G(2+\frac{J}{2})^2\G(J+1)^4}{64\pi^8\G(2J+1)^2}C(J)^{-1}=-\frac{\G(J+1)^2}{8\pi^4\G(2J+1)},
\ee
which indeed agrees (see~\eqref{eq:risminusff}) with $-\lambda_{\f\f\cO_J^{\tau=2}}^2$ given by~\eqref{eq:fffOJ} at $J\in \cJ=\{0,2,4,\cdots\}$.

\subsection{Regge trajectories of twist-4 operators}
\label{sec:twist4regges}

We now turn to the case of twist-4 operators. As discussed in the introduction, twist-4 Regge trajectories are infinitely-degenerate. For example, if we ignore regularity issues, any wavefunction of the form
\be
	\psi(\a_1,\a_2,\a_3,\a_4)=Sym\left[|\a_1-\a_2|^{-J-3}g\p{\frac{\a_1-\a_3}{\a_1-\a_2},\frac{\a_1-\a_3}{\a_1-\a_2}}\right],
\ee
where $Sym$ denotes symmetrization under $S_4$ permutations and $g$ is a generic function, satisfies the properties described in section~\ref{sec:LRpertth}.

Our first goal therefore is to find the wavefunctions $\psi$ which correspond to the physical Regge trajectories. To do so, we need to study and diagonalise the dilatation operator. This will break the degeneracy and allow us to study individual trajectories. Of course, in practice we have to limit ourselves to a finite number of loops, and in this work we focus on those trajectories for which the degeneracy is broken by the 1-loop dilatation operator.

Let us make a few comments about the computation of anomalous dimensions of light-ray operators, following~\cite{Caron-Huot:2022eqs}. We will study the light-ray operators inserted at spatial infinity,
\be
	\O_\psi(\oo, z) = \int d\alpha_1\cdots d\alpha_n\psi(\alpha_1,\cdots,\alpha_n)\phi(\alpha_1;z)\cdots\phi(\alpha_n;z),\label{eq:detectorgeneraln}
\ee
where
\begin{equation}
	\phi(\alpha;z) = \lim_{L \rightarrow \infty} L^{\Delta_{\phi}}\phi(x+Lz), \hspace{0.5cm} \alpha = -2x\cdot z.
\end{equation}
Thanks to conformal symmetry, this is without a loss of generality. However, since conformal symmetry is not manifest in perturbation theory, this choice does affect the computation of anomalous dimensions. Only the translations and Lorentz transformations are manifest in perturbation theory. This allows us to impose that $\O_\psi$ is primary\footnote{If we consider $\O_\psi(0,z)$, then the primary condition is $[K_\mu,\O_\psi(0,z)]=0$. Applying inversion, we get $[P_\mu,\O_\psi(\oo,z)]=0$, which only requires translations.} and to specify its Lorentz spin $J_L$ as \textit{exact} conditions. Our perturbative calculations will then give corrections to the eigenvalue of $D$, which is $\De_L$. The unusual feature is that, in terms of the local quantum numbers, we have $\De_L=1-J$ and $J_L=1-\De$. So, in effect we are fixing $\De$ and computing corrections to $J$.

In order to describe the one-loop dilatation operator it is convenient to go to Fourier space,
\be\label{eq:fourierdefn}
4\pi \de(\b_1+\cdots+\b_4)\tl \psi(\b_1,\cdots,\b_n)=\int d\a_1\cdots d\a_n e^{-\frac{i}{2}\a_1\b_1-\cdots -\frac{i}{2}\a_n\b_n}\psi(\a_1,\cdots,\a_n).
\ee
We have factored out the explicit momentum-conserving $\de$-function, and so $\tl\psi(\b_1,\cdots,\b_n)$ is only defined when $\sum_i \b_i=0$. The other properties of this wavefunction mirror those of $\psi$: it is permutation-invariant and
\be
	\tl\psi(\l\b_1,\cdots,\l\b_n)&=\l^{1-n-\eta}\tl\psi(\b_1,\cdots,\b_n)=\l^{1-J_L-n(1-\e/2)}\tl\psi(\b_1,\cdots,\b_n),\label{eq:psitlhomogen}\nn\\
	\tl\psi(-\b_1,\cdots,-\b_n)&=\pm \tl\psi(\b_1,\cdots,\b_n),
\ee
where we used~\eqref{eq:Jfrometa}. We denote the vector space of such wavefunctions by~$\cV_{1-J_L-n(1-\e/2)}^\pm$, where the subscript describes the degree of homogeneity and superscript the spin parity.

We define the action of the dilatation operator on the wavefunctions via
\be
	[D,\O_\psi(\oo,z)]=\O_{D\psi}(\oo,z).
\ee
In terms of the Fourier transform the one-loop action is\footnote{The same form of the dilatation operator appears in~\cite{Derkachov:1995zr}, albeit applied to polynomials only.}
\be\label{eq:oneloopdil}
	&(D\tl\psi)(\b_1,\cdots,\b_n)=(-J_L-n(1-\tfrac{\e}{2}))\tl\psi(\b_1,\cdots,\b_n)\nn\\
	&-\frac{\e}{3}\sum_{i<j}\int_0^1 dt \tl\psi(\b_1,\cdots,\b_{i-1},t(\b_i+\b_j),\b_{i+1},\cdots,\b_{j-1},(1-t)(\b_i+\b_j),\b_{j+1},\cdots,\b_n)\nn\\
	&+O(\e^2).
\ee
This result follows directly from the analysis in~\cite{Derkachov:2010zza}. There, the renormalisation of a product of $\f$ operators on a null line (i.e.\ of the integrand in~\eqref{eq:Opsidefn}) was studied. Simply integrating their non-local operator against $\psi$ and performing Fourier transform yields the dilatation operator quoted above (see in particular eq.~(4.31) in~\cite{Derkachov:2010zza}). This argument suffices at one loop; at higher orders we expect that an analysis along the lines of~\cite{Caron-Huot:2022eqs} is more convenient since the condition of being a primary can be imposed exactly. We have also verified~\eqref{eq:oneloopdil} using methods of~\cite{Caron-Huot:2022eqs}.

According to our definitions, the eigenvalue of $D$ on $\cV_{1-J_L-n(1-\e/2)}^\pm$ gives\footnote{The minus sign is due to $\O_\psi$ being at the spatial infinity.} $-\De_L$ for a given value of $J_L$. This gives
\be
	\De_L=J_L+n(1-\tfrac{\e}{2})+\g_L(J_L)+O(\e^2),
\ee
where $\g(J_L)=O(\e)$ is \textit{minus} the eigenvalue of the operator in the second line of~\eqref{eq:oneloopdil}. In terms of $\De,J$ we have
\be
	\De&=J+n(1-\tfrac{\e}{2})+\g_L(1-\De)+O(\e^2)=J+n(1-\tfrac{\e}{2})+\g_L(1-J-n)+O(\e^2)\nn\\
	&=J+n(1-\tfrac{\e}{2})+\g(J).
\ee
This shows that by neglecting higher-order terms we can describe the ``usual'' one-loop anomalous dimension $\g(J)=\g_L(1-J-n)+O(\e^2)$ as the eigenvalue of the operator $H$ defined as
\be
	&(H\tl\psi)(\b_1,\cdots,\b_n)\label{eq:Dprime}\\
	&=\frac{\e}{3}\sum_{i<j}\int_0^1 dt \tl\psi(\b_1,\cdots,\b_{i-1},t(\b_i+\b_j),\b_{i+1},\cdots,\b_{j-1},(1-t)(\b_i+\b_j),\b_{j+1},\cdots,\b_n)\nn
\ee
on the space $\cV_{J}^\pm$.

\subsubsection{Simplifying the dilatation operator}

\label{eq:simplifyingDilatationOperator}

We now consider the problem of diagonalising $H$ in the space $\cV_J$ for $n=4$.  First, we observe that for $J\in 2\Z_{\geq 0}$ $\cV_J^+$ contains  homogeneous symmetric polynomials. For example, 
\be
	\b_1^2+\b_2^2+\b_3^2+\b_4^2\in \cV_2^+.
\ee
Similarly, $\cV_J^-$ contains such polynomials for $J\in 2\Z_{\geq 0}+1$, for instance
\be
	\b_1^3+\b_2^3+\b_3^3+\b_4^3\in\cV_3^-.
\ee
Such polynomial wavefunctions $\tl\psi$ correspond to Fourier transforms $\psi$ which are products of Dirac $\de$-functions. Comparing with~\eqref{eq:Opsidefn}, it is easy to see that these wavefunctions correspond to light transforms of local operators. It is also clear from~\eqref{eq:Dprime} that $H$ maps polynomials to polynomials. The polynomial subspace of $\cV^\pm_J$ is finite-dimensional and the spectrum of $H$ on it reproduces the spectrum of one-loop anomalous dimensions of local operators.

Note that if $\tl\psi'$ is an eigenfunction of $H$,
\be\label{eq:eigenstateeqlargespace}
	H\tl\psi = \g \tl\psi,
\ee
then, unless $\g=0$, we can conclude that $\tl\psi$ is in the image of $H$. Indeed, for $\g\neq 0$ we can write $\tl\psi = H(\g^{-1}\tl\psi)$. This image is relatively small compared to $\cV_J^\pm$: it consists only of the wavefunctions which can be written as
\be\label{eq:fansatz}
	\tl \psi(\b_1,\cdots,\b_4)=\sum_{i<j} \Psi(\b_i,\b_j)
\ee
for some function $\Psi$. Indeed, the right-hand side of~\eqref{eq:Dprime} is a sum of terms like
\be
	&\int_0^1 dt\tl\psi(\b_1,t(\b_2+\b_3),(1-t)(\b_2+\b_3),\b_4)\nn\\
	&=\int_0^1 dt\tl\psi(\b_1,-t(\b_1+\b_4),-(1-t)(\b_1+\b_4),\b_4)\propto \Psi(\b_1,\b_4).
\ee
Since we are mainly interested in $H$ as a tool for breaking the degeneracy between light-ray operators, we will look for non-zero eigenvalues $\g$.  This allows us to make the ansatz~\eqref{eq:fansatz}, where the wavefunction $\Psi(p,q)$ satisfies 
\be\label{eq:Psibegin}
\Psi(A p,A q)&=A^J \Psi(p,q),\quad A>0,\\
\Psi(p,q)&=\Psi(q,p),\\
\Psi(p,q)&=\pm \Psi(-p,-q),
\label{eq:Psiend}
\ee
with $\pm$ is determined by the spin parity as usual.

The translation of the eigenstate equation~\eqref{eq:eigenstateeqlargespace} from the space of $\tl\psi$'s to the space of $\Psi$'s, however, requires certain care. This is because~\eqref{eq:fansatz} does not uniquely determine the function $\Psi$ in terms of $\tl\psi$. As we show in appendix~\ref{app:redundant}, there is a unique $\Psi(p,q)=\Psi_0(p,q)$ which maps to $\tl\psi=0$, and it is given by
\be\label{eq:kernel}
	\Psi_0(p,q)=\begin{cases}
		|p|^J(1+\tfrac{3q}{p})+|q|^J(1+\tfrac{3p}{q}), &\quad \text{even spin parity},\\
		(p+q)|p+q|^{J-1}, &\quad \text{odd spin parity}.
	\end{cases}
\ee
Let $\bK$ be the vector space spanned by $\Psi_0$. Then $H$ lifts to an operator $H'$ on the space of $\Psi$'s which is only well-defined on equivalence classes modulo $\bK$, and the equation~\eqref{eq:eigenstateeqlargespace} becomes
\be\label{eq:eigenstateeqsmallspace}
	H'\Psi=\g \Psi+\bK.
\ee
Explicitly, we can take
\be\label{eq:Dptwovar}
	(H'\Psi)(p,q)=&\frac{\e}{3}\Psi(p,q)+\frac{\e}{3}\int_0^1 dt\p{
		\Psi(t(p+q),(1-t)(p+q))}\nn\\
	&+\frac{2\e}{3}\int_0^1 dt\p{\Psi(p,-t(p+q))+\Psi(q,-t(p+q))
	}.
\ee
In principle, we could modify $H'$ by adding $\w(\Psi)\Psi_0(p,q)$ to the right-hand side above, where $\w$ is any linear functional. However, the above choice turns out to be convenient because it satisfies $H'\bK=0$ (in other words, $H'\Psi_0=0$).  This allows us to drop $\bK$ in the right-hand side of~\eqref{eq:eigenstateeqsmallspace}.

To see this, note that if we have a solution of~\eqref{eq:eigenstateeqsmallspace} with $\g\neq 0$, then by applying $\g^{-1}H'$ to~\eqref{eq:eigenstateeqsmallspace} and using $H'\bK=0$ we find
\be
	H'\Psi'=\g \Psi',
\ee
where $\Psi'=\g^{-1}H'\Psi$. Equation~\eqref{eq:eigenstateeqsmallspace} implies $\Psi'\in \Psi+\bK$, so via~\eqref{eq:fansatz} $\Psi'$ corresponds to the same wavefunction $\tl\psi$ as $\Psi$. This shows that for every eigenfunction $\tl\psi$ of $H$ with $\g\neq 0$ there exists an eigenfunction $\Psi'$ of $H'$, related to $\tl\psi$ via~\eqref{eq:fansatz}, and with the same eigenvalue $\g$ and vice versa. The converse is true also for $\g=0$ as long as $\Psi'\not\in \bK$.

This means that by studying the spectral problem for $H'$ we will find all non-zero eigenvalues of $H$, and perhaps some zero eigenvalues. In what follows we will use the notation $\Psi$ for eigenstates of $H'$,
\be
	H'\Psi=\g \Psi.
\ee

As an example, let us consider the case of light transforms of local operators. As discussed above, in this case $\tl\psi$ is a polynomial, and we can therefore look for polynomial eigenfunctions $\Psi$. It is straightforward to write a basis in the space of symmetric homogeneous degree-$J$ polynomials, and to determine the action of $H'$ on such a basis.
We report here the results at the first few even integer $J$:
\begin{description}
\item[Spin $\boldsymbol0$.] We find $\g=2\e$ and $\Psi(p,q)=1$.
\item[Spin $\boldsymbol2$.] We find the eigenvalues $\g=\frac{13}9\e$ and $\g=0$ with the corresponding eigenfunctions $\Psi^{(1)}=5(p^2+q^2)+4pq$ and $ \Psi^{(2)}= p^2+q^2+6pq$. For the zero eigenvalue, the corresponding function $\tl \psi^{(2)}$, given by \eqref{eq:fansatz}, vanishes identically upon using $\b_1+\b_2+\b_3+\b_4=0$. Indeed, $\Psi^{(2)}$ agrees precisely with $\Psi_0(p,q)$ from \eqref{eq:kernel}.
The remaining eigenvalue $\frac{13}9\e$ agrees with the known result for the unique spin-$2$ twist-$4$ operator~\cite{Kehrein:1994ff}. 
\item[Spin $\boldsymbol4$.] We find eigenvalues $\frac49\e$, $\frac{19}{15}\e$ and $0$, where again the eigenfunction corresponding to $\g=0$ is given by \eqref{eq:kernel}. The non-zero eigenvalues correspond to the known anomalous dimensions of the two twist-4 local primary operators of spin 4.
\end{description}
From spin 6 and onwards, most of the eigenvalues and eigenfunctions are no longer rational. It is convenient to present the characteristic polynomial $\det(\e^{-1}H'-\l\mathbb I)$:
\begin{align}
J=6: \quad & \l  \left(945 \l ^3-2217 \l ^2+1615 \l -371\right),
\\
J=8:\quad & \l  \left(51030 \l ^4-124443 \l ^3+100386 \l ^2-30183 \l   +2314\right),
\\
J=10:\quad & \l   ( \l -1/3) \left(561330 \l ^4-1362069 \l ^3+1102572 \l^2-335229 \l +26356\right).
\end{align}
Apart from $\l=0$, again corresponding to $\Psi_0$ of \eqref{eq:kernel}, the remaining eigenvalues agree with known results~\cite{Kehrein:1994ff,Henriksson:2022rnm}.\footnote{Starting from spin $12$, the local even-spin operator spectrum is known to contain operators with $\gamma=0$, whose degeneracy grows with spin (see \eqref{eq:genfgamma0}).} It is possible to extend this computation to very high spin, and we report data for operators with $J$ up to $700$ in section~\ref{sec:largespinlocal} below.

\subsubsection{A numerical scheme}
\label{sec:numericalscheme}

As a minimal modification of the polynomial wavefunctions that are relevant for local operators, let us consider light-ray operators for which $\Psi(p,q)$ is a continuous function. In this case, it is easy to see that it is determined by two functions of $\chi \in [0,1]$,\footnote{If $p$ and $q$ have the same sign, we recover $\Psi(p,q)$ from $\Psi_1$ by rescaling to $p+q=1$. If the signs are opposite, we recover $\Psi(p,q)$ from $\Psi_2$ by rescaling to $\max\{|p|,|q|\}=1$ and using the permutation symmetry.}
\be
\begin{aligned}
	\Psi_1(\chi)&=\Psi(\chi,1-\chi),\\
	\Psi_2(\chi)&=\Psi(1,-\chi).
\end{aligned}\label{eq:Psiidefn}
\ee
Note that $\Psi_1(\chi)=\Psi_1(1-\chi)$ and $\Psi_2(0)=\Psi_1(1)$.
We will see later that for any Regge trajectory the functions $\Psi_i$ are continuous on $[0,1]$ for sufficiently large $J$, but at small $J$ they can become singular at the endpoints of the interval (they are always analytic in the interior).

From now on we focus on even spin parity. Defining an auxiliary operator $Q'$ such that $H'=\frac{2Q'+1}{3}\e$ we find
\be\label{eq:Qprime1}
	(Q'\Psi)_1(\chi)=&(\chi^{J+1}+(1-\chi)^{J+1})\int_0^1dz \Psi_2(z)+\chi^{J+1}\int_\chi^1 dz z^{-J-2} \Psi_2(z)\nn\\
	&+(1-\chi)^{J+1}\int_{1-\chi}^1 dz z^{-J-2} \Psi_2(z)+\half\int_0^1 dz\Psi_1(z),\\
	\label{eq:Qprime2}
	(Q'\Psi)_2(\chi)=&\frac{1}{1-\chi}\int_0^{1-\chi}dz \Psi_2(z)+\frac{\chi^{J+1}}{1-\chi}\int_\chi^1 dz z^{-2-J}\Psi_1(z)+\frac{(1-\chi)^J}{2}\int_0^1 dz \Psi_1(z).
\ee

In this form the equation is suitable for numerical analysis. We introduce a discretisation parameter $N$, define a uniform lattice of points $0=\chi_0,\cdots,\chi_N=1$ with spacing $\De\chi=N^{-1}$, and functions
\be
	h_i(\chi)=\max(1-N|\chi-\chi_i|,0).
\ee
We then approximate
\be\label{eq:discrete}
	\Psi_i(\chi)=\sum_{k=0}^N \Psi_{i,k}h_k(\chi).
\ee
That is, $\Psi_i$ linearly interpolates between the values $\Psi_i(\chi_k)=\Psi_{i,k}$ at the lattice points. Evaluating $(Q'\Psi_i)(\chi_k)$, we obtain a finite-dimensional $2(N+1)\x 2(N+1)$ matrix, which can be numerically diagonalised.

In practice we find that this scheme works very well for eigenvalues $\l'$ of $Q'$ which are sufficiently away from $0$, but is unstable for smaller eigenvalues. One way to interpret this is as follows. By repeated differentiation, at a fixed value of $\l'\neq 0$, the eigenequation for $Q'$ can be turned into a decoupled system of ordinary differential equations for $\Psi_1$ and $\Psi_2$.\footnote{We omit the details since we do not rely on this result.} Each equation is a Fuchsian differential equation of order 6 with regular singularities at $\chi=0,1,\oo$. This shows that the eigenfunctions $\Psi_1,\Psi_2$ are smooth (in fact, holomorphic) except possibly at the endpoints of the interval $[0,1]$. One can analyze the possible singular behavior near these points by solving the appropriate indicial equations. It turns out that near $\chi=0$, $\Psi_1$ has solutions which behave as
\be\label{eq:psising}
	\Psi_1(\chi)\sim \chi^{J+1\pm \l'^{-1}}.
\ee
Let $\cB$ be the region of values of $\l'$ and $J$ where  at least one of these two exponents has negative real part. We do not expect our numerical scheme to work well in this region for generic eigenfunctions, since these will involve terms with the above singularities. In practice we indeed see that most numerical solutions develop singularities as the Regge trajectory approaches $\cB$.

\subsubsection{Results for anomalous dimensions and wavefunctions}
\label{sec:diagonalizationresults}

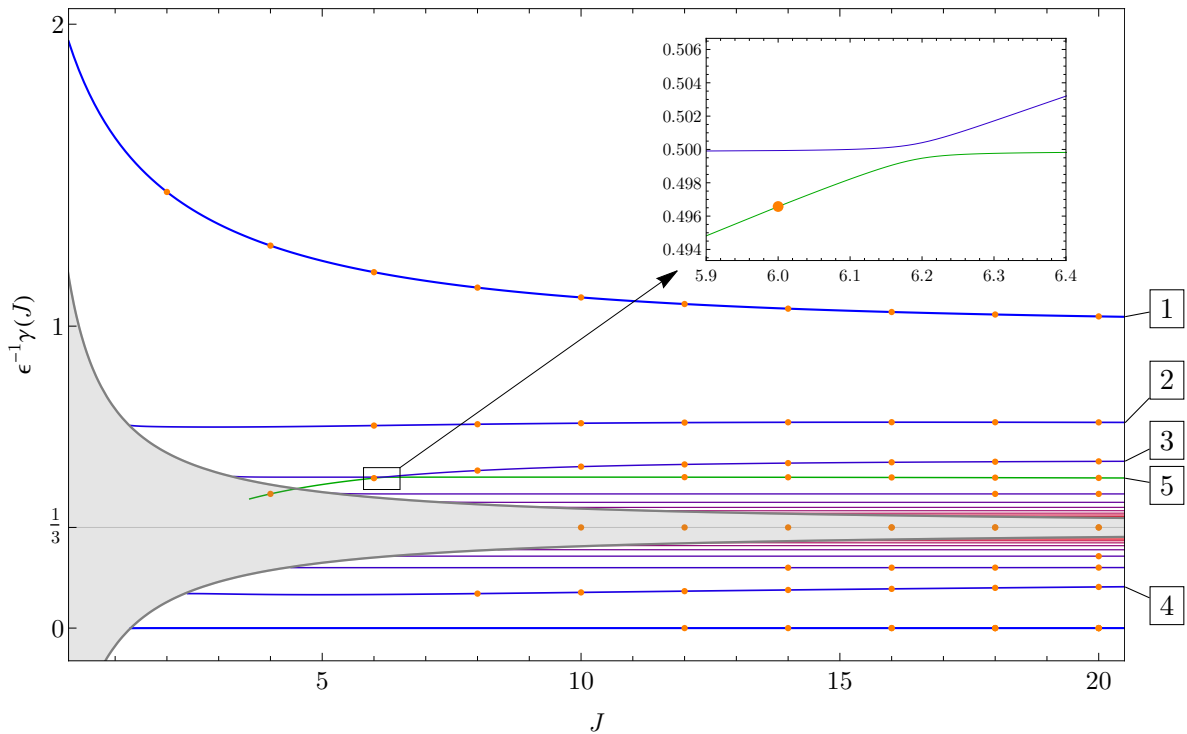
\begin{figure}[t]
	\begin{center}
		\begin{tikzpicture}
			\node[] () at (0,0) {\includegraphics[scale=.75]{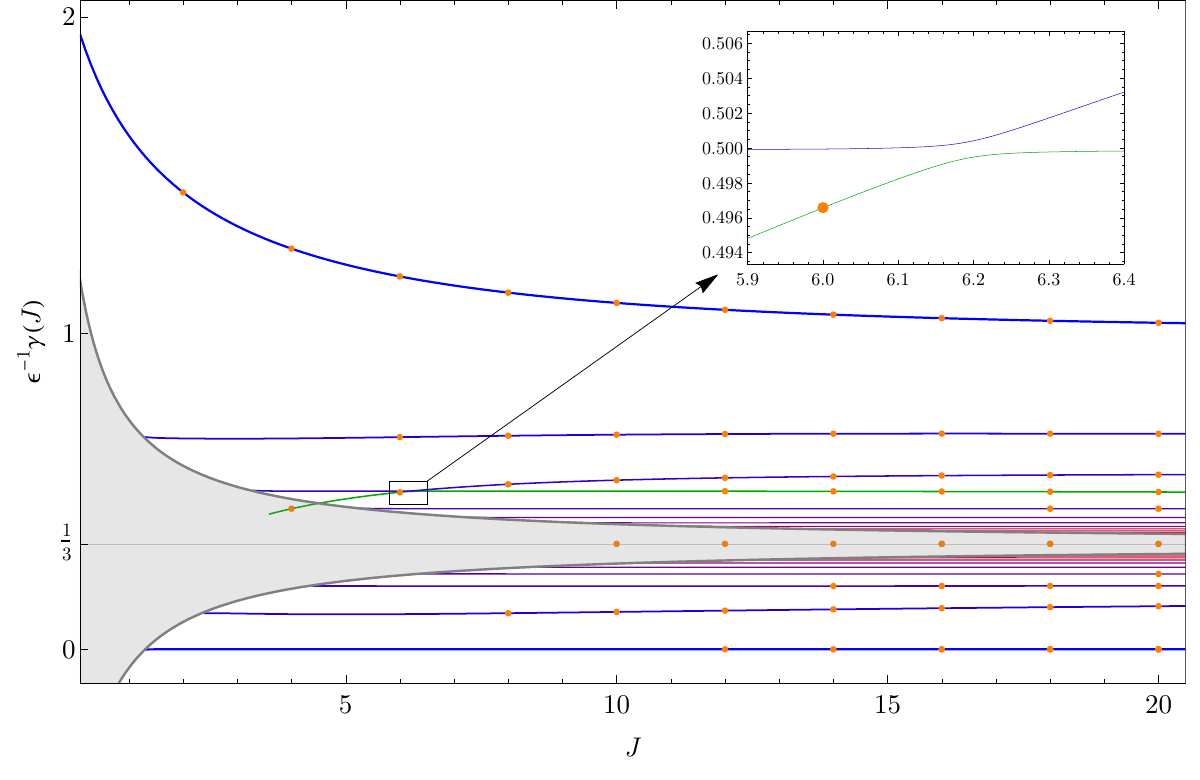}};
			\node[draw] (1) at (8,0.95) {$1$}; 
			\node[inner sep=0pt] (b1) at (7.42,0.84) {};
			\draw (1) -- (b1);

			\node[draw] (2) at (8,0) {$2$}; 
			\node[inner sep=0pt] (b2) at (7.42,-0.59) {};
			\draw (2) -- (b2);

			\node[draw] (3) at (8,-0.8) {$3$}; 
			\node[inner sep=0pt] (b3) at (7.42,-1.09) {};
			\draw (3) -- (b3);

			\node[draw] (4) at (8,-1.4) {$5$}; 
			\node[inner sep=0pt] (b4) at (7.42,-1.30) {};
			\draw (4) -- (b4);

			\node[draw] (5) at (8,-3) {$4$}; 
			\node[inner sep=0pt] (b5) at (7.42,-2.74) {};
			\draw (5) -- (b5);
		\end{tikzpicture}
		\caption{The one-loop anomalous dimension for twist-4 even-spin Regge trajectories as a function of real spin $J$. The coloured curves show Regge trajectories. The grey area is the region $\cB_+$, a slightly enlarged version of the region $\cB$ where the numerical scheme is not reliable. The orange dots represent the (light transforms of) local operators. The insert shows the behavior of the two Regge trajectories near an avoided level crossing. All shown trajectories except the one at $\g=0$ have multiplicity $1$. The numbered trajectories are referred to in the main text, and the numbering agrees with table~\ref{tab:twistfamilies}.
}
		\label{fig:reggetrajectories_main}
	\end{center}
\end{figure}

The result of the numerical diagonalisation for real values of $J>0.1$ is shown in figure~\ref{fig:reggetrajectories_main}. In terms of real $\e^{-1}\g(J)$ and $J$, the region $\cB$ is defined by the condition
\be
	\cB=\{(J,\e^{-1}\g):\, |\e^{-1}\g-\tfrac{1}{3}|<\tfrac{2}{3}(J+1)^{-1}\}.
\ee
We find that our results do not converge very well with increasing discretization parameter $N$ inside of $\cB$. In fact, the convergence slows down already slightly outside of $\cB$,\footnote{This has to do with the fact that while~\eqref{eq:psising} is non-singular just outside of $\cB$, it is still not differentiable, and this decreases the rate of convergence of the interpolation~\eqref{eq:discrete}.} and so in figure~\ref{fig:reggetrajectories_main} we plot in grey a more conservative region $\cB_+$ which is $\cB$ shifted to the right by $0.3$. The only exception is trajectory $5$, which we could reliably track into the region $\cB$ because its wavefunction happens not to develop a singularity as it enters $\cB$.

We found a single eigenfunction of $H'$ with eigenvalue $0$, which coincides with $\Psi_0$. Since $\Psi_0$ leads to $\tl\psi=0$ via~\eqref{eq:fansatz}, we discard this solution. However, there still exist infinitely many eigenstates of $H$ with eigenvalue $0$ which are not expressible through~\eqref{eq:fansatz}, and therefore we still plot a constant trajectory at $\e^{-1}\g=0$.

Outside of $\cB_+$ we find a discrete set of Regge trajectories, stable with increasing the discretisation parameter $N$, with new trajectories emerging from $\cB_+$ as $J$ increases. Reassuringly, we find trajectories which pass through all known local operators (or rather their light transforms) outside of $\cB_+$. It is reasonable to guess that if we had been able to eliminate the numerical issues within $\cB_+$, then for every $J>0$ we would find a mostly discrete set of Regge trajectories, with a single accumulation point at $\e^{-1}\g=1/3$ (and an infinite degeneracy at $\e^{-1}\g=0$). See, however, section~\ref{sec:trace} which presents some evidence to the contrary.

Another notable feature of figure~\ref{fig:reggetrajectories_main} is that we are able to detect Regge trajectories for $J$ much below the spin of their first local operator. For example, the fourth trajectory from the bottom is reliably computed for $J\gtrsim 6.5$ but has its first local operator at $J=20$. The Regge trajectory shown in green gives an example of a trajectory which has local operators for $J=4,6$, then no local operators for $J=8,10$, and then again passes through local operators at even $J\geq 12$.

\begin{figure}[t]
	\begin{center}
		\includegraphics[scale=.6]{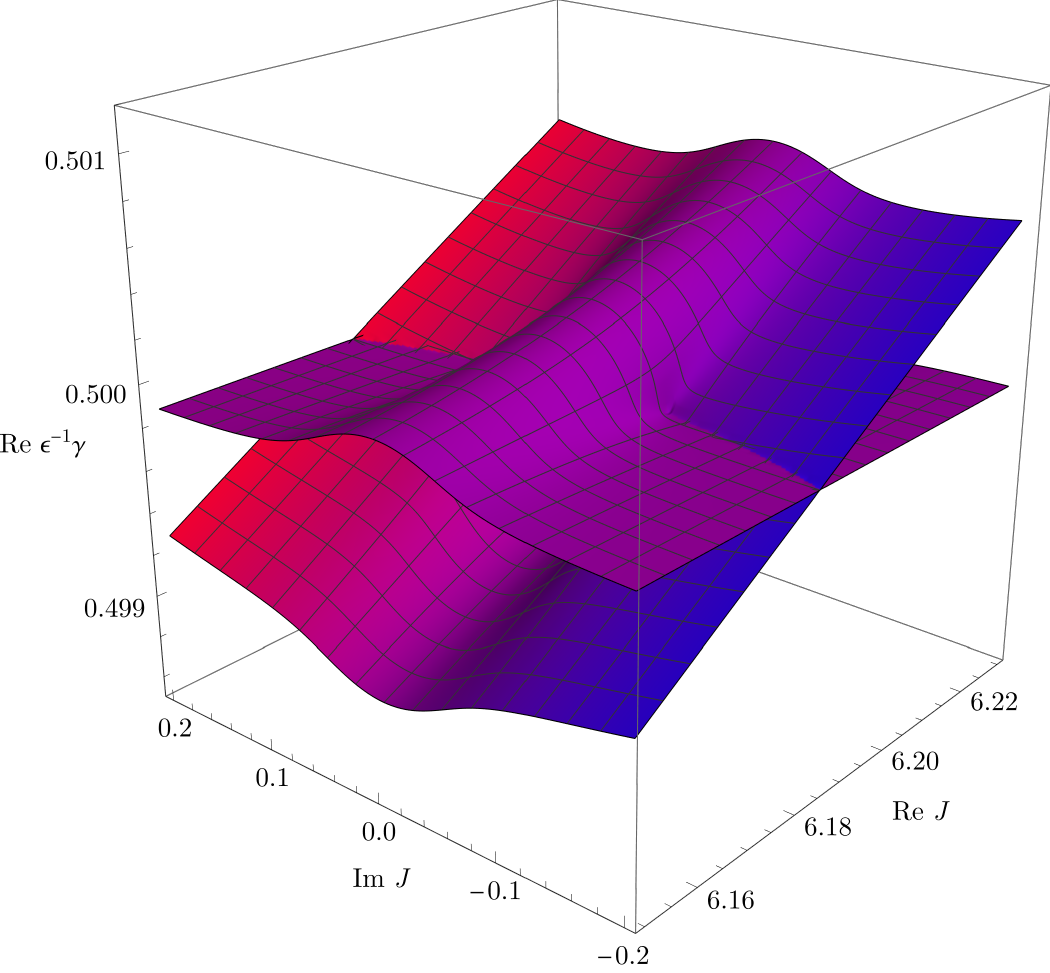}
		\caption{
			A plot of the complex Regge trajectory near the avoided intersection of real Regge trajectories 3 and 5, which now become the two branches at $\Im J=0$. The colour shows the value of $\Im \e^{-1}\g$, ranging from $-5\x10^{-3}$ (blue) to $+5\x10^{-3}$ (red). Purple represents $\Im \e^{-1}\g=0$. One of the branches is mostly purple because $\Im \e^{-1}\g$ is relatively close to $0$ due to the small overall slope of the branch. Note that at the apparent intersection the two branches have different colours and therefore do not in fact intersect in $\C^2$.
		}
		\label{fig:3dshape}
	\end{center}
\end{figure}

The green trajectory $5$ is also notable because it avoids a level crossing with the blue trajectory 3 near $J\approx 6$, as shown on the zoomed-in insert in figure~\ref{fig:reggetrajectories_main}. By diagonalising $H$ for complex values $J\in\C$ near $J\approx 6$ we can study the shape of the full complex-analytic Regge trajectories. The result for trajectories $3, 5$ is shown in figure~\ref{fig:3dshape}. As is generically expected~\cite{Caron-Huot:2022eqs}, the two real Regge trajectories turn out to be two real branches of one complex-analytic Regge trajectory. Note that we expect this full Regge trajectory to be complex-analytic at all points near $J\approx 6$, despite the apparent pair of branch points seen in figure~\ref{fig:3dshape}.\footnote{It has the same structure as $f(x,y)=-y^2+1+x^2=0$ ($x$ being the analogue of $J$ and $y$ the analogue of $\e^{-1}\g$) for $O(1)$ values of $x,y$. This defines a complex-analytic curve because the gradient $(\ptl_xf(x,y),\ptl_y f(x,y))=(2x,-2y)$ only vanishes at $x=y=0$ and $f(0,0)\neq 0$. However the solution $y=\pm \sqrt{1+x^2}$ does have apparent branch points at $x=\pm i$.} These branch points are an artifact of using $J$ as a coordinate, and occur at $J\approx 6.19\pm i0.06$.

\begin{figure}[t]
	\begin{center}
		\includegraphics[scale=.57]{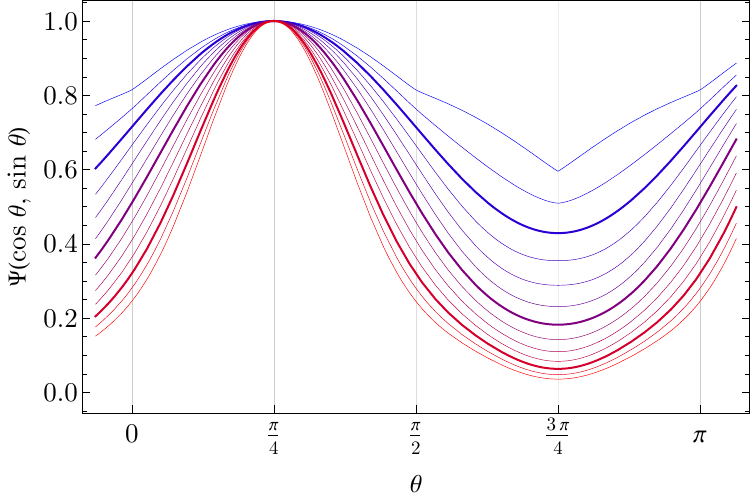}
		~
		\includegraphics[scale=.57]{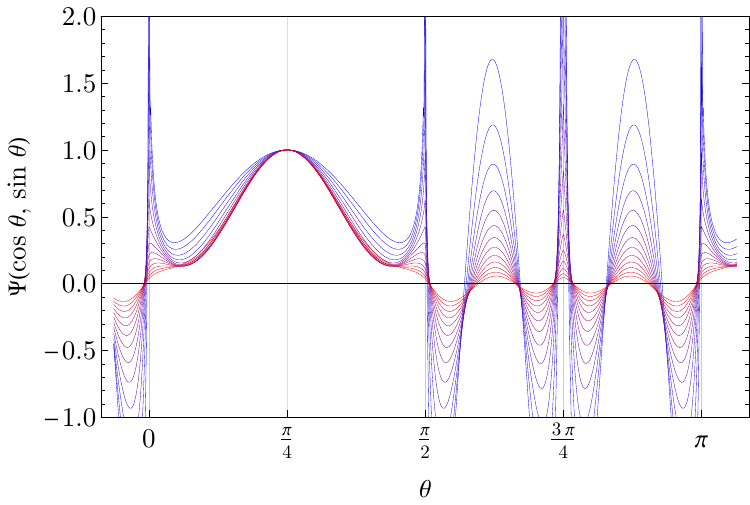}
		\caption{
			Plot of the wavefunction $\Psi(\cos\theta,\sin\theta)$ for trajectories 1 (left) and 3 (right), in the numbering of figure~\ref{fig:reggetrajectories_main}, for various values of $J$. For trajectory 1, spin runs from $J=1$ to $J=7$ in steps of $0.5$, while for trajectory $3$ it runs from $J=3$ to $J=5.4$ in steps of $0.2$. The color changes from blue to red as $J$ increases. The thick lines show polynomial (in $p,q$) wavefunctions which appear at $J\in\{2,4,6\}$ for trajectory 1. Wavefunctions are normalized to 1 at $\theta=\pi/4$.
		}
		\label{fig:wavefunctions}
	\end{center}
\end{figure}

Finally, we note that numerical diagonalization of $H'$ yields not only the anomalous dimensions, but also the wavefunctions $\Psi(p,q)$. In figure~\ref{fig:wavefunctions} we show wavefunctions for trajectories 1 and 3, for various values of spin $J$. More concretely, we plot the values of $\Psi(\cos\theta,\sin\theta)$ for $\theta\in[0,\pi]$.\footnote{Note that $\Psi(p,q)=\Psi(-p,-q)$ means that $\Psi(\cos\theta,\sin\theta)$  is $\pi$-periodic.} One can explicitly see in these plots how the wavefunction becomes more singular for values of $J$ which approach $\cB$. On the other hand, the wavefunctions become quite smooth for large $J$.

\subsection{Two-point functions and matrix elements of twist-4 operators}
\label{sec:twist4twomatrix}
\subsubsection{Two-point functions}
We now consider the two-point functions of the twist-4 light-ray operators along the Regge trajectories which we found in section~\ref{sec:twist4regges}. Specifically, we will focus on the Regge trajectories with non-zero anomalous dimension, i.e.\ those which can be expressed in terms of the ansatz~\eqref{eq:fansatz}. We will denote the wavefunctions $\psi$ in these trajectories by $\psi_{i,J}$, where $i$ labels different Regge trajectories and $J$ is the spin.\footnote{Note that there is a freedom to change the normalization $\psi_{i,J}\to A_i(J)\psi_{i,J}$ by a spin- and Regge trajectory-dependent constant $A_i(J)$.}

 The constant $C_i(J)$ which appears in~\eqref{eq:CiJdefinition} can then be expressed as a functional of the wavefunction~$\psi_{i,J}$,
 \be\label{eq:CiJCPsi}
 	C_i(J)=C(\psi_{i,J}).
 \ee
As we will show in this section, $C(\psi)$ has an integral expression
\be
	C(\psi)=\int_0^1 d\chi d\chi' \Psi_i(\chi) \Psi_j(\chi')P^{ij}(\chi,\chi')\label{eq:pairinggoal}
\ee
for some kernel $P^{ij}$, where the components $\Psi_i$ are defined in~\eqref{eq:Psiidefn}. Using the numerical results for $\Psi_i$ from section~\ref{sec:twist4regges} we will be able to determine the value of $C(\Psi)$.

We begin by considering the general two-point function
\be
	\<\O_{\psi}(X,Z)\O_{\psi'}(X',Z')\>,
\ee
where $\O$ is a twist-$n$ operator defined in~\eqref{eq:Opsidefn}. To simplify the calculation, we will set $X_i,Z_i$ to the same values as we did in section~\ref{sec:twist2two-pt},
\be
X=(0,1,0),\quad X'=(1,0,0),
\ee
with polarisations
\be
Z=(0,0,z),\quad Z'=(0,0,z').
\ee
Using~\eqref{eq:Opsidefn} and notation of section~\ref{sec:twist2two-pt}, we find
\be
	\<\O_{\psi}(X,Z)\O_{\psi'}(X',Z')\>&
	=n!\int d^n\a d^n\a'\psi(\a)\psi(\a')\prod_{i=1}^n \<\f(X_{\a_i})\f(X'_{\a'_i})\>\nn\\
	&=(4\pi^2)^{-n}n!\int d^n\a d^n\a'\psi(\a)\psi'(\a')\prod_{i=1}^n (1+\a_i\a'_i+i\e)^{-1},
\ee
where we treat $\a$ as a vector with components $\a_1,\cdots,\a_n$ and similarly for $\a'$. Using~\eqref{eq:fourierdefn} we find
\be\label{eq:inversefourier}
	\psi(\a)=(4\pi)^{1-n}\int d^n \b \de(\sum \b)e^{i\a\.\b/2} \tl \psi(\b),
\ee
where $\sum\b=\sum_{i=1}^n \b_i$. This gives
\be
&\<\O_{\psi}(X,Z)\O_{\psi'}(X',Z')\>\nn\\
&=\frac{(4\pi)^2 n!}{(64\pi^4)^{n}}\int d^n\a d^n\a'd^n\b d^n \b'\de(\sum\b)\de(\sum\b')\tl\psi(\b)\tl\psi'(\b')e^{i\a\.\b/2+i\a'\.\b'/2}\prod_{i=1}^n (1+\a_i\a'_i+i\e)^{-1}
\nn\\
&=\frac{(4\pi)^2 n!}{(64\pi^4)^{n}}\int d^n\b d^n \b'\de(\sum\b)\de(\sum\b')\tl\psi(\b)\tl\psi'(\b')\prod_{i=1}^n \cA(\b_i,\b_i'),
\ee
where
\be
	\cA(\b,\b')&=\int d\a d\a' e^{i\a\b/2+i\a'\b'/2} (1+\a\a'+i\e)^{-1}=\int d\a d\a' |\a|^{-1} e^{i\a\b/2+i\a'\b'/(2\a)} (1+\a'+i\e)^{-1}\nn\\
	&=2\pi i \int d\a \theta(-\a/\b') |\a|^{-1} e^{i\a\b/2-i\b'/(2\a)}=2\pi i \int_0^\oo d\a \a^{-1}e^{-i\a\b\b'/2+i/(2\a)}\nn\\
	&=4\pi iK_0(\sqrt{\b\b'-i0}),
\ee
where $K_0$ is the modified Bessel function of the second kind. The two-point function becomes
\be
&\<\O_{\psi}(X,Z)\O_{\psi'}(X',Z')\>\nn\\
&=\frac{(4\pi)^2 n! i^n}{(16\pi^3)^{n}}\int d^n\b d^n \b'\de(\sum\b)\de(\sum\b')\tl\psi(\b)\tl\psi'(\b')\prod_{i=1}^n K_0\left(\sqrt{\b_i\b'_i-i0}\right).
\ee

The next step is to specialise to $n=4$ and reduce the above expression according to the ansatz~\eqref{eq:fansatz}. The wavefunctions $\psi,\psi'$ each consist of 6 terms according to~\eqref{eq:fansatz}, so we get 36 terms total. After some relabelling of integration variables, they split into
\begin{enumerate}
	\item 6 terms containing $\Psi(\b_1,\b_2)\Psi'(\b'_1,\b'_2)$,
	\item 24 terms containing $\Psi(\b_1,\b_2)\Psi'(\b'_2,\b'_3)$,
	\item 6 terms containing $\Psi(\b_1,\b_2)\Psi'(\b'_3,\b'_4)$.
\end{enumerate}
Anticipating that the integration over the overall scale will produce a factor of $\vol\SO(1,1)$, we can write
\be
	&\<\O_{\psi}(X,Z)\O_{\psi'}(X',Z')\>\nn\\
	&=\vol (\SO(1,1))\int d\chi d\chi'\p{6 P_1(\chi,\chi')+24 P_2(\chi_1,\chi_2)+6 P_3(\chi_1,\chi_2)}\Psi(\chi,1-\chi)\Psi'(\chi',1-\chi'),
\ee
where $P_1,P_2,P_3$ correspond to the 3 different types of terms described above, and the factor $\vol\SO(1,1)$ is introduced for future convenience.

Let us look at a term of the first type. It is given by
\be
	&\frac{3}{2^9\pi^{10}}\int d^4\b d^4 \b'\de(\sum\b)\de(\sum\b')\Psi(\b_1,\b_2)\Psi(\b'_1,\b'_2)\prod_{i=1}^4 K_0(\sqrt{\b_i\b'_i-i0})\nn\\
	&=\frac{3}{2^{13}\pi^{10}}\int d\b_1d\b_2d\b_3 d\b'_1d\b'_2d\b'_3\Psi(\b_1,\b_2)\Psi(\b'_1,\b'_2)\nn\\
	&\hspace{2cm}\x \cY_1(\b_1\b'_1)\cY_1(\b_2\b'_2)\cY_1(\b_3\b'_3)\cY_1((\b_1+\b_2+\b_3)(\b'_1+\b'_2+\b'_3)),
\ee
where
\be
	\cY_\De(x)=2(-x+i0)^{\frac{\De-1}{2}}K_{\De-1}(\sqrt{x-i0})=\int_0^\oo d\a \a^{-\De}e^{i/(2\a)-i\a x/2}.
\ee
Using the above integral representation for $\cY_\De(x)$, one can prove the identity
\be
	&\int dx \cY_{\De_1}(a x)\cY_{\De_2}(b(x+c))\nn\\
	&=4\pi |a|^{-1}\theta(-ba^{-1})(-ba^{-1})^{\De_2-1}(1-ba^{-1})^{-\De_1-\De_2+1}\cY_{\De_1+\De_2}((b-a)c),
\ee 
a special case of which is
\be
	\int d\b d\b' \cY_{\De_1}(\b\b')\cY_{\De_2}((\b-\g)(\b'-\g'))=4\pi \frac{\G(\De_1)\G(\De_2)}{\G(\De_1+\De_2)}\cY_{\De_1+\De_2}(\g\g').
\ee
The latter identity allows one to perform the $\b_3,\b'_3$ integrations. After a change of variables to isolate the integral over the overall scale (which produces $\vol(\SO(1,1))$) one can straightforwardly show
\be\label{eq:P1}
	P_1(\chi,\chi')=\frac{3}{2^{10}\pi^9}\int_{-\oo}^{+\oo} dr |r|^{J+1}\cY_1(r \chi\chi')\cY_1(r(1-\chi)(1-\chi'))\cY_2(r).
\ee

Using similar techniques, for kernels $P_2$ and $P_3$ one can show
\be
	P_2(\chi,\chi')&=|1-\chi|^{-J-2}\int_0^{\frac{\chi-1}{\chi}}dz z^{J+1}P_1(z,\chi'),\nn\\
	P_3(\chi,\chi')&=\theta(0<\chi<1)\int_{-\oo}^{+\oo} dz P_1(z,\chi').
\ee
It is then easy to verify
\be
	&\<\O_{\psi}(X,Z)\O_{\psi'}(X',Z')\>\nn\\
	&=\vol \SO(1,1)\int d\chi d\chi'\p{6 P_1(\chi,\chi')+24 P_2(\chi_1,\chi_2)+6 P_3(\chi_1,\chi_2)}\Psi(\chi,1-\chi)\Psi'(\chi',1-\chi')\nn\\
	&=18\vol\SO(1,1)\int d\chi d\chi' P_1(\chi,\chi') (H'\Psi)(\chi,1-\chi)\Psi'(\chi',1-\chi'),
\ee
where $H'$ is as defined in~\eqref{eq:Dptwovar}.

Comparing to~\eqref{eq:standard2ptL} we find
\be
	\<\O_{\psi}(X,Z)\O_{\psi'}(X',Z')\>=\<\psi,\psi'\>\vol(\SO(1,1))\<\wL[\cO](X,Z)\wL[\cO](X,Z)\>_0,
\ee
where
\be
	\<\psi,\psi'\>&\equiv 9\pi^{-1}i(2J+3)\p{H'\Psi,\Psi'},\label{eq:pairingrelation}\nn\\
	\p{\Psi,\Psi'}&\equiv \int d\chi d\chi' P_1(\chi,\chi') \Psi(\chi,1-\chi)\Psi'(\chi',1-\chi').
\ee
In terms of $C(\psi)$ (see~\eqref{eq:CiJCPsi}) we have then
\be
	C(\psi)=\<\psi,\psi\>.\label{eq:CpsiPairing}
\ee

Note that both pairings $\<\.,\.\>$ and $\p{\.,\.}$ are symmetric -- the first due to the symmetry of the two-point function, and the second simply by definition.  This is only consistent with~\eqref{eq:pairingrelation} if $H'$ is symmetric with respect to both pairings,
\be
	\<H'\psi,\psi'\>&=\<\psi,H'\psi'\>,\\
	\p{H'\Psi,\Psi}&=\p{\Psi,H'\Psi'}.
\ee
This in turn implies that eigenfunctions with different eigenvalues of $H'$ are orthogonal, as we would expect based on conformal symmetry.

\paragraph{Numerical evaluation} In order to evaluate the pairings $\<\.,\.\>$, we rewrite~\eqref{eq:pairingrelation} in terms of $\Psi_1(\chi)$ and $\Psi_2(\chi)$, so that it takes the form~\eqref{eq:pairinggoal}. We evaluate the kernels $P^{ij}(\chi,\chi')$ numerically, and then~\eqref{eq:pairinggoal} can be evaluated for the eigenfunctions of $H'$ computed in section~\ref{sec:twist4regges}.

It turns out that the integrals in~\eqref{eq:pairinggoal} receive important contributions from the boundaries of the integration region, where the kernels have $\log$ singularities. This requires the kernels $P^{ij}(\chi,\chi')$ to be evaluated on a grid of values $(\chi,\chi')$ which gets denser near these boundaries. This issue becomes worse as $J$ gets larger, and is the main factor limiting the range of $J$ for which we were able to compute the two-point functions reliably.

Furthermore, each evaluation of $P^{ij}(\chi,\chi')$ requires one to compute an integral of three Bessel-$K$ functions -- a ``triple-$K$ integral'' -- in~\eqref{eq:P1}.  Such integrals have been studied extensively in the context of CFT three-point functions in momentum space in~\cite{Bzowski:2013sza}. In some cases (for integer $J$, see~\cite{Bzowski:2020lip}) they can be computed analytically, but in our case even the simplest examples give extremely complicated expressions. In the end, we resort to numerical evaluation of these integrals.

The procedure is relatively straightforward; we only highlight one subtle aspect. The most general triple-$K$ integral that we encounter is
\be
	I_{\a,\{0,0,1\}}(a,b,1)=\int_0^\oo dx x^\a K_0(ax)K_0(bx)K_1(x),
\ee
where $a, b\in \R_{>0}\cup i\R_{>0}$. When $a$ or $b$ are large and imaginary, the integral becomes highly oscillatory at large $x$ due to a factor $e^{-(a+b+1)x}$ in the asymptotic. In this case the convergence can be significantly improved by integrating $x$ in the direction of $1+a^*+b^*$ instead of the real line. Specifically, if we write $a+b+1=e^{i\theta}k$ with $k>0$ and $\theta\in (-\pi/2,\pi/2)$, then by a contour deformation argument we can show
\be
	I_{\a,\{0,0,1\}}(a,b,1)=e^{-i\theta(\a+1)}\int_0^\oo dy y^\a K_0(ae^{-i\theta}y)K_0(be^{-i\theta}y)K_1(e^{-i\theta}y),
\ee
where the integrand decays as $e^{-ky}$ at large $y$ and doesn't have strong oscillations.

\subsubsection{Matrix elements}
\label{sec:matrixelements}

To compute the matrix elements, we follow the same strategy as in the case of twist-2 operators in section~\ref{sec:twist2matrixelts}. We will compute the matrix elements between the states created by $\f$ and $\f^2$ operator insertions.\footnote{Note that $\phi^2$ here is not the canonically normalised operator, but instead defined directly as the normal-ordered product $\phi^2(x)=:\phi(x)^2:$, satisfying $\langle \phi^2(x)\phi^2(y)\rangle =\frac{2}{16\pi^4}(x-y)^{-4}$.}
 
We begin with two insertions of $\f^2$. The analogue of equation~\eqref{eq:twist2matrixeltintermediate} is
\be
&\<0|\f^2(X_1)\O_\psi(X_3,Z)\f^2(X_2)|0\>\nn\\
&=\frac{4!}{(4\pi^2)^4}\int \frac{ d\a_1d\a_2d\a_3d\a_4\psi(\a_1,\a_2,\a_3,\a_4)}{(-\a_1+i\e)(-\a_2+i\e)(-2z\.x_1-\a_3x_1^2+i\e)(-2z\.x_1-\a_4x_1^2+i\e)}.
\ee
Using
\be
	\int d\a \frac{e^{i\a\b/2}}{-\a+i\e}=-2\pi i\theta(\b),\quad \int d\a \frac{e^{i\a\b/2}}{-2z\.x_1-\a x_1^2+i\e}=\frac{2\pi i e^{-i z\.x_1 x_1^{-2}\b}}{x_1^2}\theta(-\b)
\ee
and~\eqref{eq:inversefourier} we find
\be
&\<0|\f^2(X_1)\O_\psi(X_3,Z)\f^2(X_2)|0\>=\frac{3}{(2\pi)^7}\int_{{\b_1,\b_2>0}\atop{\b_3,\b_4<0}}  d^4\b \de(\sum\b)\tl\psi(\b)e^{-i z\.x_1 x_1^{-2}(\b_3+\b_4)} x_1^{-4}.
\ee
We introduce new variables via $\b_1=xr,\b_2=(1-x)r,\b_3=-ty,\b_4=-t(1-y)$ with $r,t>0$ and $x,y\in [0,1]$. The integral becomes
\be
&\<0|\f^2(X_1)\O_\psi(X_3,Z)\f^2(X_2)|0\>\nn\\
&=\frac{3}{(2\pi)^7x_1^{4}}\int dr dt dx dy\, r t \de(r-t)\tl\psi(rx,r(1-x),-ty,-t(1-y))e^{i z\.x_1 x_1^{-2}t} \nn\\
&=\frac{3}{(2\pi)^7x_1^{4}}\p{\int_0^\oo dr r^{2+J} e^{i z\.x_1 x_1^{-2}r}} \int_0^1 dx dy \tl\psi(x,1-x,-y,y-1) \nn\\
&=\frac{3}{(2\pi)^7x_1^{4}}\p{\frac{z\.x_1}{x_1^2}}^{-J-3}e^{i\pi(J+3)/2}\G(J+3) \int_0^1 dx dy \tl\psi(x,1-x,-y,y-1).
\ee
At the same time, we find from~\eqref{eq:standard3ptL}, in this configuration
\be
	\<0|\f^2(X_1)\wL[\cO](X_3,Z)\f^2(X_2)|0\>_0=-2\pi i\frac{\G(2J+3)}{\G(J+2)^2}\frac{(-2z\.x_1)^{-J-3}}{(-x_1^2)^{-J-1}}, 
\ee
which implies
\be
	\frac{\<0|\f^2\O_\psi\f^2|0\>}{\<0|\f^2\wL[\cO]\f^2|0\>_0}=\frac{3e^{i\pi J/2}2^{J+3}}{(2\pi)^8}\frac{\G(J+3)\G(J+2)^2}{\G(2J+3)}\int_0^1 dx dy \tl\psi(x,1-x,-y,y-1).
\ee

Using the ansatz~\eqref{eq:fansatz} we can write (assuming even spin parity\footnote{For odd spin parity, this integral vanishes, as can be easily checked. This is expected, since there are no odd-spin operators in $\f^2\x\f^2$ OPE.})
\be
	\int_0^1 dx dy \tl\psi(x,1-x,-y,y-1)&=2\int_0^1 dx\Psi(x,1-x)+8\int_{0<y<x<1} dx dy\Psi(x,-y)\nn\\
	&=2\int_0^1 dx\Psi_1(x)+8\int_{0<y<x<1} dx dy x^J\Psi_2(y/x)\nn\\
	&=2\int_0^1 dx\Psi_1(x)+\frac{8}{J+2}\int_{0}^1 dy\Psi_2(y).
\ee
So, we finally obtain
\be\label{eq:ratiophi2final}
	\frac{\<0|\f^2\O_\psi\f^2|0\>}{\<0|\f^2\wL[\cO]\f^2|0\>_0}=\frac{6e^{i\pi J/2}2^{J+3}}{(2\pi)^8}\frac{\G(J+3)\G(J+2)^2}{\G(2J+3)}\p{\int_0^1 dx\Psi_1(x)+\frac{4}{J+2}\int_{0}^1 dy\Psi_2(y)}.
\ee

Now we consider the matrix elements of twist-4 operators with a pair of insertions of $\phi$. For local operators, it is known that $\lambda_{\phi\phi\cO}=O(\epsilon)$ where $\cO$ has twist 4, and the same holds for light-ray operators. Using the equation of motion $\square \phi=g\phi^3/3!$ we can relate matrix elements $\<0|\f\O_\psi\f|0\>$ to $\<0|\f\O_\psi\f^3|0\>$, which we can compute in the free theory. 

At the fixed point the leading-order equation of motion takes the form
\begin{equation}
\square \phi(x) = \frac{2(2\pi)^2\epsilon}{9}\phi^3,
\end{equation}
where $\square=\ptl^2$ is the Laplacian. 
This means that we can write
\begin{equation}
\frac{\langle0|\phi\O_\psi\phi|0\rangle}{\langle0|\phi\wL[\cO]\phi|0\rangle_0}
=
\frac{\langle0|\phi\O_\psi\square\phi|0\rangle}{\langle0|\phi\wL[\cO]\square\phi|0\rangle_0} = \frac{2(2\pi)^2\epsilon}{9} \frac{\langle0|\phi\O_\psi\phi^3|0\rangle}{\langle0|\phi\wL[\cO]\square\phi|0\rangle_0} .
\end{equation}
Using that 
\be
	\langle0|\phi\wL[\cO]\square\phi|0\rangle_0=4(J+2)\langle0|\phi\wL[\cO]\phi^3|0\rangle_0,
\ee 
as follows from~\eqref{eq:standard3ptL} and~\eqref{eq:standard3ptLdistinct}, we find
\begin{equation}
\frac{\langle0|\phi\O_\psi\phi|0\rangle}{\langle0|\phi\wL[\cO]\phi|0\rangle_0}
=
\frac{(2\pi)^2\epsilon}{18(J+2)}
\frac{\langle0|\phi\O_\psi\phi^3|0\rangle}{\langle0|\phi\wL[\cO]\phi^3|0\rangle_0}.
\label{eq:relationphi-phi3-final}
\end{equation}

Thus, to give results for the matrix elements of our light-ray operators in the state created by $\phi$, we can evaluate the matrix elements between states created by the insertions of $\phi$ and $\phi^3$. The computation is completely analogous to that of $\phi^2$ and $\phi^2$ above, and the final expression reads
\begin{equation}
	\frac{\langle0|\phi\O_\psi\phi^3|0\rangle}{\langle0|\phi\wL[\cO]\phi^3|0\rangle_0}=\frac{3e^{i\pi J/2}2^{J+3}}{(2\pi)^8}\frac{\Gamma(J+3)^2\Gamma(J+1)}{\Gamma(2J+3)}\int_0^1 dx\int_0^{1-x} dy \tl\psi(-1,1-x-y,x,y).
\end{equation}
Expressed in terms of $\Psi_1$ and $\Psi_2$, this gives
\begin{equation}\label{eq:finalPhiPhi^3}
\frac{\langle0|\phi\O_\psi\phi^3|0\rangle}{\langle0|\phi\wL[\cO]\phi^3|0\rangle_0}
=
\frac{9e^{i\pi J/2}2^{J+3}}{(2\pi)^8}\frac{\Gamma(J+3)^2\Gamma(J+1)}{\Gamma(2J+3)}
\left(
\frac1{J+2}\int_0^1dx\Psi_1(x)+\int_0^1dx(1-x)\Psi_2(x)
\right).
\end{equation}

\subsubsection{Numerical results and residues $r_i(J)$}
\label{sec:resultsOPEcoefs}

\begin{figure}[t]
	\centering
	{
	\setlength{\tabcolsep}{0.05cm}
	\begin{tabular}{ccc}
	\includegraphics[scale=0.5]{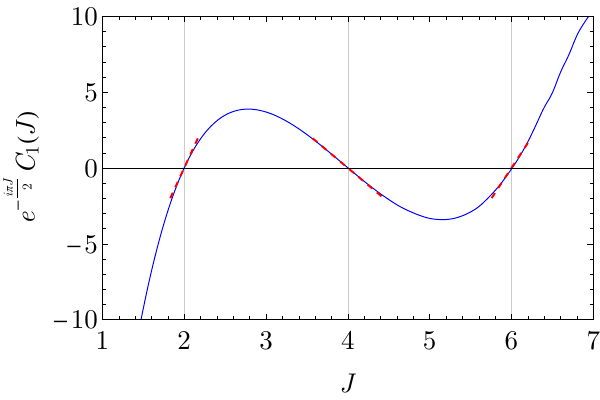}  & \includegraphics[scale=0.5]{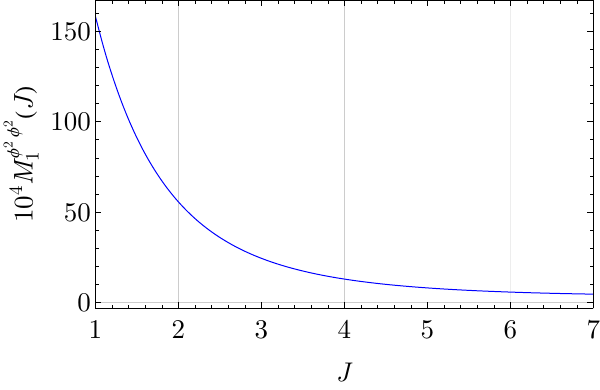}  & \includegraphics[scale=0.5]{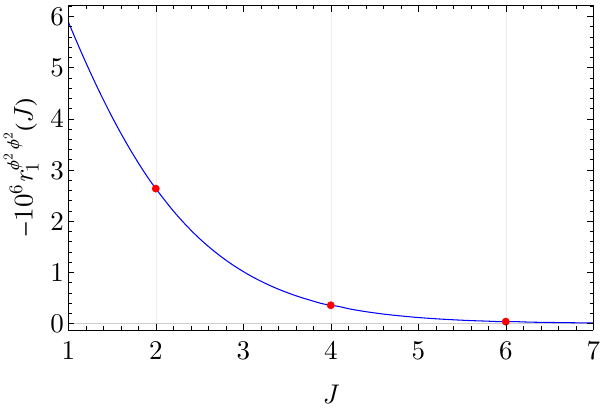} \\
	\includegraphics[scale=0.5]{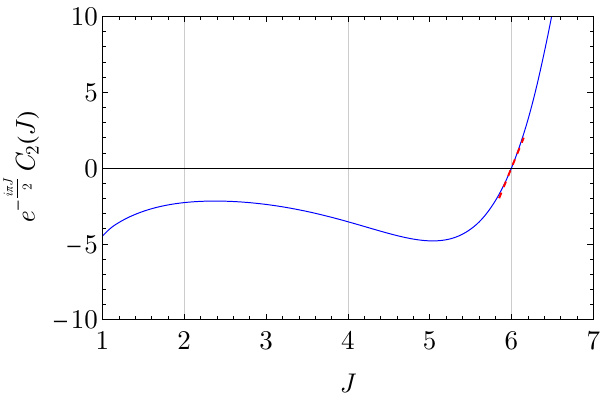} & \includegraphics[scale=0.5]{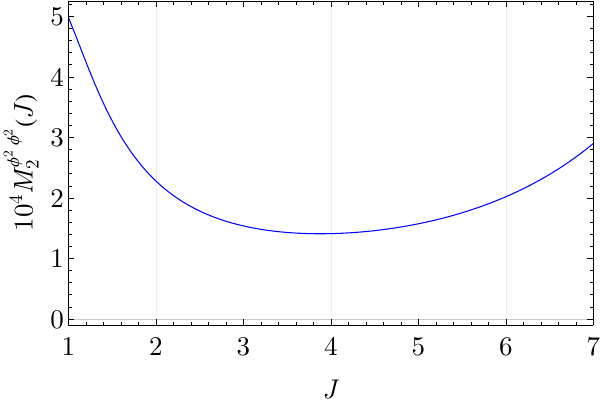} & \includegraphics[scale=0.5]{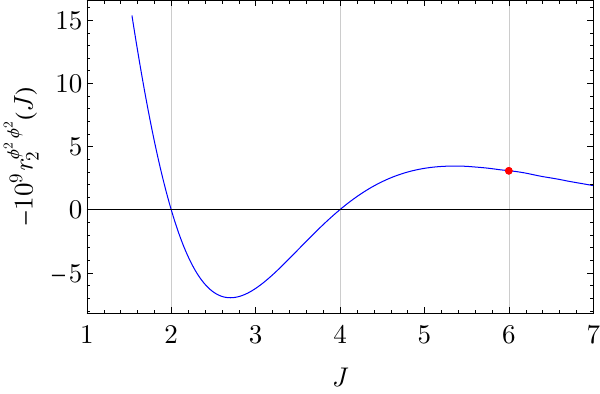} \\
	\includegraphics[scale=0.5]{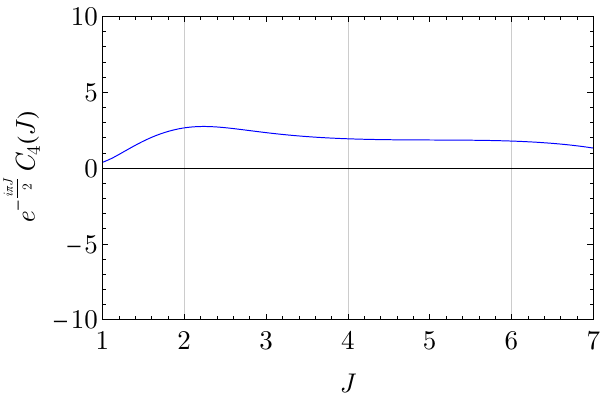} & \includegraphics[scale=0.5]{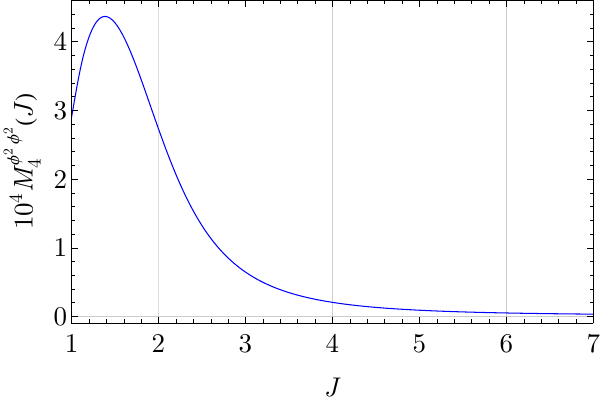} & \shifttext{-0.3cm}{\raisebox{-0.1cm}{\includegraphics[scale=0.53]{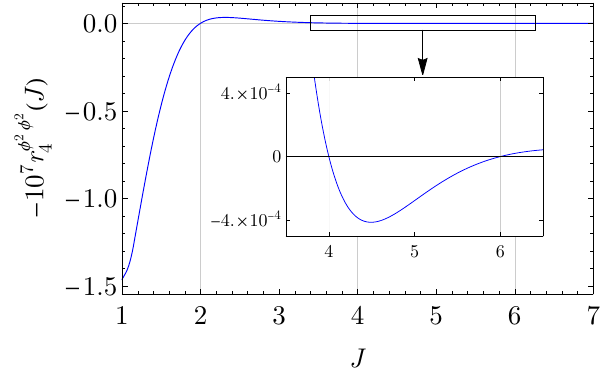}}}
	\end{tabular}
	}
	\caption{
		Various properties of Regge trajectories 1,2,4 in figure~\ref{fig:reggetrajectories_main}. The first column shows the two-point functions $C_i(J)$, the second column shows the matrix elements $M_i^{\f^2\f^2}(J)$, and the third column shows the residues $r_i^{\f^2\f^2}(J)$. Note that in some cases a rescaling has been applied, as indicated by the labels on the ordinate axes. Red dashed lines at zeros of two point functions indicate the slope predicted by~\eqref{eq:CiJderivative}. Red dots in the residue plots indicate the values predicted by~\eqref{eq:risminusff}.
	}\label{fig:grid124}
\end{figure}

\begin{figure}[t]
	\centering
	{
		\setlength{\tabcolsep}{0.05cm}
		\begin{tabular}{ccc}
			\includegraphics[scale=0.5]{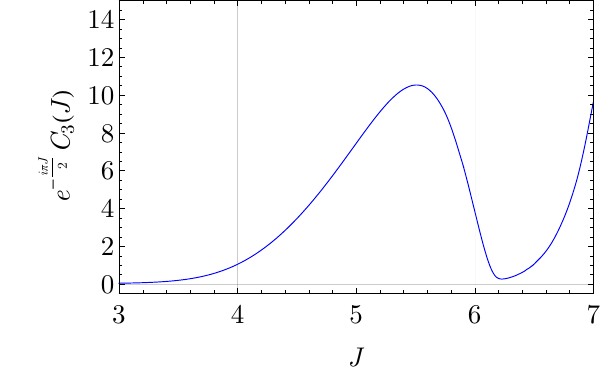}  & \includegraphics[scale=0.5]{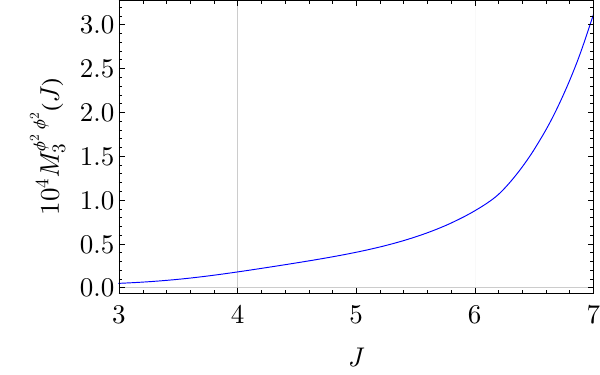}  & \includegraphics[scale=0.5]{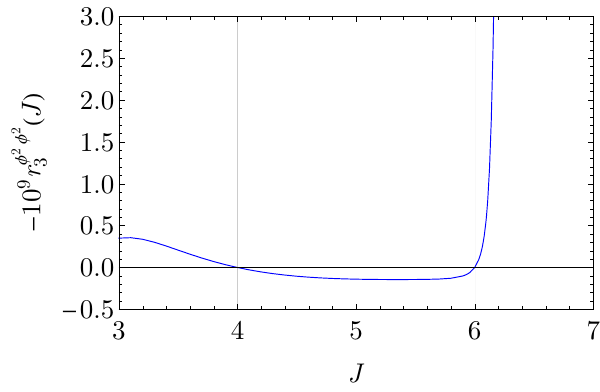} \\
			\includegraphics[scale=0.5]{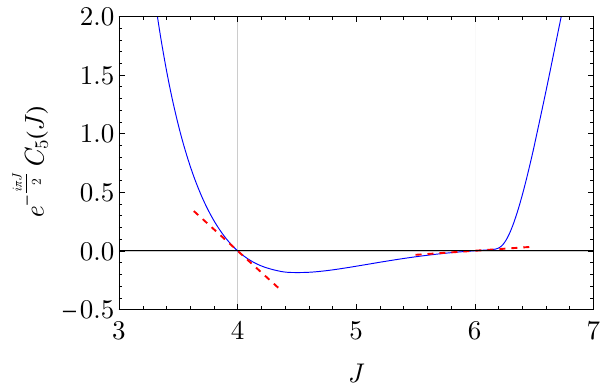} & \includegraphics[scale=0.5]{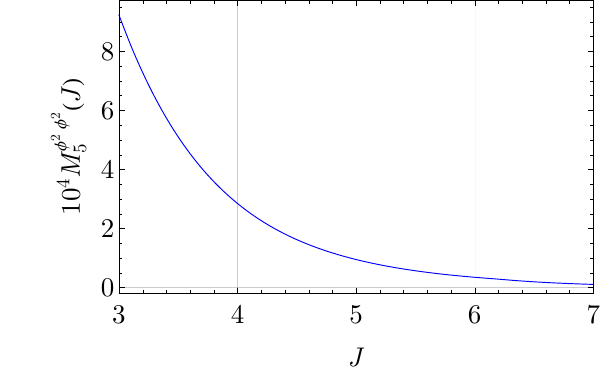} & \includegraphics[scale=0.5]{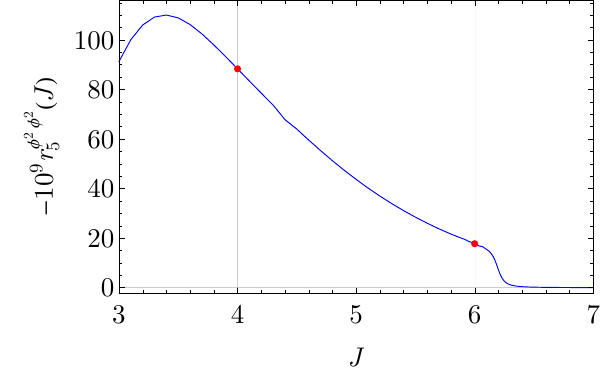} 
		\end{tabular}
	}
	\caption{
		Various properties of Regge trajectories 3,5 in figure~\ref{fig:reggetrajectories_main}. See the caption to figure~\ref{fig:grid124} for the description.
	}\label{fig:grid56}
\end{figure}

\begin{figure}[t]
	\centering
	{
		\setlength{\tabcolsep}{0.05cm}
		\begin{tabular}{ccc}
			\includegraphics[scale=0.5]{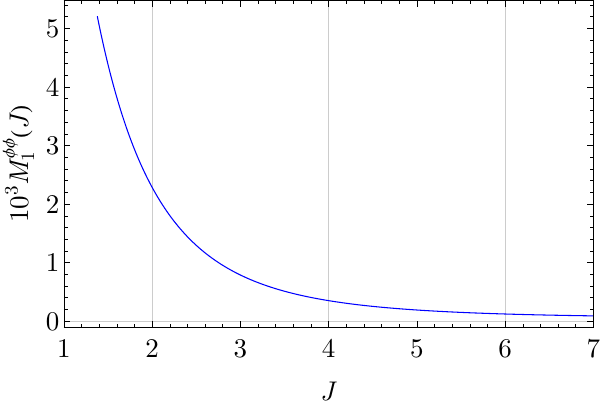}  &  \includegraphics[scale=0.5]{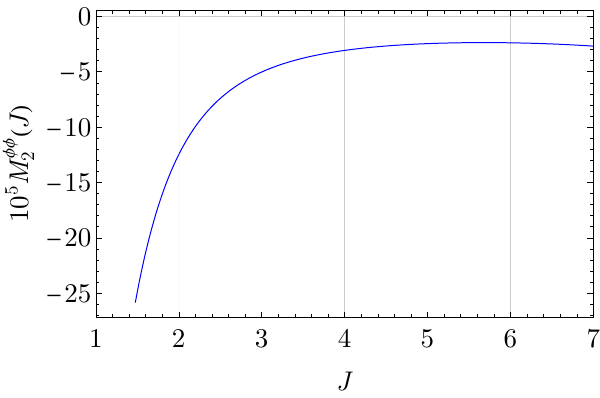} & \includegraphics[scale=0.5]{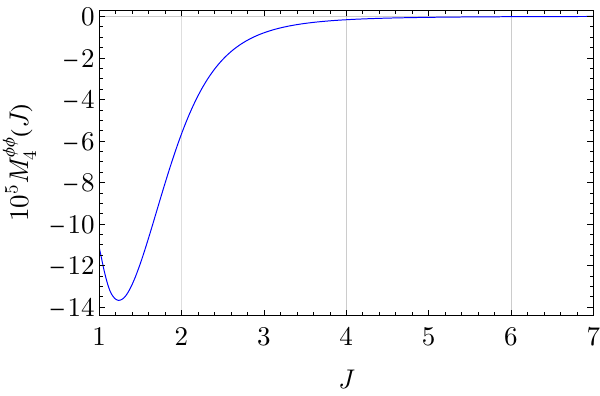} \\
			\includegraphics[scale=0.5]{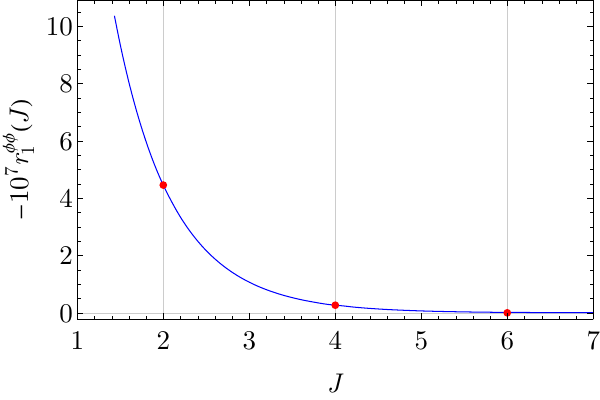} &   \includegraphics[scale=0.5]{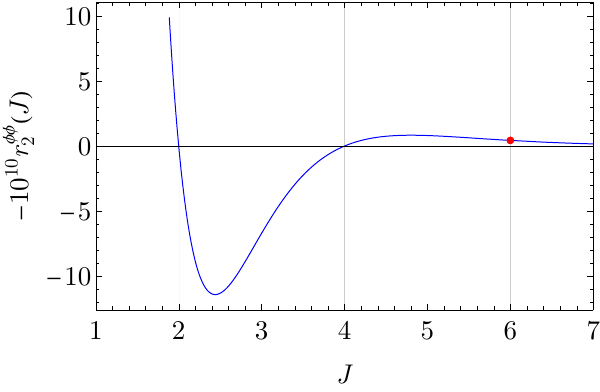}  & \shifttext{-0.3cm}{\raisebox{-0.1cm}{\includegraphics[scale=0.53]{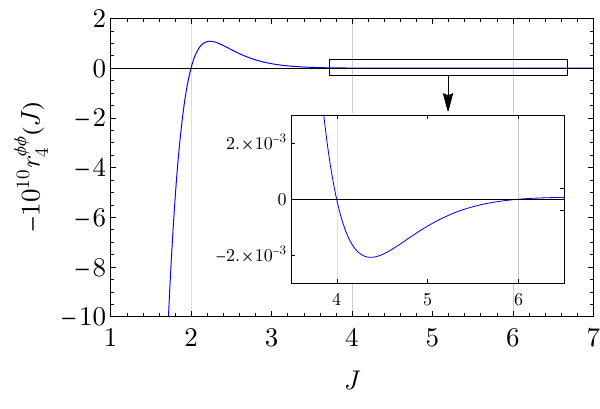}}}
		\end{tabular}
	}
	\caption{
		Similar to figures~\ref{fig:grid124} and~\ref{fig:grid56} but for matrix elements $M_i^{\f\f}(J)$ and residues $r_i^{\f\f}(J)$.
	}\label{fig:grid124ff}
\end{figure}

\begin{figure}[t]
	\centering
	\includegraphics[scale=0.8]{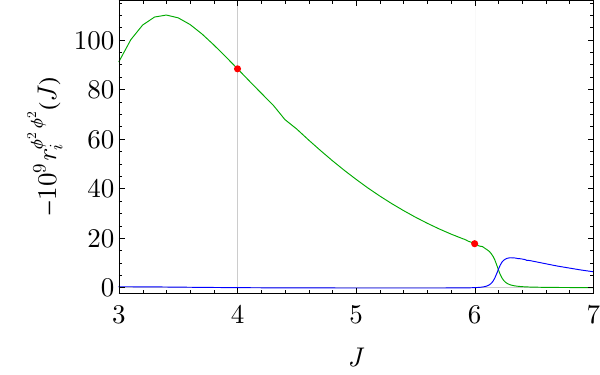}
	\caption{
		The residue $r_i(J)$ for trajectories 3 and 5, with trajectory 5 shown in green as in figure~\ref{fig:reggetrajectories_main}.
	}\label{fig:residue56}
\end{figure}

Using the methods described in the previous subsections we have computed the two-point functions, the matrix elements, and the corresponding residues $r_i(J)$ for the Regge trajectories labelled $1-5$ in figure~\ref{fig:reggetrajectories_main}. Our results are shown in figure~\ref{fig:grid124} and~\ref{fig:grid124ff} for trajectories $1,2,4$ and in figures~\ref{fig:grid56} and~\ref{fig:residue56} for trajectories $3,5$.

For the two-point functions, we plot the functions $C_i(J)=C(\psi_{i,J})$, defined in~\eqref{eq:CiJdefinition} and computed by~\eqref{eq:CpsiPairing}. The explicit expressions show that for real $J$ and real wavefunctions $\psi_{i,J}$ the function $C_i(J)$ has phase $e^{i\pi J/2}$. We therefore plot the real combination $e^{-\frac{i\pi J}{2}}C_i(J)$. Due to the numerical subtleties described above, we have only been able to reliably compute $C_i(J)$ up to spin $J\lesssim7$.

The two-point function plots show the expected behavior: the function $C_i(J)$ vanishes when there is a local operator present and is non-zero otherwise. For example, trajectory 2 doesn't contain local operators at $J=2,4$ but passes through a local operator at $J=6$ (see figure~\ref{fig:reggetrajectories_main}). Correspondingly, we see $C_2(2),C_2(4)\neq 0$ and $C_2(6)=0$ in figure~\ref{fig:grid124}.

Furthermore,~\eqref{eq:CiJderivative} allows us to predict the slope of $C_i(J)$ whenever there is a local operator on a Regge trajectory. For this, we examine the wave-function $\psi_{i,J}$ at such points to determine the value of the constants $\cN_{i,J}$ from their definition~\eqref{eq:NiJdefinition}. We represent the resulting slope by red dashed lines in~\ref{fig:grid124} and~\ref{fig:grid56}. There is a perfect agreement in all cases.

For the matrix elements, we plot the real combinations
\be
M^{\f^2\f^2}_i(J)&=e^{-i\pi J/2}\frac{\<0|\f^2\O_{\psi_{i,J}}\f^2|0\>}{\<0|\f^2\wL[\cO]\f^2|0\>_0},\\
M^{\f\f}_i(J)&=e^{-i\pi J/2}\frac{\<0|\f\O_{\psi_{i,J}}\f|0\>}{\<0|\f\wL[\cO]\f|0\>_0},
\ee
where the ratios in the right-hand side are computed using~\eqref{eq:ratiophi2final},~\eqref{eq:relationphi-phi3-final} and~\eqref{eq:finalPhiPhi^3}. The matrix elements for various trajectories are shown in figures~\ref{fig:grid124},~\ref{fig:grid124ff},~\ref{fig:grid56}.

Notably, the functions $M_i^{\f^2\f^2}(J)$ and $M_i^{\f\f}(J)$ do not appear to have any special features near even integer values of $J$. Using $M^{xx}_i(J)$ ($x\in \{\f,\f^2\}$) and $C_i(J)$ we can compute the residues $r_i(J)$ using~\eqref{eq:rJfinalFromDerivation}, which takes the form
\be
	r_i^{xx}(J)=\frac{2\sin \frac{\pi J}{2}}{\pi}\frac{M^{xx}_i(J)^2}{e^{-\frac{i\pi J}{2}}C_i(J)}.
\ee
In figures~\ref{fig:grid124},~\ref{fig:grid124ff} and~\ref{fig:grid56} the residues $r_i^{xx}(J)$ are shown together with the discrete points predicted from the OPE coefficients of local operators via~\eqref{eq:risminusff}. As expected, the analytic functions $r_i^{xx}(J)$ pass through these values whenever a local operator is present on the Regge trajectory. Whenever a local operator is absent for even integer $J$, we find $r_i^{xx}(J)=0$.

It is also interesting to look at the residues for trajectories 3 and 5 on the same plot, see figure~\ref{fig:residue56}. In this plot, we can see that the residues ``exchange'' places near the avoided intersection of these trajectories (shown in figure~\ref{fig:reggetrajectories_main} and discussed further in section~\ref{sec:diagonalizationresults}). This effect can also be seen to some extent in the plots of the two-point functions $C_i(J)$, but these are harder to interpret since they depend on the arbitrary normalisation of the wavefunctions. 

\section{Double-twist interpretation of twist-4 operators}
\label{sec:more}

In this section, we discuss the interpretation of the twist-4 operators as double-twist operators constructed from pairs of operators with twists less than 4.
Such twist-additivity has been derived non-perturbatively to hold in any CFT \cite{Fitzpatrick:2012yx,Komargodski:2012ek,Caron-Huot:2017vep,Pal:2022vqc}. The results of these papers imply the following structure of twist-four operators:
\begin{enumerate}
\item For every pair of operators $\cO_1$ and $\cO_2$ with $\tau_1+\tau_2=4+O(\epsilon)$, there should be a twist-4 trajectory $[\cO_1,\cO_2]_{0,J}$ with\footnote{The notation $[\cO_1,\cO_2]_{n,J}$ refers to primary operators with the schematic form $\cO_1 \square^n\partial^{\mu_1}\ldots \partial^{\mu_{j}}\cO_2$, where $j=J-J_1-J_2$. }
\begin{equation}
\label{eq:Jtoinftygamma1gamma2}
\lim_{J\to\infty} \gamma_{J}\to \gamma_1+\gamma_2.
\end{equation}
Moreover, since $2\Delta_\phi+2\approx 4$, we also expect to see a subleading trajectory $[\phi,\phi]_{1,J}$, with $\gamma_J\to \epsilon$, where the value $\epsilon$ follows from comparing our definition of the anomalous dimension $\tau_J=4-2\epsilon+\gamma_J$ with the expected asymptotic value $2\Delta_\phi+2=4-\epsilon+O(\epsilon^2)$.
\item The corrections to \eqref{eq:Jtoinftygamma1gamma2} should be proportional to a negative power of spin $J^{-\tau_{\mathrm{min}}}$ determined by the twist $\tau_{\mathrm{min}}$ of the lowest-twist non-trivial operator in either the joint OPEs $\cO_1\times \cO_1$ and $\cO_2\times \cO_2$, or the $\cO_1\x\cO_2$ OPE. In all our cases, this operator is $\Z_2$-even and therefore is at least a twist-2 operator. Thus we expect
\begin{equation}
\label{eq:largeSpinHyp}
\gamma_J = \gamma_1+\gamma_2+\frac{c}{J^2}+\ldots
\end{equation}
where $c$ could depend logarithmically on the spin.
\item The trajectory $[\cO_1,\cO_2]_{0,J}$ should contain local operators for even integer spins $J\geq J_{\mathrm{min}}$ for some $J_{\mathrm{min}}\geq J_1+J_2$, where $J_1$ and $J_2$ are the spins of $\cO_1$ and $\cO_2$.
\item For a pair $\cO_1$, $\cO_2$ of lower-twist operators, the contribution from twist-4 operators in OPE $\cO_1\times\cO_2$ should be dominated by the double-twist operators $[\cO_1,\cO_2]_{0,J}$, in the limit of large $J$.
\end{enumerate}
The operators $\cO_1$ and $\cO_2$ above can be chosen from the set of operators with twist $\leq3$. Such operators have been explicitly characterised~\cite{Kehrein:1992fn}, as we will now summarise. At twist 1, we only have the operator $\phi$, with $\gamma=O(\epsilon^2)$. At twist 2, we have the operator $:\phi^2:$ with $\gamma=\frac\epsilon3+O(\epsilon^2)$, and the (broken) currents \eqref{eq:OJdefn0} with $\gamma_J=O(\epsilon^2)$ and spin $J=2,4,\ldots$. At twist 3, we have operators $\cO^{\tau=3}_\ell$ with spins $\ell=2,3,4,\ldots$\footnote{
The operator $:\phi^3:$ is not present in the primary spectrum of the interacting theory.
} and \be\label{eq:anomt3}\gamma_\ell=\left(\frac13+\frac{2(-1)^\ell}{3(\ell+1)}\right)\e+O(\epsilon^2).\ee Moreover, starting at $\ell\geq 6$, there are additional twist-3 operators with $\gamma=O(\epsilon^2)$.\footnote{They are degenerate with a degeneracy $d_\ell$ that grows with spin and is characterised by the generating formula $\sum d_\ell q^\ell=\frac{q^6}{(1-q^2)(1-q^3)}=q^6+q^8+q^9+q^{10}+q^{11}+2q^{12}+\ldots$.}

The first of the points above, twist additivity for anomalous dimensions, has been carefully examined in the context of anomalous dimensions in $\phi^4$ theories. For twist-2 operators, it was discussed already in \cite{Callan:1973pu}, following a conjecture by Parisi. The works \cite{Kehrein:1995ia,Derkachov:1995zr} proposed general statements of twist-additivity for anomalous dimensions, giving rise to a hierarchial structure of anomalous dimensions, with accumulation points of accumulation points. This stucture was then proven in \cite{Derkachov:1996ph} for one-loop anomalous dimensions in scalar $\phi^4$ theory.\footnote{The argument can be extended at least to theories that are conformal at one-loop.}

The reminder of this section is aimed at addressing the remaining points. In section~\ref{sec:largeJ} we derive the $J\to\infty$ limit of anomalous dimensions of light-ray operators, and recover all expected eigenvalues. In section~\ref{sec:largespinlocal}, we turn to the corrections. A non-rigorous computation allows for the determination of $c$ in \eqref{eq:largeSpinHyp} from the integral form of the dilatation operator. We confirm these results numerically by generating anomalous dimensions of local operators up to spin 700 and perform some fits of the tail. We also comment on the third point, noting that the spin $J_{\mathrm{min}}$ of the first local operator of the double-twist trajectory is significantly higher than $J_1+J_2$ for all except the first few trajectories.

In section~\ref{sec:trace} we compute various traces of this dilatation operator which suggest that there might be Regge trajectories that are missing from our numerical analysis. If these trajectories exist, they do not contain local operators and do not have double-twist interpretations.
Finally, in section~\ref{sec:opes} we consider the OPE coefficients of twist-4 local operators in the OPEs of various pairs of lower-twist operators. We find, numerically, that the results for OPE coefficients agree with our expectations. More precisely, we find evidence that out of the total contribution from twist-4 operators in a given OPE $\cO_1\times \cO_2$, the double-twist family $[\cO_1,\cO_2]_{0,J}$ amounts for a fractional part that tends to 1 as $J\to\infty$. 
Some conventions and further explicit results for twist-4 operators are collected in appendix~\ref{app:moreLocalData}.

\subsection{Large-spin analysis from the integral operator}
\label{sec:largeJ}

In this subsection we analyse the dilatation operator $H$, defined in section~\ref{sec:twist4regges}, at large spin $J$. For simplicity, we will focus on even spin parity.
Note that the large-spin analysis of perturbative local operator spectrum has been peformed in~\cite{Kehrein:1995ia,Derkachov:1995zr,Derkachov:1996ph}, here we are interested in the non-local spectrum encoded by $H$. 

As discussed in section~\ref{sec:twist4regges}, $H$ has an (infinite) number of eigenfunctions with eigenvalue $0$, which span its kernel. The non-zero eigenvalues and the corresponding eigenfunctions can be obtained by considering the spectral problem for the operator $Q'$ defined in equations~\eqref{eq:Qprime1} and~\eqref{eq:Qprime2}, which we reproduce here for convenience,
\be\label{eq:Qprime1duplicate}
(Q'\Psi)_1(\chi)=&(\chi^{J+1}+(1-\chi)^{J+1})\int_0^1dz \Psi_2(z)+\chi^{J+1}\int_\chi^1 dz z^{-J-2} \Psi_2(z)\nn\\
&+(1-\chi)^{J+1}\int_{1-\chi}^1 dz z^{-J-2} \Psi_2(z)+\half\int_0^1 dz\Psi_1(z),\\
\label{eq:Qprime2duplicate}
(Q'\Psi)_2(\chi)=&\frac{1}{1-\chi}\int_0^{1-\chi}dz \Psi_2(z)+\frac{\chi^{J+1}}{1-\chi}\int_\chi^1 dz z^{-2-J}\Psi_1(z)+\frac{(1-\chi)^J}{2}\int_0^1 dz \Psi_1(z).
\ee
Recall that
\be
	H'=\frac{2Q'+1}{3}\e.
\ee
We therefore focus on the spectral problem for $Q'$.

Consider~\eqref{eq:Qprime1duplicate} for $\chi\in(0,1)$ and for large $\Re J$. Assuming that the wavefunctions $\Psi_i(\chi)$ tend to finite limits, we find that the first term is exponentially supressed. The second term can be approximated to the leading order as
\be
	\chi^{J+1}\int_\chi^1 dz z^{-J-2} \Psi_2(z)\approx \chi^{J+1}\Psi_2(\chi)\int_\chi^1 dz z^{-J-2}=\frac{\Psi_2(\chi)}{J+1}(1-\chi^{J+1}),
\ee
where we used that the $z$ integral is dominated by $z\approx \chi$. Therefore, this term is still subleading in $1/J$ expansion. Similar arguments apply to the third term in~\eqref{eq:Qprime1duplicate} and we find, to the leading order in $J$
\be
	(Q'\Psi)_1(\chi)=\half\int_0^1 dz\Psi_1(z)+O(J^{-1}).
\ee
A similar analysis of~\eqref{eq:Qprime2duplicate} shows
\be
	(Q'\Psi)_2(\chi)=\frac{1}{1-\chi}\int_0^{1-\chi}dz \Psi_2(z)+O(J^{-1}).
\ee

The leading order spectral problem for $Q'$ then becomes
\be
	\l \Psi_1(\chi)&=\half\int_0^1 dz\Psi_1(z),\label{eq:psi1leading}\\
	\l \Psi_2(\chi)&=\frac{1}{1-\chi}\int_0^{1-\chi}dz \Psi_2(z).\label{eq:psi2leading}
\ee
Supposing for the moment that $\l\not\in\{0,\half\}$, we conclude from the first equation that $\Psi_1(\chi)=0$. Indeed, $\l\neq 0$ implies that $\Psi_1$ is a constant, and $\l\neq \half$ implies that this constant must be $0$.

To solve for $\Psi_2(\chi)$, we multiply the second equation by $(1-\chi)$ and differentiate, yielding
\be\label{eq:trick1}
	-\l\Psi_2(\chi)+\l(1-\chi)\Psi'_2(\chi)=-\Psi_2(1-\chi).
\ee
Changing $\chi\to 1-\chi$ we get
\be\label{eq:trick2}
	-\l\Psi_2(1-\chi)+\l \chi \Psi'_2(1-\chi)=-\Psi_2(\chi).
\ee
Plugging in the expression for $\Psi_2(1-\chi)$ implied by~\eqref{eq:trick1}, we find a local differential equation
\be\label{eq:trick3}
	(1-\chi)\chi \Psi''_2(\chi)+(1-3\chi)\Psi_2'(\chi)-(1-\l^{-2})\Psi_2(\chi)=0.
\ee
This is a hypergeometric equation which at $\chi=0$ has solutions that behave as $\chi^0$ and $\chi^0\log\chi$. At $\chi=1$ it has solutions which behave as $(1-\chi)^{-1}+\# \log(1-\chi)$ and $(1-\chi)^{0}$. One can check that the both solutions solve~\eqref{eq:trick1} (and~\eqref{eq:trick2}), but in general not~\eqref{eq:psi2leading}. Indeed, consider the $\chi\to1$ limit of~\eqref{eq:psi2leading}. In the right-hand side, the $\chi^0$ solution will produce a finite limit, while $\chi^0\log\chi$ will produce a logarithmic singularity. However, neither solution produces a $(1-\chi)^{-1}$ singularity, and therefore only the solution which behaves as $(1-\chi)^0$ can solve~\eqref{eq:psi2leading}.

The $\chi\to 0$ limit of~\eqref{eq:trick2} then implies that the solution of the type $\chi^0\log\chi$  at $\chi=0$ is not allowed (since the left-hand side does not have logarithmic terms). Therefore, $\Psi_2$ is given by the solution analytic at $\chi=0$ which is, up to an inessential normalisation,
\be
	\Psi_2(\chi)={}_2F_1(1-\l^{-1},1+\l^{-1},1,\chi).
\ee
Plugging this into~\eqref{eq:psi2leading} we get
\be\label{eq:psi2leadingansatz}
	\l {}_2F_1(1-\l^{-1},1+\l^{-1},1,\chi)={}_2F_1(1-\l^{-1},1+\l^{-1},2,1-\chi).
\ee
We have 
\be
	{}_2F_1(1-\l^{-1},1+\l^{-1},2,1-\chi)=-\frac{\log\chi}{\G(1-\l^{-1})\G(1+\l^{-1})}+\ldots,
\ee
so the equation can only be satisfied if $\l^{-1}=k$ for some integer $k\neq 0$. In this case, both sides of~\eqref{eq:psi2leadingansatz} are polynomials, and we find that~\eqref{eq:psi2leadingansatz} is satisfied if $k=(-1)^{\ell}(\ell+1)$ with $\ell$ a non-negative integer.

Above, we assumed that $\l\not \in \{0,\half\}$. If $\l=0$, then differentiating~\eqref{eq:psi2leading} we find $\Psi_2(\chi)=0$, and~\eqref{eq:psi1leading} is solved by any $\Psi_1$ which satisfied $\int_0^1 d\chi \Psi_1(\chi)=0$, giving an infinitely-degenerate eigenvalue. If $\l=\half$ then~\eqref{eq:psi1leading} implies, up to normalisation, $\Psi_1(\chi)=1$. On the other hand, the above analysis for $\Psi_2(\chi)$ remains valid for $\l=\half$. Since we didn't find any solutions for $\Psi_2(\chi)$ with $\l=\half$, $\Psi_2(\chi)=0$ in this case.

To summarise, we find the following spectrum for $Q'$ at large $\Re J$:
\begin{enumerate}
	\item $\l=\thalf$, $\e^{-1}\g=\frac{2}{3}$: a simple eigenvalue with $\Psi_1(\chi)=1$ and $\Psi_2(\chi)=0$. This trajectory can be interpreted as the double-twist trajectory $[\f^2,\f^2]_{0,J}$.
	\item $\l=(-1)^{\ell} (\ell+1)^{-1}$, $\e^{-1}\g=\frac{1}{3}+\frac{2(-1)^\ell}{3(\ell+1)}$: A simple eigenvalue with $\Psi_1(\chi)=0$ and 
	\be
	\Psi_2(\chi)={}_2F_1(-\ell,\ell+2,1,\chi)=P^{(0,1)}_\ell(1-2\chi),
	\ee
	where $\ell\in\Z_{\geq 0}$ and $P^{(a,b)}_\ell(z)$ is the Jacobi polynomial. For $\ell\geq 2$, this can be interpreted as the double-twist trajectory $[\f,\cO_\ell^{\tau=3}]_{0,J}$, where $\cO_\ell^{\tau=3}$ the unique spin-$\ell$ twist-3 primary which has a non-zero one-loop anomalous dimension \eqref{eq:anomt3}. The case $\ell=0$ can be interpreted as the $[\f,\f]_{1,J}$ double-twist family. The case $\ell=1$ gives the large-$J$ limit of $\Psi_0$ (see~\eqref{eq:kernel}) and should therefore be discarded from the physical spectrum.\footnote{Correspondingly, recall that there is no spin-$1$ twist-3 primary since $\f^2\ptl_\mu\f=\frac{1}{3}\ptl_\mu(\f^3)$.} \item $\l=0$, $\e^{-1}\g=\frac{1}{3}$: An infinitely-degenerate eigenvalue with $\Psi_2(\chi)=0$ and $\Psi_1(\chi)$ any function subject to $\int_0^1 d\chi \Psi_1(\chi)=0$. These can be interpreted as the double-twist trajectories $[\f^2,\cO]_{0,J}$, where $\cO$ is any of the infinitely-many twist-2 primaries (of spin unrelated to $J$) with vanishing one-loop anomalous dimension.
\item $\e^{-1}\g=0$: there also exist infinitely many trajectories with $\g=0$, which are not in the spectrum of $H'$ but are in the spectrum of $H$. These include families $[\f,\cO]_{0,J}$ where $\cO$ is a twist-3 primary with vanishing 1-loop anomalous dimension, as well as families $[\cO,\cO']_{0,J}$, where $\cO$ and $\cO'$ are twist-2 primaries with vanishing 1-loop anomalous dimensions.
\end{enumerate}
For the counting of operators with $\gamma=0$ and $\gamma=\frac13$, see appendix~\ref{app:rationalGammas}.

\subsection{Numerical local operator data and corrections to the infinite spin limit}
\label{sec:largespinlocal}

In this section we present the results for the computation of anomalous dimensions of local twist-4 operators of even spin up to 700. The computation proceeds exactly as outlined in section~\ref{eq:simplifyingDilatationOperator}, and uses the reduced dilatation operator $H'$ of \eqref{eq:Dptwovar} acting on polynomial wave functions $\Psi(p,q)$.
This computation captures all local operators, except those with $\gamma=0$. However, we know precisely how many such local operators there are (see equation \eqref{eq:genfgamma0} in the appendix), and we can add them to the spectrum by hand.

\paragraph{Large spin fits}
\begin{table}[t]
	\caption{Trajectories of twist-4 operators, and spins of the local operators in the trajectories, not including the trajectories with anomalous dimensions $\gamma=\frac13$ or $\gamma=0$. Here $\mathcal J^2=(J+1)(J+2)$. The operators $\cO_{\ell}^{\tau=3}$ are the twist-three operators of spin $\ell$ with non-vanishing $O(\epsilon)$ anomalous dimensions. The general-$k$ expression for the large-$J$ expansion is valid for $k\neq 2$. Here we take the avoided crossing between trajectories 3 and 5 into account.}\label{tab:twistfamilies}
	{
		\def\arraystretch{1.6}
		\begin{tabular}{|c|l|c|c|}\hline
			$k$ & \multicolumn{1}{c|}{Spin} & Large-$J$ expansion  & Double-twist
			\\\hline\hline
			$1$ & $0,\,2,\,4,\ldots$ & $ 1+\frac{16}{\mathcal J^2}+\frac{-1280+192S_1(J+1)}{\mathcal J^4}+\ldots $ & $[\phi,\phi]_{1,J}$
			\\\hline
			$2$ & $6,\,8,\,10,\ldots$  & $\frac23+\frac{-88+32S_1(J+1)}{3\mathcal J^2}+\ldots$  & $[\phi^2,\phi^2]_{0,J}$
			\\\hline
			$3$ & $8,\,10,\ldots$ & $\frac59+\frac0{\mathcal J^2}-\frac{1280}{\mathcal J^4}+\ldots$ & $[\phi,\cO_2^{\tau=3}]_{0,J}$
			\\\hline
			$4$ & $8,\,10,\,\ldots$ & $\frac16-\frac{224}{9\mathcal J^2}+\ldots$ & $[\phi,\cO_3^{\tau=3}]_{0,J}$
			\\\hline
			$5$ & $4,\,6,\,12,\ldots$ & $\frac7{15}+\frac{560}{9\mathcal J^2}+\ldots$ & $[\phi,\cO_4^{\tau=3}]_{0,J}$
			\\\hline
			$6$ & $14,\,16,\ldots$ & $\frac29+\frac{72}{\mathcal J^2}+\ldots$ & $[\phi,\cO_5^{\tau=3}]_{0,J}$
			\\\hline
			\vdots  & \multicolumn{3}{c|}{}
			\\\hline
			$k$ & $2\!\left\lfloor\frac{3(k-1)}2\right\rfloor\!,\ldots\!$   &  $\frac13-\frac{2}{3m}-\frac{8(-1)^mm(m-2)(m+3)}{3(m+2)\mathcal J^2}+\cdots,\,m=(-1)^kk$ & $[\phi,\cO_{k-1}^{\tau=3}]_{0,J}$
			\\\hline
		\end{tabular}
	}
\end{table}

By using the anomalous dimensions of operators at large spin, we can make a precise numerical fit of the large-spin expansion of each family.
By matching with simple rational expressions in $\mathcal J^2=(J+1)(J+2)$ and $S_1(J+1)$,\footnote{$S_1(J+1)=\psi(J+2)+\gamma_E$ denotes the analytic continuation of the harmonic numbers.} we have also been able to identify the $1/\mathcal J^2$ correction to the anomalous dimension for a number of families, and the $1/\mathcal J^4$ correction for the first two families.\footnote{We identified these coefficients by performing numerical fits matching with anomalous dimensions computed for the 51 highest spins ($J=600, 602, \ldots,700$).} These results are presented in table~\ref{tab:twistfamilies}.

For $k\neq 2$ we have been able to derive the $1/J^{2}$ correction analytically, via an extension of the analysis in section~\ref{sec:largeJ}. Since this derivation is rather technical and we couldn't fully justify all the steps, we do not present it in this paper. The analytic result is presented in table~\ref{tab:twistfamilies} and we merely note that it agrees with our numerics in all cases that we have verified.

\paragraph{Large-spin asymptotic region}

\begin{figure}
  \centering
\includegraphics[scale=0.72]{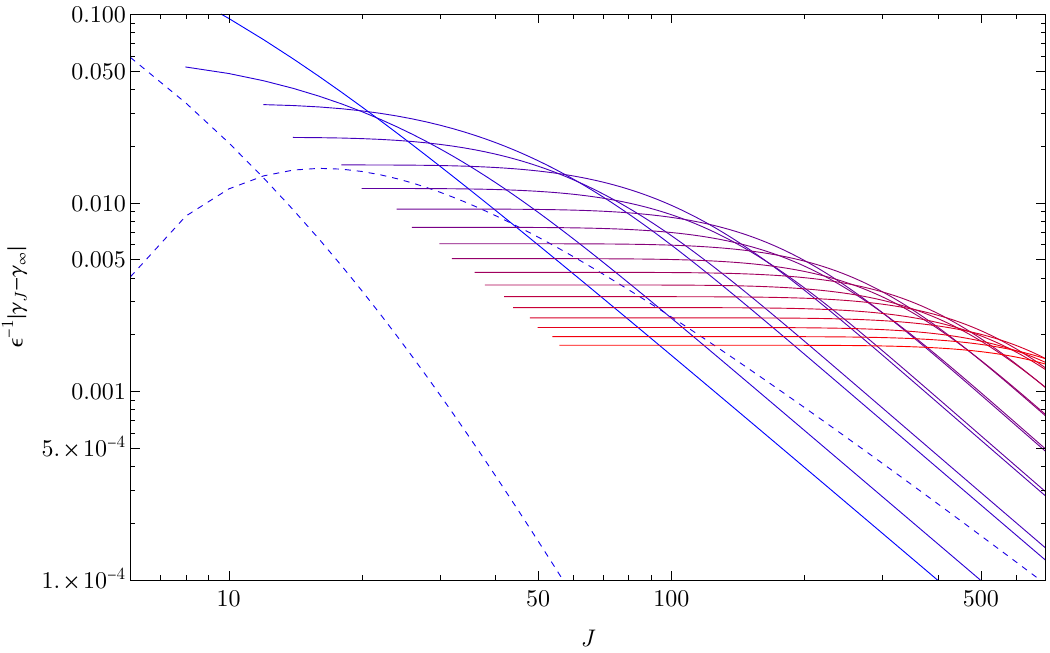}
  \caption{Plot of anomalous dimensions minus the asymptotic values for the first 20 families, with small $k$ shown in blue and large $k$ in red. The values are interpolated for presentation purposes. Actual data points are only present at even integer spins. The trajectories appear to intersect in this plot only because they have been shifted by $\g_\oo$. The dashed trajectories are $k=2,3$ which are the only trajectories for which the leading correction is not $J^{-2}$ (cf. table~\ref{tab:twistfamilies}).}\label{fig:gamma-asymall}
\end{figure}

After establishing the large-spin expansions of the anomalous dimensions, it is interesting to ask what spin is large enough to be in the asymptotic region where the large-$J$ expression is valid. To this end, we plot the value of $\gamma_J-\gamma_\infty$ for a representative sample of families, see figure~\ref{fig:gamma-asymall}. 
In this figure, shown in log-log scale, the $J^{-2}$ dependence at large spin is clearly visible as a approximately parallel lines with negative slope at larger spins for all trajectories except two.

From figure~\ref{fig:gamma-asymall}, we see that it requires larger and larger spin until we reach the large-spin asymptotic region. We can make a rough estimate of this point by looking at the inflection point in $\gamma_J$ as a function of $J$. Only for families with $k\leq 18$, the inflection point is at $J<700$. By looking at the spin of the inflection point, we find roughly a quadratic dependence on $k$.

\paragraph{Spin of first local operator}
A related issue is the question of the lowest spin of a local operator in a given double-twist trajectory.  Naively, one might expect that for a pair of operators $\cA$ and $\cB$ with spins $J_A$ and $J_B$ their (local) double-twists should exist for $J\geq J_A+J_B$. In the case at hand, for $k>2$ this gives $J\geq k-1$ (cf.\ table~\ref{tab:twistfamilies}). Instead, we find empirically that the smallest local spin in trajectory $k$ is $J=2\lfloor{\frac{3(k-1)}{2}}\rfloor$. This is greater than $k-1$ already at $k=3$, and asymptotically at large $k$ gives $3k+O(1)$. From the perturbation theory perspective we can explain this phenomenon by permutation symmetries between the fundamental field that enter into $\cA$ and $\cB$.

Overall, we see that while all our Regge trajectories have a double-twist interpretation, it only sets in at a spin $J$ that grows with $k$.

\paragraph{Small-spin asymptotic}

From figure~\ref{fig:gamma-asymall}, we see the emergence of a region at small spin, where the anomalous dimension in a given trajectory is approximately constant. 
From considering trajectories with $k$ index from $\approx 10$ and upwards, it's possible to determine rational expressions for this small-spin asymptote. Specifically, we find empirically that
\begin{equation}\label{eq:smallspinasymptote}
\lim_{k\to \infty} \gamma_{J_{\mathrm{min}}}^{(k)}-\left(\frac13-\frac{2 (-1)^k}{3 (k-1)}\right)=0.
\end{equation}
One can then observe that all trajectories with $k\geq6$ are confined in the interval between the small-spin and the large spin asymptotes,
\begin{equation}
\gamma_J^{(k)}\in\left[ \frac13-\frac{2 (-1)^k}{3 (k-1)}, \,
\frac13+\frac{2 (-1)^{k+1}}{3 k}\right], \quad J\in \mathcal J_{k},
\end{equation}
and that the local operator anomalous dimensions move monotonically between the two asymptotes as $J$ increases. In fact, this observation was made already in \cite{Derkachov:1995zr}.

\subsection{The reduced trace}
\label{sec:trace}
In this section we study the trace of the reduced operator $Q'$ (recall $H'=(2Q'+1)\e/2$), given by equations~\eqref{eq:Qprime1duplicate} and \eqref{eq:Qprime2duplicate}, as well as of its square $(Q')^2$. These traces allow us to check whether we found all eigenstates of $Q'$. As we will see, this gives some indications that there might exist Regge trajectories which do not have double-trace interpretations.

First, we compute the traces. This can be done by writing down the kernel corresponding to operator $Q'$ via
\be
	(Q'\Psi)_i(\chi)=\sum_{j=1}^2 \int_0^1 d\chi Q'_{ij}(\chi,\chi') \Psi_j(\chi').
\ee
Using~\eqref{eq:Qprime1duplicate} and~\eqref{eq:Qprime2duplicate} we find
\be
\label{eq:kernelsggff}
Q'_{11}(\chi,\chi')&=\frac12,\qquad\qquad Q'_{22}(\chi,\chi')=\frac1{1-\chi}\Theta(1-\chi-\chi'),\nn\\
Q'_{12}(\chi,\chi')&= \frac{\chi^{J+1}}{\chi'^{J+2}} \Theta (\chi' -\chi)+\frac{ (1-\chi)^{J+1}}{\chi'^{J+2}} \Theta (\chi-(1-\chi'))+\chi^{J+1}+(1-\chi)^{J+1},\nn\\
Q'_{21}(\chi,\chi')&=\frac{(1-\chi)^J}2+\frac{ \chi^{J+1}}{(1-\chi)\chi'^{J+2}}.
\ee
The traces can then be written as
\be
	\tr Q'&=\sum_i \int_0^1 d\chi Q'_{ii}(\chi,\chi),\\
	\tr (Q')^2&=\sum_{ij} \int_0^1 d\chi d\chi' Q'_{ij}(\chi,\chi')Q'_{ji}(\chi',\chi).
\ee
These integrals can be performed explicitly and they yield
\be
	\label{eq:traceQprimDirect}
	\tr Q'&=\log2+\frac12,\\
	\label{eq:traceQprim2Direct}
	\tr (Q')^2&=\frac14+\zeta(2)+\frac4{(J+1)(J+2)}+2\psi^{(1)}(J+2)+2\left(\psi^{(1)}(\tfrac{J+2}2)-\psi^{(1)}(\tfrac{J+3}2)\right).
\ee
It might be possible to compute higher power traces, although we have not attempted this.

These values allow us to make some interesting observations. First of all, the values of the traces at large $J$ agree with the expectation based on the large-$J$ diagonalisation in section~\ref{sec:largeJ}.\footnote{Note that it is the actual spectrum of $Q'$ and not the physical spectrum (which is derived from it) which is important here. In particular, we have to include the $\ell=1$ case.} Indeed, the latter yields at $J=\oo$
\be
	\tr Q'&=\frac{1}{2}+\sum_{\ell=0}^\infty\frac{(-1)^{\ell}}{\ell+1}=\frac1{2}+\log 2,\\
	\tr (Q')^k&=\frac{1}{2^k}+\sum_{\ell=0}^\infty\frac{(-1)^{\ell k}}{(\ell+1)^k}=\frac1{2^k}+\left(1-\frac{1-(-1)^k}{2^k}\right)\zeta(k), \qquad k=2,3,\ldots.
\ee
This agrees with~\eqref{eq:traceQprimDirect} and~\eqref{eq:traceQprim2Direct} at large $J$.

This provides strong evidence that we have found the entire asymptotic $J=\oo$ spectrum, modulo possible $\l=0$ eigenvalues, in section~\ref{sec:largeJ}. However, there are several indications that we have not found the full spectrum at finite $J$.

First of all, we can attempt to compute these traces numerically. Of course, our numerics in section~\ref{sec:numericalscheme} are not able to detect eigenvalues in region $\cB$. However, in section~\ref{sec:largespinlocal} we observed that  at a given $J$, we can label Regge trajectories by $k$ so that in the limit of large $k$ the anomalous dimension appears to be given by~\eqref{eq:smallspinasymptote}, and the asymptotic appears to be reached very quickly and outside of $\cB$, at least for large enough values of $J$. We can then take the largest eigenvalues from the numerics of section~\ref{sec:numericalscheme} and the 
rest from~\eqref{eq:smallspinasymptote}, which can be summed up analytically. Doing so, we find values which do not agree with equations~\eqref{eq:traceQprimDirect} and~\eqref{eq:traceQprim2Direct}.

Perhaps more convincingly, we can compute the traces for odd spin trajectories. Rederiving the operator $Q'$ in that case yields a kernel that differs from~\eqref{eq:kernelsggff} by a minus sign in front of terms which contain $\chi'^{-J-2}$. The resulting traces are
\be
\tr Q'_\text{odd}&=\log2+\frac12,\\
\tr (Q'_\text{odd})^2&=\frac14+\zeta(2)-2\psi^{(1)}(J+2).
\ee
Notably, $\tr (Q'_\text{odd})^2$ is not constant. This is in conflict with the observation~\cite{Derkachov:1995zr} that the anomalous dimensions in odd-spin case are spin-independent rational numbers (which we have also verified numerically using the methods of section~\ref{sec:numericalscheme}).

This suggests that the numerical diagonalisation of section~\ref{sec:numericalscheme} is missing some eigenstates (which should be confined to a neighborhood of region $\cB$). It would be interesting to understand these missing eigenstates and analyze whether they define legitimate light-ray operators. If so, these would be Regge trajectories which do not contain any local operators.

\subsection{OPE coefficients of local operators}
\label{sec:opes}

Our final topic is the OPE coefficients, where we would like to study the coefficients of twist-4 operators in the OPE of two operators of lower twist. To this end, we construct local operators and compute the three-point functions by Wick contractions.

In the free theory, local operators are constructed out of normal-ordered products of fields with partial derivatives.~\cite{Kehrein:1992fn,Kehrein:1994ff,Kehrein:1995ia,Hogervorst:2015akt,Liendo:2017wsn,Henriksson:2022rnm}. For traceless-symmetric operators we use the compact one-loop dilatation operator given in~\cite{Kehrein:1995ia}
\begin{equation}
\hat V_1=\frac16\sum_{l=0}^\infty \frac1{l+1}\sum_{k,j=0}^l a_j^\dagger a_{l-j}^\dagger a_ka_{l-k},
\end{equation}
where the creation and annihilation operators act on expressions $\mathcal O= \Phi_{j_1}\cdots \Phi_{j_m}$, and $a_j^\dagger|0\rangle =\Phi_j|0\rangle = \frac1{j!}(z\cdot \partial)^j\phi(x)|0\rangle$. It is straightforward to write down the action of $\hat V_1$ on a basis of opertors with total spin $j_1+\ldots+j_m=J$, and the resulting eigenfunctions are the scaling operators of spin $J$. To determine which scaling operators are primaries, we apply the generator $K_\mu$ of special conformal transformation, and use the Gram--Schmidt algorithm to determine a basis in the space of primaries that's orthonormal with respect to the two-point pairing \eqref{eq:standard2pt}.\footnote{This step is needed when considering the degenerate cases $\gamma=0$ and $\gamma=\frac13$.} The result is explicit forms of twist-4 operators of spin up to some moderately large spin (in practice, we stop at spin $30$). Using these, the OPE coefficients can be extracted by computing the three-point functions using Wick contractions. We stress that the OPE coefficients are computed in the free theory, but using operators constructed by the one-loop dilatation operator.

We consider the OPE coefficients of all local twist-4 operators of even spin $J\leq30$ in the following OPEs:
\begin{align}
\nonumber
&\phi\times\phi,&& \phi^2\times \phi^2, && \phi^2\times T,&& \phi^2\times C,&& T\times T,
\\
&\phi\times \cO_2^{\tau=3},&& \phi\times \cO_3^{\tau=3},&& \phi\times \cO_4^{\tau=3},&& \phi\times \cO_5^{\tau=3},&&\phi\times \cO_6^{\tau=3},&&\phi\times \cO_{6,\gamma=0}^{\tau=3}  ,
\label{eq:collectionOfOPEs}
\end{align}
where $T$ is the stress-tensor, $C$ is the spin-4 operators in the twist-2 family, $\cO_\ell^{\tau=3}$ are the twist-3 operators with spin $\ell$ and non-zero one-loop anomalous dimension \eqref{eq:anomt3}, and $\cO_{6,\gamma=0}^{\tau=3}$ is the twist-3 operator with spin 6 and $\gamma=0$. Explicit expressions for these operators are given in appendix~\ref{eq:explicitForms}.

Thus we compute the canonically normalised free-theory OPE coefficients $\lambda_{\cO_1\cO_2\cO_3}$, where $\cO_3$ is a twist-$4$ operator of spin $J_3=J$, and where $\cO_1$ and $\cO_2$ are chosen from the collection in \eqref{eq:collectionOfOPEs}.\footnote{For $\cO_1$ and $\cO_3$ with non-zero spin, we expect, generically, more than one three-point structure and therefore not a unique OPE coefficient. It turns out, however, that in the ``extremal'' case at hand, where $\tau=3=\tau_1+\tau_2$, only one three-point structure survives, and we can give meaning to a unique OPE coefficient. See appendix~\ref{app:OPEdefs} for details.
}
 We define the total OPE coefficient as the sum
\begin{equation}
\overline{\lambda^{(0)}_{\cO_1\cO_2}(J)^2} = \sum_{i=1}^{d_J}(\lambda^{(0)}_{\cO_1\cO_2\cO_{i,J}})^2
\end{equation}
over the $d_J$ operators of twist $4$ with spin $J$, which are degenerate in the free-theory limit, and $\cO_{i,J}$ belongs to trajectory $i$. 
From now on, we suppress the superscript $^{(0)}$ and remember that we always work with leading-order OPE coefficients. We can then consider the contribution from operators in the various trajectories, by studying the ratio
\begin{equation}
\nu_{\cO_1\cO_2}(i,J)= (\lambda_{\cO_1\cO_2\cO_{i,J}})^2 \big/\,\overline{\lambda_{\cO_1\cO_2}(J)^2}.
\end{equation}

\begin{figure}
  \centering
\includegraphics[width=130mm]{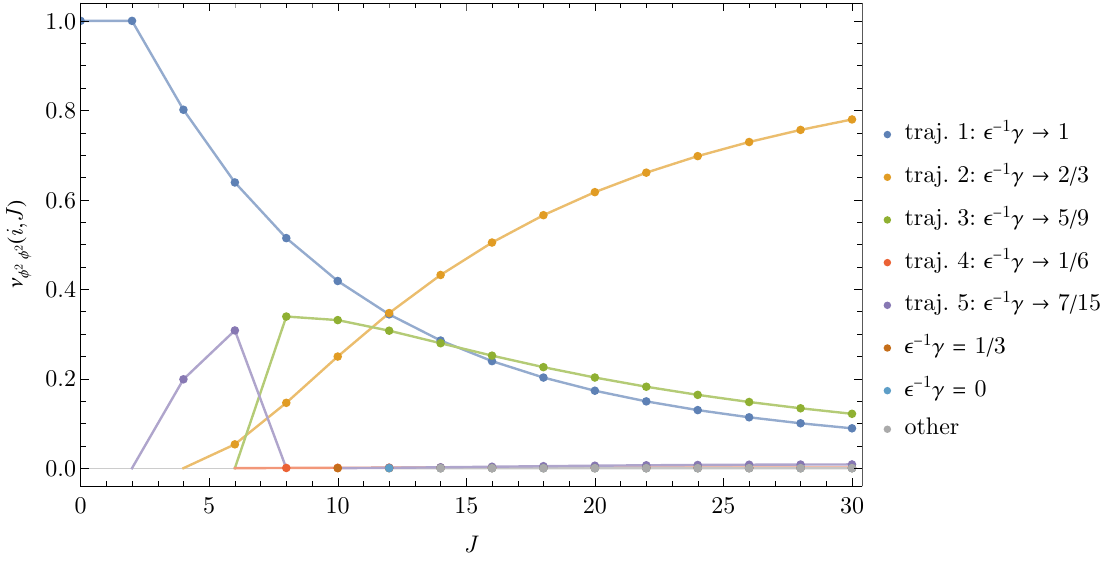}
  \caption{Ratios of OPE coefficients in the $\phi^2\times \phi^2$ OPE. The connecting lines are only to guide the eye -- the interpolation for complex spin was discussed in section~\ref{sec:resultsOPEcoefs}. 
}\label{fig:OPEratios22}
\end{figure}

\begin{figure}
  \centering
\includegraphics[width=130mm]{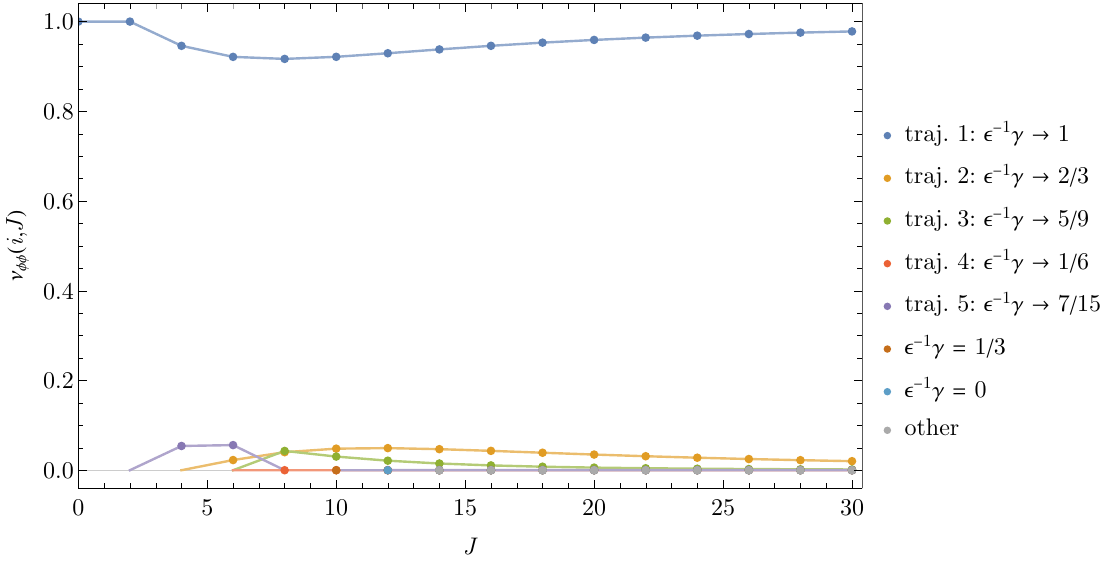}
  \caption{Ratios of OPE coefficients in the $\phi\times \phi$ OPE.}\label{fig:OPEratios11}
\end{figure}

We report the ratios $\nu_{\cO_1\cO_2}(i,J)$ from the first five trajectories, as well as the share from operators with eigenvalues $0$ and $\frac13$, and from remaining operators. The colour-coding is the same as in figure~\ref{fig:trajectories}. In figure~\ref{fig:OPEratios22} we consider the case of the $\phi^2\times\phi^2$ OPE, in figure~\ref{fig:OPEratios11} the $\phi\times\phi$ OPE.\footnote{These are derived by computing the OPE coefficients $\lambda_{\phi\phi^3\cO}$, and using the equation-of-motion relation between $\phi$ and $\phi^3$, discussed in section~\ref{sec:matrixelements}.} The remaining cases are reported in appendix~\ref{app:moreLocalData}, together with further explicit results.

In all cases, we find that for large enough spin, the family whose anomalous dimension $\gamma_J$ approaches $\gamma_1+\gamma_2$ of the operators $\cO_1$ and $\cO_2$ dominates the OPE. For instance, in the $\phi^2\times\phi^2$ OPE it is the family 3 with $\gamma\to\frac23$ that dominates (figure~\ref{fig:OPEratios22}). The $\phi\times\phi$ OPE is dominated by operators with $\gamma\to1$ (figure~\ref{fig:OPEratios11}). This is consistent with the interpretation of these operators as the subleading trajectory $[\phi,\phi]_{1,J}$ with $\tau\to 2\tau_\phi+2=4-\epsilon+O(\epsilon^2)$.

Another way to reach the same conclusion is to study the weighted average of anomalous dimensions. By matching with simple expressions, it is possible to find closed-form formulas for 
\begin{equation}
\overline{\lambda_{\cO_1\cO_2}(J)^2\gamma(J)} = \sum_{i=1}^{d_J}(\lambda_{\cO_1\cO_2\cO_{J,i}})^2\gamma_{i,J}.
\end{equation}
Let us give explicit results for the case $\varphi^2\times\varphi^2$, where $\varphi^2$ is canonically normalised version of $\phi^2$. 
We find the expressions
\begin{align}
\overline{\lambda^2_{\varphi^2\varphi^2}(J)}&=\frac{\Gamma(2+J)^2}{\Gamma(2J+3)}\left(2(J+1)(J+2)+8\right),
\\
\overline{\lambda^2_{\varphi^2\varphi^2}(J)\gamma(J)}&=\frac{\Gamma(2+J)^2}{3\Gamma(2J+3)}\left(4 ( J+1)(J+2)+32+32 S_1(J+1)\right)\epsilon ,
\label{eq:agamma22}
\end{align}
which (expectedly) agree with results computed by decomposing the free-theory and one-loop correlators~\cite{Dolan:2000ut,Henriksson:2020jwk,Bertucci:2022ptt}.\footnote{It is also straightforward to check the asymptotic agreement with the OPE coefficients of a generalised free field~\cite{Fitzpatrick:2011dm}, which for the case at hand takes the form $\frac{2\Gamma(2+J)^2(J+1)(J+2)}{\Gamma(2J+3)}$.} The dominance in the OPE of operators with $\gamma\to\frac23\epsilon$ can now be seen by noting that
\begin{equation}
\overline{\lambda_{\varphi^2\varphi^2}(J)^2\gamma(J)}\Big/\overline{\lambda_{\varphi^2\varphi^2}(J)^2}\to\frac23\varepsilon .
\end{equation}

\section{Discussion}
\label{sec:discussion}

In this paper we have outlined a self-consistent picture of Regge trajectories for higher-twist operators in general conformal field theories. In this picture, all local operators belong to Regge trajectories, and there is an infinite number of Regge trajectories of bounded twist. For a finite integral spin, most of these Regge trajectories decouple in the sense that the corresponding residues in the function $C(\De,J)$ derived from any four-point function vanish. This decoupling is explained by equation~\eqref{eq:mainformulaIntro}, which contains an explicit prefactor that vanishes for integer spin. If a Regge trajectory contains a local operator, the two-point function in the denominator of~\eqref{eq:mainformulaIntro} vanishes, cancelling the zero in the prefactor, and allowing the residue to be non-zero.

We have verified this picture by performing a detailed one-loop analysis of twist-4 local operators in the Wilson--Fisher CFT in $d=4-\e$. Using numerics, we have been able to largely diagonalise the one-loop dilatation operator for a general spin $J\in\C$ and determine the explicit shape of the twist-4 Regge trajectories. We have confirmed explicitly that all twist-4 local operators appear to lie on Regge trajectories. Moreover, we have obtained the explicit eigenfunctions of the light-ray operators, which allowed us to verify~\eqref{eq:mainformulaIntro} and the decoupling mechanism in several non-trivial cases.

By tracking the twist-4 Regge trajectories to large $J$ we saw that they all can be identified with double-twist trajectories of lower-twist operators. In particular, we did not see any evidence that genuinely ``triple-twist'' or ``quadruple-twist'' Regge trajectories exist. The double-twist behavior of the Regge trajectories is however delayed to very large spin $J$ as the spin of the constituent lower-twist operators grows. It is therefore interesting to ask if there is any universal behavior that can be predicted for smaller values of $J$, and we expect that these questions should be answered using higher-point analogues of light-cone bootstrap.

Although we have identified Regge trajectories which pass through all local operators, in section~\ref{sec:trace} we computed traces of the dilatation operator and found that their the values cannot be accounted for using only our numerical spectrum. This suggests the possibility that there might exist twist-4 Regge trajectories which do not contain any local operators and which are not captured by our numerics. While equation~\eqref{eq:mainformulaIntro} makes sure that such Regge trajectories decouple from local OPEs, they would still contribute to Lorentzian observables such as event shapes~\cite{Kologlu:2019mfz,Chang:2020qpj,Korchemsky:2019nzm,Dixon:2019uzg} or the Regge limit of four-point functions~\cite{Costa:2012cb,Cornalba:2007fs}.

It is therefore an important open problem to obtain the full picture of the spectrum of the operator $H$ defined in~\eqref{eq:Dprime}, and in particular of any eigenstates in the region $\cB$ in which the wavefunctions become singular. It would be especially interesting to study the spectrum in the left half-plane of complex $J$, where the twist-4 trajectories intersect with shadow trajectories of twist-2 (and higher-twist) operators, as well as with infinitely-many horizontal trajectories discussed in~\cite{Caron-Huot:2022eqs}.

Answering these questions will bring us closer to understanding the perturbative structure of the Chew--Frautschi plot in the Wilson--Fisher theory in $\Z_2$-even parity-even sector. More broadly, one can consider other symmetry sectors. For example it would be interesting to study the twist-3 Regge trajectories in $\Z_2$-odd sector. We didn't consider it in this paper because the degeneracy between twist-3 Regge trajectories isn't broken at one-loop (except for the $[\f,\f^2]_{0,J}$ trajectory). However, at two loops the degeneracy should be broken and one should be able to see the full structure of leading-twist $\Z_2$-odd trajectories.\footnote{We note in passing that numerical studies such as~\cite{Simmons-Duffin:2016wlq} are mostly sensitive to those local operators which couple to external primaries with large OPE coefficients. For example, among all twist-4 trajectories that we identified, only $[\f,\f]_{1,J}$ and $[\f^2,\f^2]_{J}$ can be seen unambiguously in~\cite{Simmons-Duffin:2016wlq} who studied four-point functions involving $\f$ and $\f^2$ only (at $d=3$). 
However, although our results from figures~\ref{fig:OPEratios22} and~\ref{fig:OPEratios11} show there is a range of spin where three twist-four operators have significant OPE coefficients in both $\phi\times\phi$ and $\phi^2\times\phi^2$ at leading order in $\epsilon$, the numerical bootstrap study \cite{Henriksson:2022gpa} only observed two of these operators at spins in this range (with a possible exception at spin 6), even at small $\epsilon$.
} In the case of $\mathrm{O}(N)$-symmetric $\phi^4$ theory, the action of the two-loop dilatation operator has been determined on a subclass of local operators that are completely traceless-symmetric with respect to $\mathrm{O}(N)$ \cite{Kehrein:1995ia}, and it would be interesting to study the corresponding action on light-ray operators.

\section*{Acknowledgments}

The authors would like to thank Simon Caron-Huot, Aleix Gimenez-Grau, Matthijs Hogervorst, Alexandre Homrich, Mark van Loon, Pedro Liendo, Alexander Manashov, Jeremy Mann, Erik Panzer, Slava Rychkov, David Simmons-Duffin, Andreas Stergiou and Pedro Viera for useful discussions. The work of PK was supported by the UK Engineering and Physical Sciences Research council grant number EP/X042618/1; and the
Science Technology \& Facilities council grant number ST/X000753/1. 
The work of JH was supported by the ERC Starting Grant 758903, and by the ERC Starting Grant 853507.

\appendix

\newpage
\section{An integral for twist-2 two-point function}
In this appendix we compute the integral~\eqref{eq:integralforapp}. We have
\label{app:integral}
\be
&\<\O_J(X,Z)\O_J(X',Z')\>\nn\\
&=2\int d\a_1 d\a_2 d\a'_1 d\a'_2 \psi_J(\a_1,\a_2)\psi_J(\a'_1,\a'_2)\<\f(X_{\a_1})\f(X'_{\a'_1})\>\<\f(X_{\a_2})\f(X'_{\a'_2})\>\nn\\
&=\frac{1}{8\pi^4}\int d\a_1 d\a_2 d\a'_1 d\a'_2 \frac{\psi_J(\a_1,\a_2)\psi_J(\a'_1,\a'_2)}{(1+\a_1\a_1'+i\e)(1+\a_2\a_2'+i\e)}\nn\\
&=\frac{1}{8\pi^4\G(-\tfrac{J+2}{2})^2}\int d\a_1 d\a_2 d\a'_1 d\a'_2 \frac{|\a_1-\a_2|^{-J-1}|\a'_1|^{-J-1}}{(1+\a_1(\a_1'+\a_2')+i\e)(1+\a_2\a_2'+i\e)}.
\ee
We have shifted $\a_1\to\a_1+\a_2$ and $\a'_1\to\a'_1+\a'_2$ to get to the last line. The integral over $\a'_2$ has poles at
\be
\a'_2=\frac{-1-i\e}{\a_2},\qquad \a'_2=\frac{-1-i\e}{\a_1}-\a'_1.
\ee
Unless $\a_1$ and $\a_2$ have opposite signs, these poles are on the same side of integration contour and the integral vanishes. So, we get
\be
&\int d\a_1 d\a_2 d\a'_1 d\a'_2 \frac{|\a_1-\a_2|^{-J-1}|\a'_1|^{-J-1}}{(1+\a_1(\a_1'+\a_2')+i\e)(1+\a_2\a_2'+i\e)}\nn\\
&=\int_{\a_1>0,\a_2<0} d\a_1 d\a_2 d\a'_1 d\a'_2 \frac{|\a_1-\a_2|^{-J-1}|\a'_1|^{-J-1}}{\a_1\a_2(\a_1^{-1}+\a_1'+\a_2'+i\e)(\a_2^{-1}+\a_2'-i\e)}\nn\\
&\quad +\int_{\a_1<0,\a_2>0} d\a_1 d\a_2 d\a'_1 d\a'_2 \frac{|\a_1-\a_2|^{-J-1}|\a'_1|^{-J-1}}{\a_1\a_2(\a_1^{-1}+\a_1'+\a_2'-i\e)(\a_2^{-1}+\a_2'+i\e)}\nn\\
&=-4\pi i\int_{\a_1>0,\a_2>0} d\a_1 d\a_2 d\a'_1 \frac{(\a_1+\a_2)^{-J-1}|\a'_1|^{-J-1}}{\a_1\a_2(\a_1^{-1}+\a_2^{-1}+\a_1'+i\e)}\nn\\
&=\frac{8\pi^2 e^{i\pi J}}{1-e^{i\pi J}} \int_{\a_1>0,\a_2>0} d\a_1 d\a_2 {(\a_1\a_2)^J(\a_1+\a_2)^{-2J-2}}\nn\\
&=\frac{8\pi^2 e^{i\pi J}}{1-e^{i\pi J}} \frac{2^{-2J-1}\sqrt\pi \G(J+1)}{\G(J+\frac{3}{2})}\int_0^{+\oo} \frac{d\a_2}{\a_2}.
\ee

\section{Local twist-4 operators and free-theory three-point functions}
\label{app:moreLocalData}

The purpose of this appendix is to give a comprehensive presentation of explicit results that we have produced for twist-four operators in the $\epsilon$-expansion. Section~\ref{app:OPEdefs} sets some conventions, section~\ref{app:rationalGammas} presents the complete set of twist-four operators with rational anomalous dimensions. Section~\ref{eq:explicitForms} gives the explicit form of some operators considered in this paper. The main results of the appendix are presented in section~\ref{eq:explicitOPEs}, and are in support of the argument of section~\ref{sec:opes} in the main text, where we discussed the dominance in OPE coefficients of double-twist operators. Finally, section~\ref{app:weightedAverages} contains results for weighted averages of anomalous dimensions, which may be useful input for future analytic bootstrap computations. 
In this appendix, $\gamma_J$ is taken to mean the one-loop anomalous dimension, with a factor $\epsilon$ understood.

\subsection{Definitions and conventions}
\label{app:OPEdefs}

We begin by reviewing some conventions. The two-point function of a spinning operator is given by \eqref{eq:standard2pt}.
The three-point function of two scalar and one spinning operator are given by the appropriate generalisation of \eqref{eq:standardstructure}:
\begin{equation}
\langle \cO_1(x_1) \cO_2(x_2) \cO_3(x_3,z)\rangle =
\lambda_{\cO_1\cO_2\cO_3}\frac{(2z\cdot x_{23}x_{13}^2-2z\cdot x_{13}x_{23}^2)^{J}}{x_{12}^{\Delta_1+\Delta_2-\Delta_3+J}
		x_{13}^{\Delta_1+\Delta_3-\Delta_2+J}
		x_{23}^{\Delta_2+\Delta_3-\Delta_1+J}}.
\end{equation}
For the three-point functions involving more than one spinning operator, there is in general more than one allowed three-point structure, and therefore more than one OPE coefficient. This fact is perhaps best known from the case of conserved global symmetry currents and stress-tensors. Reference~\cite{Cuomo:2017wme} gave an automatised way to generate these structures in $d=4$. 

In principle, we therefore would need to specify more than one OPE coefficient for the case of three-point functions of multiple spinning operators. However, by explicit computation we find that for the case we are interested in, where the equation $\tau_3=\tau_1+\tau_2$ holds exactly, the resulting three-point function is always proportional to a single three-point structure. Thus we can write
\begin{align}
\label{eq:lambdaextremal}
&\langle \cO_1(x_1,z_1) \cO_2(x_2,z_2) \cO_3(x_3,z_3)\rangle 
\\&\quad =
\lambda^{[\text{extremal}]}_{\cO_1\cO_2\cO_3}
\frac{(-2z_1\cdot I(x_{13})\cdot z_3)^{J_1}
(-2z_2\cdot I(x_{23})\cdot z_3)^{J_2}
(2z\cdot x_{23}x_{13}^2-2z\cdot x_{13}x_{23}^2)^{-J_{12;3}}
}{x_{12}^{\Delta_{12;3}-J_{12;3}}x_{13}^{\Delta_{13;2}-J_{13;2}}x_{23}^{\Delta_{23;1}-J_{23;1}}},
\nonumber
\end{align}
where $J_{ij;k}=J_i+J_j-J_j$ and $\Delta_{ij;k}=\Delta_i+\Delta_j-\Delta_k$ and the definition of $I(x)$ is given after \eqref{eq:standard2pt}. 
We introduced the label $^{[\text{extremal}]}$ to indicate that the twist-relation $\tau_3=\tau_1+\tau_2$ is exactly satisfied.
This ``extremality condition'' will not survive in perturbation away from the free theory.\footnote{Note that although we are using the operator basis found using perturbative computations, the OPE coefficients are evaluated in strictly in the $d=4$ free theory using Wick contractions.}
In what follows, we will use $\lambda_{\cO_1\cO_2\cO_3}$ as short for $\lambda^{[\text{extremal}]}_{\cO_1\cO_2\cO_3}$.

\subsection{Rational anomalous dimensions}
\label{app:rationalGammas}

In this section, we review all occurences of rational values of the one-loop anomalous dimensions of twist-4 operators. They are known from previous work~\cite{Kehrein:1994ff}.

\paragraph{Even spin} At the first few spins, all anomalous dimensions are rational: $\gamma=2$ at $J=0$, $\gamma=\frac{13}9$ at $J=2$ and $\gamma=\frac49,\gamma=\frac{19}{15}$ at $J=4$.

There are even-spin operators with anomalous dimension $\gamma=\frac13$, whose degeneracies $d_J$ are given by the generating formula
\begin{equation}
\label{eq:genfgamma13}
\sum_J d_\ell q^J=\frac{q^{10}}{(1-q^2) (1-q^6)}=q^{10}+q^{12}+q^{14}+2 q^{16}+2 q^{18}+2 q^{20}+\ldots.
\end{equation}
There are also even-spin operators with anomalous dimension $\gamma=0$, with degeneracy given by the generating formula
\begin{equation}
\label{eq:genfgamma0}
\frac{q^{12}}{(1-q^2) (1-q^4) (1-q^6)}=q^{12}+q^{14}+2 q^{16}+3 q^{18}+4 q^{20}+\ldots.
\end{equation}

\paragraph{Odd spin} For odd spin, all anomalous dimensions are rational. 
Firstly, at every odd spin $J\geq3$, we have one operator with $\gamma=1$. This is not present in the spectrum of the interacting theory, since it becomes related to the divergence of the broken twist-2 currents of even spins $J\geq4$.

For every $\ell=2,3,4,\ldots$, we have for every odd spin $J\geq 3+2\lfloor \frac{3\ell}2\rfloor$ exactly one operator with $\gamma=\frac13+\frac{2(-1)^\ell}{3(\ell+1)}$.

There are odd-spin operators with anomalous dimension $\gamma=\frac13$, whose degeneracy is given by the generating formula 
\begin{equation}
\label{eq:genfgamma13odd}
\frac{q^7}{\left(q^2-1\right)^2 \left(q^4+q^2+1\right)}=q^7+q^9+q^{11}+2q^{13}+2q^{15}+\ldots.
\end{equation}
There are also odd-spin operators with anomalous dimension $\gamma=0$, whose degeneracy is given by the generating formula
\begin{equation}
\label{eq:genfgamma0odd}
\frac{q^{15}}{\left(1-q^2\right) \left(1-q^4\right)
   \left(1-q^6\right)}=q^{15}+q^{17}+2q^{19}+\ldots.
\end{equation}

\subsection{Explicit form of some local operators}
\label{eq:explicitForms}

In this section we write explicit form of the unit-normalised versions of the operators appearing in \eqref{eq:collectionOfOPEs}, in terms of the fundametal field $\phi$ appearing in the Lagrangian. They are constructed to satisfy the normalisation \eqref{eq:standard2pt}. To write the symbols of these operators, we use the letter $\varphi$.
In particular, for pure powers of the field $\phi$, we have
\begin{equation}
\label{eq:definingCanonicallyNormalisedOps}
\varphi^p(x)=(2\pi)^p\frac1{\sqrt{p!}}:\phi(x)^p:\,.
\end{equation}
For the spinning operators, we give the explicit form of some local operators in terms of a polarisation vector $z$:
\begin{align}
\hat T(x,z)&=\frac{4\pi^2}{\sqrt{12}}\left(
{\phi } {\phi ''}-2 {\phi '}^2
\right),
\\
C(x,z)&=\frac{4\pi^2}{\sqrt{20160}}\left(
{\phi } {\phi ^{(4)}}+18 {\phi ''}^2-16  {\phi '}{\phi ^{(3)}}
\right),
\\
\cO_2^{\tau=3}(x,z)&=\frac{8\pi^3}{4}\left(
{\phi }^2 {\phi ''}-2 {\phi } {\phi '}^2
\right),
\\
\cO_3^{\tau=3}(x,z)&=\frac{8\pi^3}{\sqrt{1260}}\left(
{\phi }^2 {\phi ^{(3)}}+12 {\phi '}^3-9 {\phi } {\phi '} {\phi ''}
\right),
\\
\cO_4^{\tau=3}(x,z)&=\frac{8\pi^3}{144}\left(
{\phi }^2{\phi ^{(4)}} +18 {\phi } {\phi ''}^2-16 {\phi } {\phi '} {\phi ^{(3)}}
\right),
\\
\cO_5^{\tau=3}(x,z)&=\frac{8\pi^3}{\sqrt{3326400}}\left(
 {\phi }^2{\phi ^{(5)}}-25 {\phi ^{(4)}} {\phi } {\phi '}+20 {\phi ^{(3)}} {\phi
   } {\phi ''} +160 {\phi ^{(3)}} {\phi '}^2-180 {\phi '} {\phi ''}^2
\right),
\\
\cO_6^{\tau=3}(x,z)&=\frac{8\pi^3}{\sqrt{197683200}}\Big(
{\phi }^2 {\phi ^{(6)}}-\frac{460}{3} {\phi } (\phi ^{(3)})^2-36   {\phi } {\phi ^{(5)}} {\phi '}
+70 {\phi^{(4)}} {\phi '}^2+210 {\phi ''}^3
\nonumber\\
&
\qquad \qquad \qquad\qquad +190 {\phi } {\phi ^{(4)}} {\phi ''}-280 {\phi ^{(3)}} {\phi '} {\phi ''}
\Big),
\\
\cO_{6,\gamma=0}^{\tau=3}(x,z)&=\frac{8\pi^3}{\sqrt{311040}}\left(
4 {\phi } (\phi ^{(3)})^2+18 {\phi ''}^3-3 {\phi } {\phi ^{(4)}} {\phi ''}
+6 {\phi^{(4)}} {\phi '}^2-24 {\phi ^{(3)}} {\phi '} {\phi ''}
\right),
\end{align}
where normal-ordering is understood, and $\phi^{(n)}=(z\cdot \partial)^n\phi(x)$. 
The twist-2 operators $\varphi^2$, $\hat T$ and $C$ agree with \eqref{eq:OJdefn0} for $J=0,2,4$.
Note that $\hat T(x,z)=\frac{-2}{\sqrt{C_T}}T(x,z)$ for the stress-tensor.

\subsection{OPE coefficients of local operators}
\label{eq:explicitOPEs}

In this section, we present the results of the study of OPE coefficients of twist-four families in the different OPEs given in \eqref{eq:collectionOfOPEs}. For each pair of lower-twist operators $\cO_1,\cO_2$ from the collection \eqref{eq:collectionOfOPEs}, we present a plot of the contribution $\nu_{\cO_1\cO_2}(i,J)$ for the different trajectories. The cases $\varphi^2\times\varphi^2$ and $\varphi\times\varphi$ were discussed in section~\ref{sec:opes} in the main text, figures~\ref{fig:OPEratios22} and~\ref{fig:OPEratios11}). 

As to further clarify our normalisation, we give a few low-lying OPE coefficients in table~\ref{tab:opecoefs}. In that table we used the explicit forms of the operators given in the previous section.
\begin{table}
\caption{OPE coefficients involving low-lying operators. We give $\lambda_{\cO_1\cO_2\cO_3}$.  For the spinning OPEs, we give the value of the unique (extremal) OPE coefficient defined by \eqref{eq:lambdaextremal}. For $\cO_1$ and $\cO_2$ scalars, the normalisation agrees with~\cite{Simmons-Duffin:2016wlq,Henriksson:2022rnm}. For the general case, we give the unique OPE coefficient $\lambda^{[\text{extremal}]}_{\cO_1\cO_2\cO_3}$, defined in \eqref{eq:lambdaextremal}.
The local operators are always defined up to a sign; we choose signs so that as many as possible of the low-lying OPE coefficents are positive.}\label{tab:opecoefs}
{
\begin{center}
\def\arraystretch{1.6}
\begin{tabular}{|c|c|c|c|c|}\hline
$\cO_1\times\cO_2$ \textbackslash\  $\cO_3$ & $\varphi^4$ ($\gamma=2$) & $\partial^2\varphi^4$ ($\gamma=\frac{13}9$)& $\partial^4\varphi^4$ ($\gamma=\frac49$) & $\partial^4\varphi^4$ ($\gamma=\frac{19}{15}$)
\\\hline
$\varphi\times \varphi$   &   $\epsilon\sqrt{\frac{1}{54}}$  &   $\epsilon\sqrt{\frac{1}{1440}}$  &   $-\epsilon\sqrt{\frac{1}{419580}}$  &   $\epsilon\sqrt{\frac{1}{23976}}$  
\\\hline
$\varphi^2\times\varphi^2$   &   $\sqrt{6}$  &   $\sqrt{\frac85}$  &   $\sqrt{\frac{125}{2331}}$  &   $\sqrt{\frac{8}{37}}$  
\\\hline
$\varphi^2\times \hat T$   &   $0$  &   $\sqrt{\frac{5}{3}}$  &   $-\sqrt{\frac{686}{555}}$  &   $\sqrt{\frac{48}{37}}$  
\\\hline
$\varphi^2\times C$   &   $0$&$0$  &      $0$  &   $\sqrt{\frac{37}{35}}$
\\\hline
$\hat T\times \hat T$   &   $0$&$0$   &    $\sqrt{\frac{140}{37}}$  &   $\sqrt{\frac{8}{37}}$ 
\\\hline
$\varphi\times \cO_2^{\tau=3}$   &   $0$  &   $\sqrt{\frac{5}{2}}$  &   $-\sqrt{\frac{448}{1665}}$  &   $\sqrt{\frac{18}{37}}$  
\\\hline
$\varphi\times\cO_3^{\tau=3}$   &   $0$  &   $0$  &   $\sqrt{\frac{2916}{925}}$  &   $-\sqrt{\frac{32}{1295}}$  
\\\hline
$\varphi\times\cO_4^{\tau=3}$   &   $0$  &   $0$  &   $0$  &   $\sqrt{\frac{37}{18}}$  
\\\hline
\end{tabular}
\end{center}
}
\end{table}
For the $\varphi\times\varphi$ OPE, we use the relation between $\varphi$ and $\varphi^3$ as was discussed in section~\ref{sec:matrixelements}. There we derived \eqref{eq:relationphi-phi3-final}, which using canonically-normalised operators translates to
\begin{equation}
\label{eq:phi-phi3-relation-app}
\lambda_{\varphi\varphi\cO_{J}}=\frac{\epsilon}{3\sqrt6(J+2)} \lambda_{\varphi\varphi^3\cO}.
\end{equation}

\subsubsection{$\varphi^2\times \hat T$ OPE}

\begin{figure}
  \centering
\includegraphics[width=130mm]{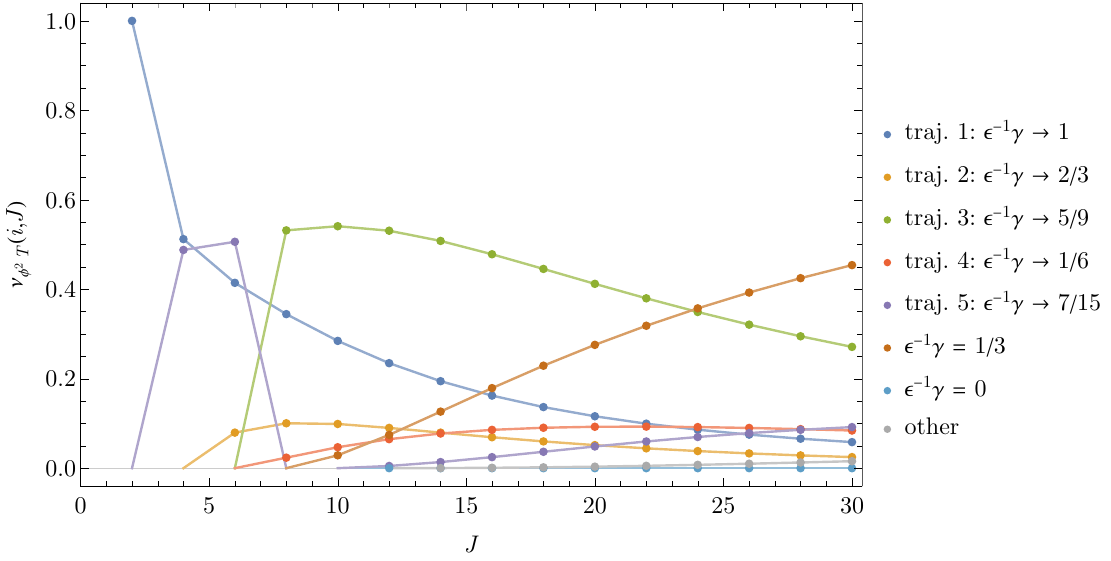}
  \caption{Ratios of OPE coefficients in the $\varphi^2\times \hat T$ OPE}\label{fig:OPEratios2T}
\end{figure}

In the $\varphi^2\times \hat T$ OPE, we find that
\begin{equation}
\overline{\lambda^2_{\varphi^2\hat T}(J)}=\frac{\Gamma(J)\Gamma(J+4)}{120 \Gamma (2 J+3)}(J^6+9 J^5+65 J^4+255 J^3-306 J^2-1944 J+2880).
\end{equation}
The associated plot of the OPE coefficents for the ratios $\nu_{\varphi^2T}(i,J)$ is given in figure~\ref{fig:OPEratios2T}. We see that combined contribution from the operators with asymptotic value $\gamma_J\to\frac13$ dominates the OPE for large enough spin.

We can also compute the average with insertion of anomalous dimensions:
\begin{align}
\nonumber
\overline{\lambda^2_{\varphi^2\hat T}(J)\gamma(J)}&=\frac{ \Gamma   (J)\Gamma(J+4)}{360\Gamma (2 J+3)}\Big((J^6+9 J^5-255 J^4-1665 J^3-3106 J^2-1704 J+2880)+\\&\quad\qquad\qquad\qquad\qquad+160 J (J+1)(J+2)(J+3)S_1(J+1)\Big),
\end{align}
from which we immediately see that
\begin{equation}
\lim_{J\to\infty}\overline{\lambda^2_{\varphi^2\hat T}(J)\gamma(J)}\Big/\overline{\lambda^2_{\varphi^2\hat T}(J)}=\frac13.
\end{equation}

\subsubsection{Other OPEs}

For the remaining OPEs, we do not give explicit formulas. Instead we refer to the relevant figures where we plot the averages. 

\paragraph{$\boldsymbol{\varphi^2\times C}$ OPE}

\begin{figure}
  \centering
\includegraphics[width=130mm]{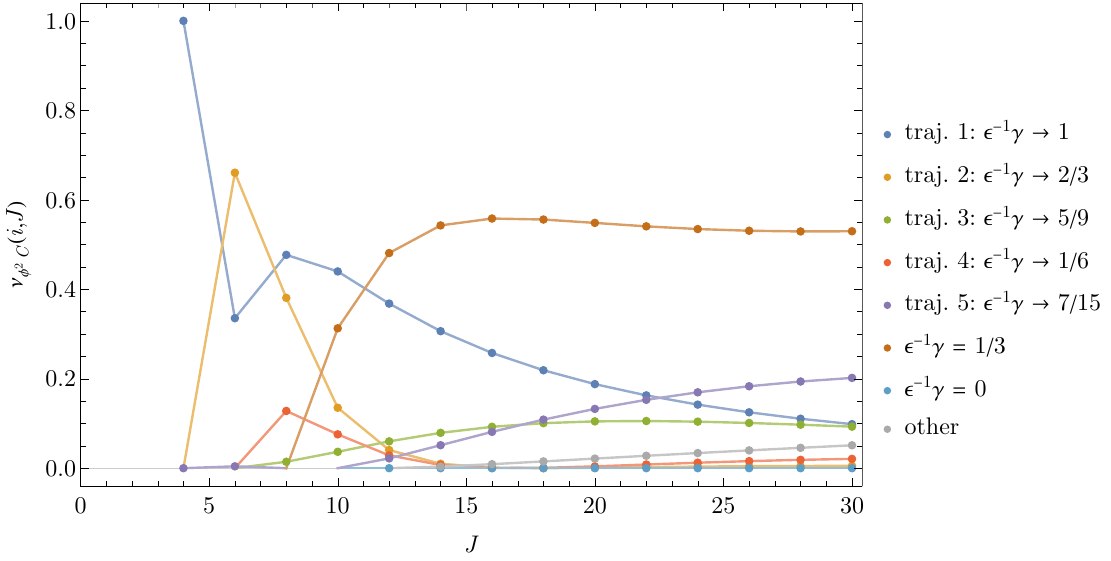}
  \caption{Ratios of OPE coefficients in the $\varphi^2\times C$ OPE, where $C$ is the twist-2 operator of spin 4.}\label{fig:OPEratios2C}
\end{figure}

The associated plot of the OPE coefficents for the various ratios $\nu_{\varphi^2C}(i,J)$ is given in figure~\ref{fig:OPEratios2C}. Here we expect to see that, just like for the stress-tensor, the combined contribution from the operators with asymptotic value $\gamma_J\to\frac13$ should dominate the OPE for large enough spin. This is plausible from the figure, but it appears spin is not large enough to see this.

\paragraph{$\boldsymbol{\hat T\times\hat T}$ OPE}

\begin{figure}
  \centering
\includegraphics[width=130mm]{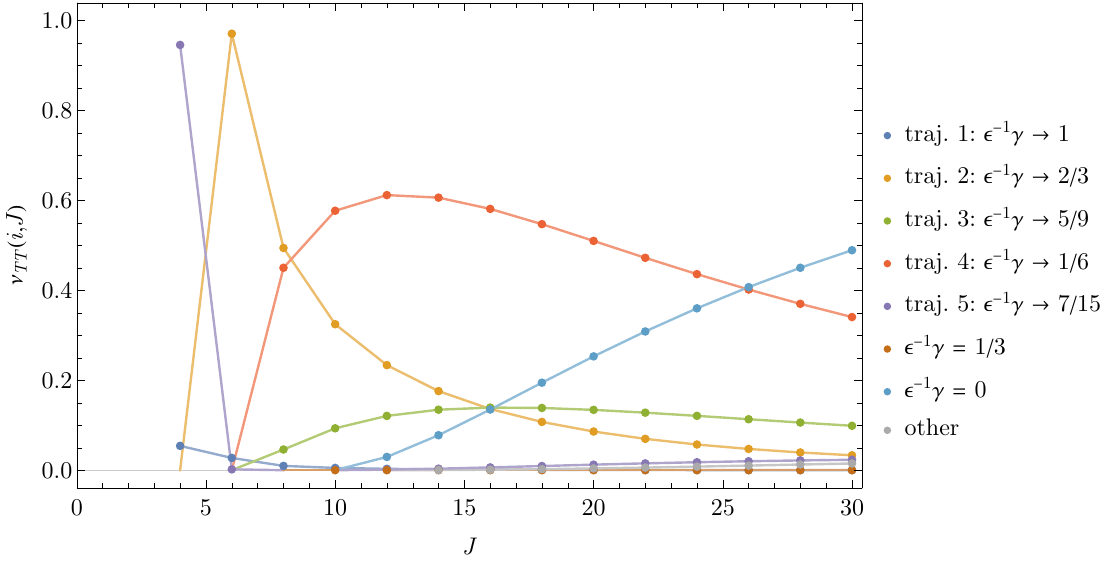}
  \caption{Ratios of OPE coefficients in the $\hat T\times\hat T$ OPE.}\label{fig:OPEratiosTT}
\end{figure}

The associated plot of the OPE coefficents for the various ratios $\nu_{\hat T\hat T}(i,J)$ is given in figure~\ref{fig:OPEratiosTT}. We see that combined contribution from the operators with asymptotic value $\gamma_J\to0$ dominates the OPE for large enough spin.

\paragraph{$\boldsymbol{\varphi\times\cO_\ell^{\tau=3}}$ OPE}

\begin{figure}
  \centering
\includegraphics[width=130mm]{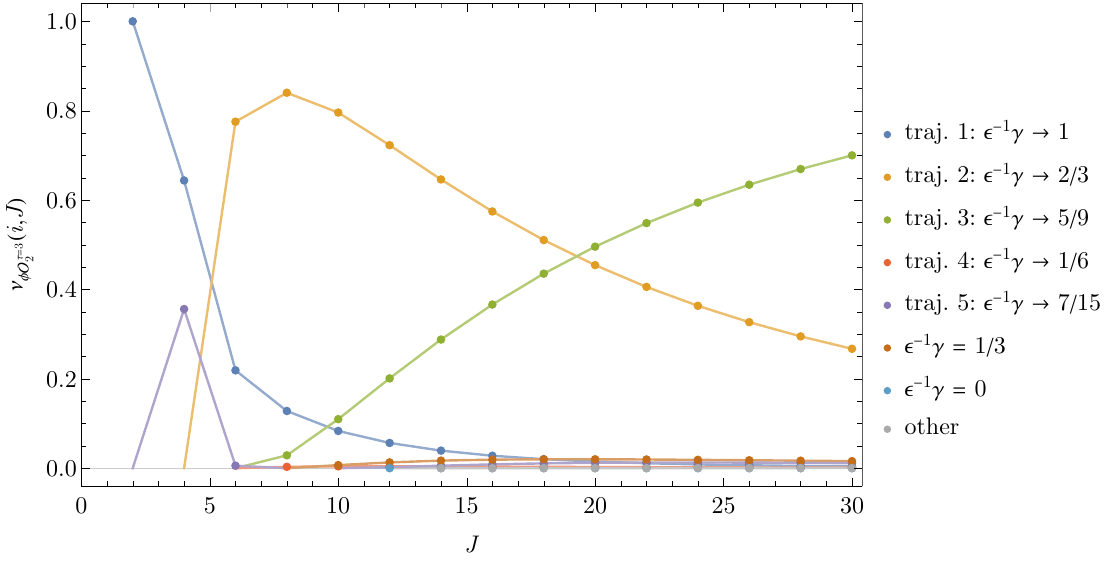}
  \caption{Ratios of OPE coefficients in the $\varphi\times \cO_2^{\tau=3}$ OPE.}\label{fig:OPEratios1O2}
\end{figure}

\begin{figure}
  \centering
\includegraphics[width=130mm]{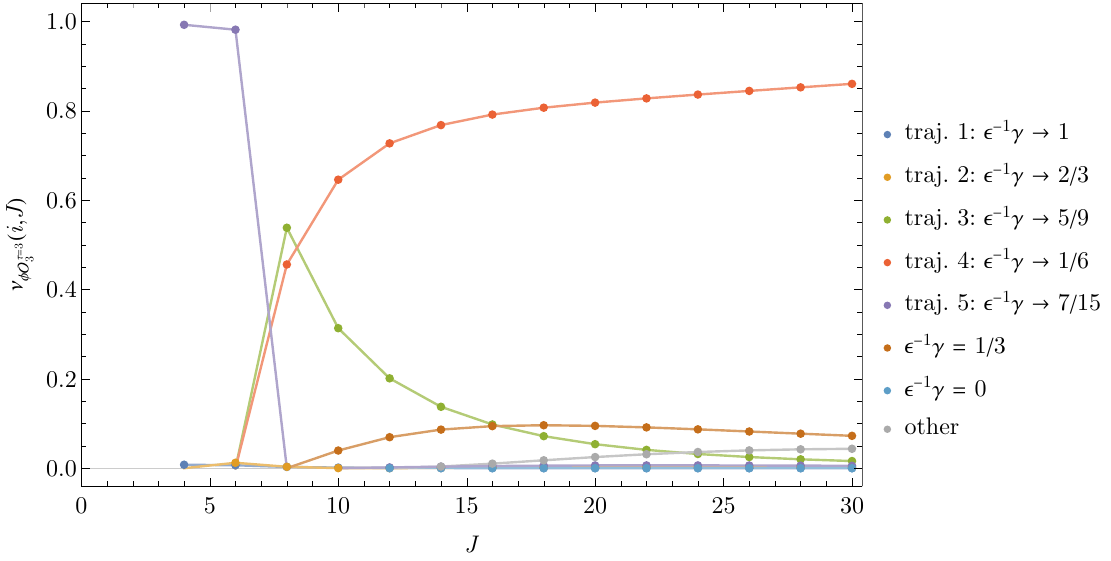}
  \caption{Ratios of OPE coefficients in the $\varphi\times \cO_3^{\tau=3}$ OPE.}\label{fig:OPEratios1O3}
\end{figure}

\begin{figure}
  \centering
\includegraphics[width=130mm]{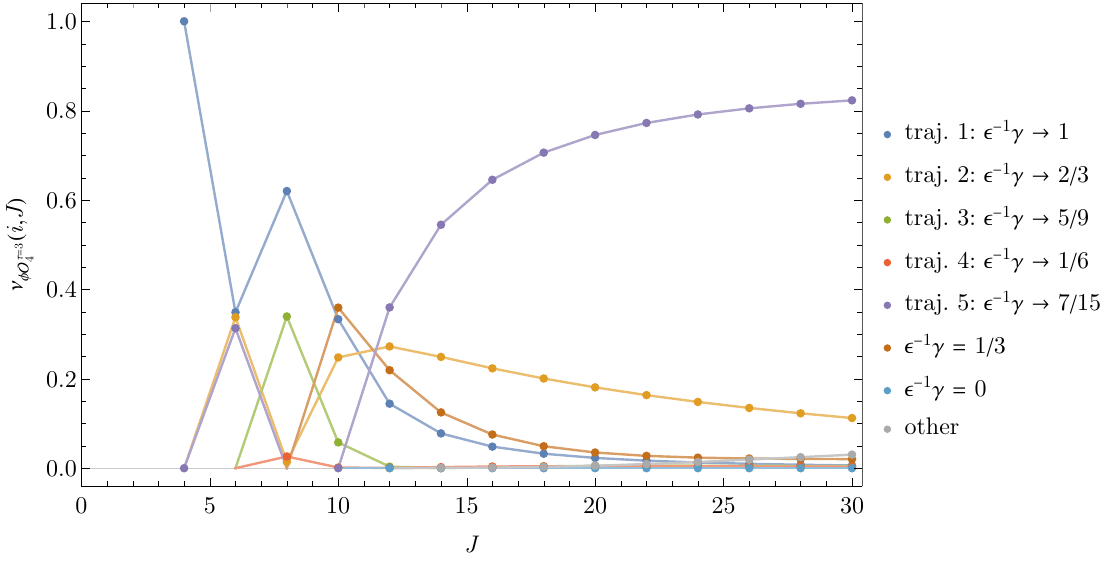}
  \caption{Ratios of OPE coefficients in the $\varphi\times \cO_4^{\tau=3}$ OPE.}\label{fig:OPEratios1O4}
\end{figure}

\begin{figure}
  \centering
\includegraphics[width=130mm]{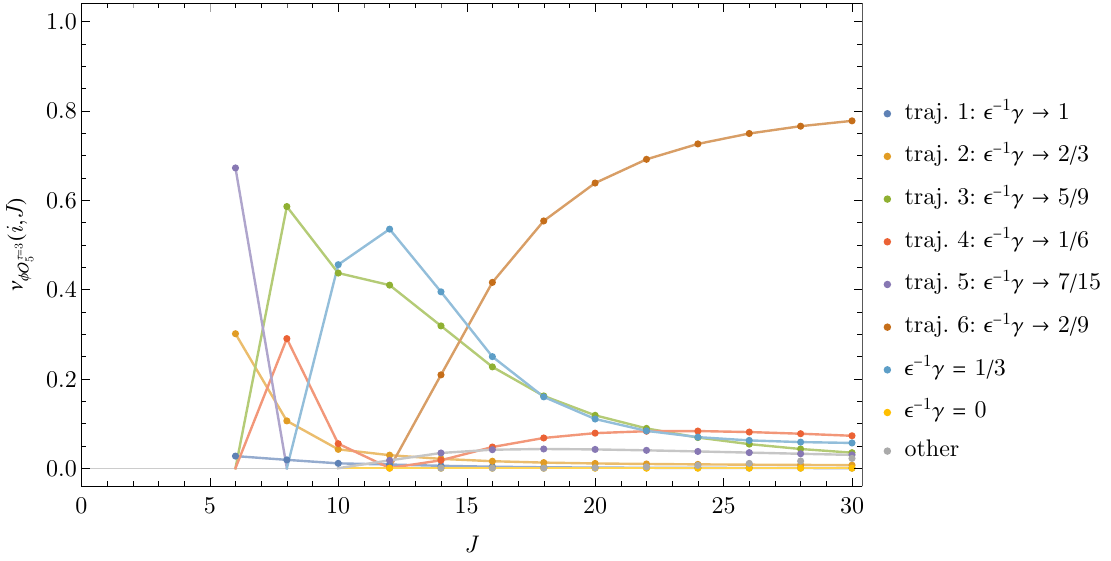}
  \caption{Ratios of OPE coefficients in the $\varphi\times \cO_5^{\tau=3}$ OPE.}\label{fig:OPEratios1O5}
\end{figure}

\begin{figure}
  \centering
\includegraphics[width=130mm]{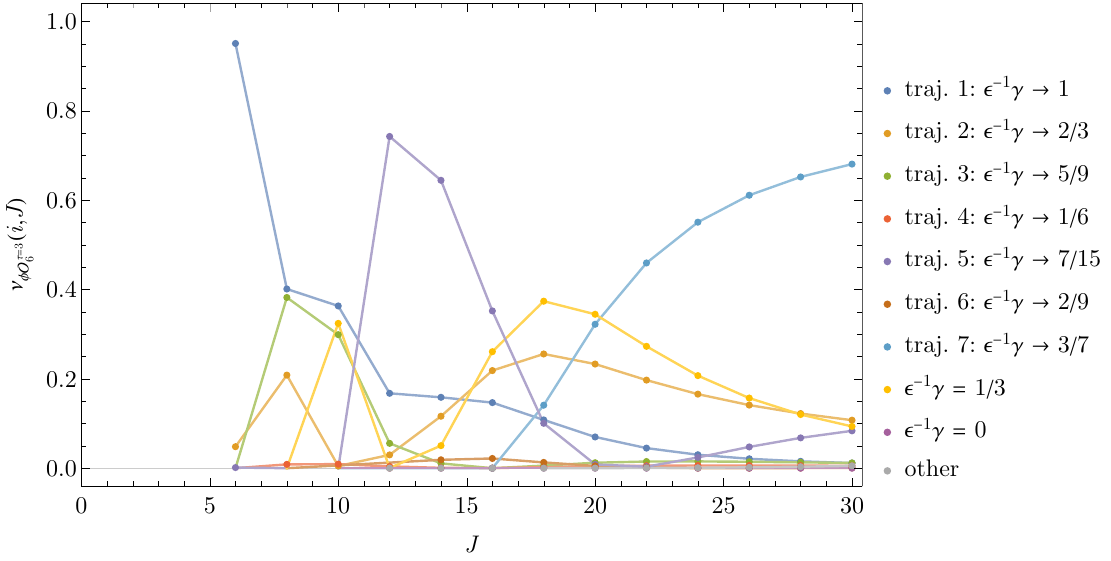}
  \caption{Ratios of OPE coefficients in the $\varphi\times \cO_6^{\tau=3}$ OPE.}\label{fig:OPEratios1O61}
\end{figure}

Now we consider the case $\varphi\times\cO_\ell^{\tau=3}$, where $\cO_\ell^{\tau=3}$ is the twist-three operator with $\gamma_\ell=\frac13+\frac{2(-1)^\ell}{3(\ell+1)}$. 
The associated plots of the OPE coefficents for the various ratios $\nu_{\varphi\cO_\ell}(i,J)$ are given in figures \ref{fig:OPEratios1O2}, \ref{fig:OPEratios1O3}, \ref{fig:OPEratios1O4}, \ref{fig:OPEratios1O5},  \ref{fig:OPEratios1O61} for $\ell=2,3,4,5,6$. From these figures we see that according to our expectations, trajectory $\ell+1$ with asymptotic value $\gamma_J\to \frac13+\frac{2(-1)^\ell}{3(\ell+1)}$ dominates the OPE for large enough spin.

\begin{figure}
  \centering
\includegraphics[width=130mm]{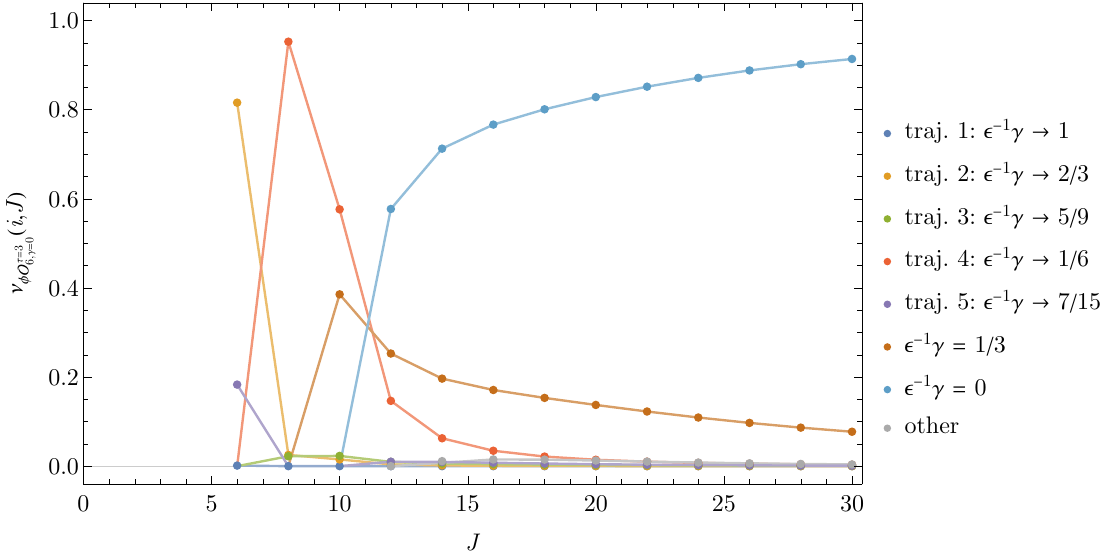}
  \caption{Ratios of OPE coefficients in the $\varphi\times \cO_{6,\gamma=0}^{\tau=3}$ OPE.}\label{fig:OPEratios1O60}
\end{figure}

It is interesting to consider the case at spin six, where there are two twist-3 operators: $\cO_6^{\tau=3}$ with $\gamma=\frac37$,  and $\cO_{6,\gamma=0}^{\tau=3}$ with $\gamma=0$. In the latter case we expect that the OPE coefficients of twist-4 operators in the $\phi\times\cO_{6,\gamma=0}^{\tau=3}$ OPE should be dominated by those with zero one-loop anomalous dimensions. This is indeed what we find in figure~\ref{fig:OPEratios1O60}.

\subsection{Weighted averages of anomalous dimensions}
\label{app:weightedAverages}

In this section we give our final set of explicit results. 
By inserting powers of the anomalous dimensions, we can find ancillary results for some weighted averages. Interestingly, for integer spins these are always rational, even though the involved OPE coefficients and anomalous dimensions generically are irrational numbers. Wherever possible, we match with expressions derived in the analytic conformal bootstrap~\cite{Alday:2017zzv,Bertucci:2022ptt} and find perfect agreement. In the remaining cases, our results might be useful as input in higher-order analytic bootstrap computations. In this section, we restore all factors of $\epsilon$, and we use the relation \eqref{eq:phi-phi3-relation-app} between $\lambda_{\varphi\varphi^3\cO}$ and $\lambda_{\varphi\varphi\cO}$.

In the $\varphi$ four-point function we find
\begin{align}
\overline{\lambda^2_{\varphi\varphi}(J)}&=\frac{\Gamma(2+J)^2}{\Gamma(2J+3)}\left(\frac1{108}+\frac1{18\mathcal J^2}\right)\epsilon^2,
\\
\overline{\lambda^2_{\varphi\varphi}(J)\gamma(J)}&=\frac{\Gamma(2+J)^2}{\Gamma(2J+3)}\left(\frac1{108}+\frac7{54\mathcal J^2}\right)\epsilon^3,
\\
\overline{\lambda^2_{\varphi\varphi}(J)\gamma(J)^2}&=\frac{\Gamma(2+J)^2}{\Gamma(2J+3)}\left(\frac{1}{108}+\frac{37}{162\mathcal J^2}
+\frac{8}{81 \mathcal J^4}\right)\epsilon^4,
\\
\overline{\lambda^2_{\varphi\varphi}(J)\gamma(J)^3}&=\frac{\Gamma(2+J)^2}{\Gamma(2J+3)}\left(\frac{1}{108}+\frac{167}{486 \mathcal J^2}+\frac{80}{243\mathcal J^4}+\frac{32S_1(J+1)}{243\mathcal J^4}\right)\epsilon^5,
\end{align}
where $\mathcal J^2=(J+1)(J+2)$. 
The first three results were found in~\cite{Alday:2017zzv}.\footnote{The method used in~\cite{Alday:2017zzv} to find these results assumed consistency with a transcendentality principle, which was not rigorously justified. Here we can confirm by explicit computations that the results of~\cite{Alday:2017zzv} are indeed correct.}

In the $\varphi^2$ four-point function we find
\begin{align}
\overline{\lambda^2_{\varphi^2\varphi^2}(J)} &=\frac{\Gamma(2+J)^2}{\Gamma(2J+3)}\left(2\mathcal J^2+8\right),
\\
\label{eq:gammaver22}
\overline{\lambda^2_{\varphi^2\varphi^2}(J)\gamma(J)}&=\frac{\Gamma(2+J)^2}{3\Gamma(2J+3)}\left(4 \mathcal J^2+32+32 S_1(J+1)\right)\epsilon,
\\
\overline{\lambda^2_{\varphi^2\varphi^2}(J)\gamma(J)^2}&=\frac{\Gamma(2+J)^2}{9\Gamma(2J+3)}\left(8 \mathcal J^2+\frac{64}{ \mathcal J^2}+96+160S_1(J+1)-128S_{-2}(J+1)\right)\epsilon^2,
\\
\overline{\lambda^2_{\varphi^2\varphi^2}(J)\gamma(J)^3}&=\frac{\Gamma(2+J)^2}{27\Gamma(2J+3)}\bigg(16\mathcal J^2+\frac{448}{\mathcal J^2}+256-768S_{-2}(J+1)+512S_{-3}(J+1)
\nonumber\\
&\qquad\qquad\qquad\qquad-1024S_{-2,1}(J+1)+\left(544+\frac{512}{\mathcal J^2}\right) S_1(J+1)\bigg)\epsilon^3.
\end{align}
The result \eqref{eq:gammaver22} was found in~\cite{Henriksson:2020jwk}, see also~\cite{Bertucci:2022ptt}.

Finally, we also consider the mixed four-point function $\langle\varphi\varphi\varphi^2\varphi^2\rangle$, in which case we find
\begin{align}
\overline{\lambda_{\varphi\varphi}(J)\lambda_{\varphi^2\varphi^2}(J)}&=\frac{2\Gamma(2+J)^2}{3\Gamma(2J+3)}\epsilon
\\
\label{eq:gammaver1122}
\overline{\lambda_{\varphi\varphi}(J)\lambda_{\varphi^2\varphi^2}(J)\gamma(J)}&=\frac{8\Gamma(2+J)^2}{9\Gamma(2J+3)}\left(1+\frac{1}{\mathcal J^2}\right)\epsilon^2,
\\
\overline{\lambda_{\varphi\varphi}(J)\lambda_{\varphi^2\varphi^2}(J)\gamma(J)^2}&=\frac{\Gamma(2+J)^2}{27\Gamma(2J+3)}\left(28+\frac{56}{ J^2}+\frac{32 S_1(J+1)}{\mathcal  J^2}\right)\epsilon^3,
\\
\overline{\lambda_{\varphi\varphi}(J)\lambda_{\varphi^2\varphi^2}(J)\gamma(J)^3}&=\frac{\Gamma(2+J)^2}{81\Gamma(2J+3)\!}\!\left(92+\frac{264}{\mathcal J^2}+\frac{64}{\mathcal J^4}+\frac{256 S_1(J+1)}{ \mathcal J^2}-\frac{128 S_{-2}(J+1)}{\mathcal J^2}\right)\!\epsilon^4\!.
\end{align}
The result \eqref{eq:gammaver1122} was found in~\cite{Bertucci:2022ptt}.

In all of the above expressions, $\mathcal J^2=(J+1)(J+2)$ and the nested harmonic numbers are defined by
\begin{equation}
S_{a_1,a_2,\ldots}(n)=\sum_{k_1=1}^n\frac{(\mathrm{sgn}\,a_1)^{k_1}}{k_1^{|a_1|}}S_{a_2,\ldots}(k_1)\,,
\end{equation}
with $S_{\{\,\}}(n)=1$, see \emph{e.g.}~\cite{Albino:2009ci}.

\section{General CRT transformations}
\label{app:CRT}

This appendix follows and extends the discussion in section 4.2.3 of~\cite{Kologlu:2019bco}. Any CFT has an anti-unitary CRT symmetry $J$ which satisfies $J^2=1$ and acts on local bosonic operators as
\be
	J\cO(x,z)J^{-1}=[\cO(\bar x,\barz)]^\dagger,
\ee
where $\bar x=(-x^0, -x^1,x^2,\cdots)$ and similarly for $\barz$. One can notice that this transformation doesn't commute with most conformal symmetries (for example, it singles out the directions $x^0$ and $x^1$). As discussed in~\cite{Kologlu:2019bco}, the invariant way of viewing $J$ is as being defined by two points $A$ and $B$ on the Lorentzian cylinder. In the equation above, both $A$ and $B$ are at the past null infinity:\footnote{In~\cite{Kologlu:2019bco} the convention is that they are at future null infinity. This convention doesn't affect anything in our discussion.} $A$ can be reached by following $x^0=x^1\to -\oo$ and $B$ by $x^0=-x^1\to -\oo$. We can consider the CRT transformation for a generic choice of these points $A\approx B$, which we denote by $J_{AB}$.

The action of $J_{AB}$ on local operators is convenient to describe using embedding space. First, we define the linear map $R_{AB}$ by
\be
	R_{AB}(X)=X-2\frac{(X\.X_A)}{(X_A\.X_B)}X_B-2\frac{(X\.X_B)}{(X_A\.X_B)}X_A,
\ee
where $X_A$ and $X_B$ are the embedding-space coordinates of $A$ and $B$. Then
\be
	J_{AB}\cO(X,Z)J_{AB}^{-1}=[\cO(R_{AB}(X),R_{AB}(Z))]^\dagger.
\ee

Consider now the action of $J_{AB}$ on the light-transform $\wL[\cO](X_A,Z)$~\cite{Kravchuk:2018htv},
\be
	J_{AB}\wL[\cO](X_A,Z)J_{AB}^{-1}&=\int d\a J_{AB}\cO(Z-\a X_A,-X_A)J_{AB}^{-1}\nn\\
	&=[\int d\a \cO(Z-2(Z\.X_B)(X_A\.X_B)^{-1}X_A+\a X_A,X_A)]^\dagger\nn\\
	&=[\int d\a \cO(Z-\a X_A,X_A)]^\dagger\nn\\
	&=(-1)^{J_\cO}(\wL[\cO](X_A,Z))^\dagger,
\ee
where $J_\cO$ is the Lorentz spin of $\cO$.  The light-ray operators, being analytic continuations of the light-transforms of local operators, satisfy a similar equation,
\be
	J_{AB}\O_{i,J}(X_A,Z)J_{AB}^{-1}=\pm \O_{i,J}(X_A,Z)^\dagger,
\ee
where $+$ sign should be used for even-spin trajectories and $-$ sign for odd-spin trajectories. This generalises the spin signature property of~\cite{Kravchuk:2018htv} to generic CRT transformations.

\section{Special CRT transformation as a complex dilatation}
\label{app:ipiD}

We derive the form of $R_{AB}$ for the special case when $A$ is at the origin $x=0$ of Minkowski space and $B$ is at the spatial infinity. This corresponds to $X_A=(1,0,0)$ and $X_B=(0,1,0)$, where we list the coordinates as $(X^+,X^-,X^\mu)$. Acting on $X=(1,x^2,x^\mu)$ we get
\be
	R_{AB}(X)=(1,x^2,x^\mu)-(0,2x^2,0)-(2,0,0)=(-1,-x^2,x^\mu).
\ee
Now, note that
\be
	e^{\l D}\cO(X,Z)|0\>=\cO(e^{\l D}X, e^{\l D}Z)|0\>,
\ee
where $D$ is the anti-Hermitian dilatation operator and
\be
	e^{\l D}(X^+,X^-,X^\mu)=(e^{-\l}X^+,e^{\l} X^-, X^\mu).
\ee
It follows that if we choose $\l=\pm \pi i$ then formally
\be
	e^{\l D}\cO(X,Z)|0\>=\cO(R_{AB}(X),R_{AB}(Z))|0\>.
\ee
However, for this to make sense, we need make sure that as we move $\l$ from $0$ to $\pm 2\pi i$, the state $e^{\l D}\cO(X,Z)|0\>$ remains normalisable. This is not trivial, since $iD$ is not a positive-definite operator -- inversion conjugates it to minus itself.

Consider the state $\cO(X)|0\>$ (we omit the polarization vector for simplicity). This is a Hilbert-space valued function of $X$ on $X^2=0$, satisfying
\be\label{eq:homogeneity}
	\cO(\l X)|0\>=\l^{-\De}\cO(X)|0\>
\ee
for $\l>0$. Effectively, it is defined on the projective null cone $C$. Note however that $C$ contains only two copies of the Poincare patch, while the Lorentzian cylinder contains infinitely many. We need to make the choice of which patches $C$ covers, and we will assume that it is the ``standard'' or first Poincare patch, parameterised by $X=(1,x^2,x)$, and the one in its future, parameterised by $X=(-1,-x^2,-x)$, which we will refer to as the second.

Consider now the problem of analytically continuing $\cO(X)|0\>$ in $X$. We know that it admits an analytic continuation to $X=(1,x^2,x)$ with $\Im x>0$ by standard energy positivity arguments. Using~\eqref{eq:homogeneity} we can then construct an analytic continuation to $X=(X^+, X^-, X^\mu)$ with small $\arg X^+$ and $\Im (X^\mu/X^+)>0$. Extending the range of $\arg X^+$ we can reach $\arg X^+=\pm i\pi$, which overlaps with the second patch which has $X^+<0$. On the other hand, we do know that for $X$ in the first Poincare patch~\cite{Kravchuk:2018htv}
\be
	\cO(-X)|0\>=e^{i\pi \De}\cO(X)|0\>,
\ee
which we can interpret as~\eqref{eq:homogeneity} being extended to $\arg\l\in [-\pi,\pi)$ and therefore defining $\cO(X)|0\>$ as an analytic function of $X\in \C^{2,d}$ on the domain
\be
	\cD=\{X|X^2=0,X^+\neq 0, \arg X^+\in[-\pi,\pi),\Im(X^\mu/X^+)>0\}.
\ee

Suppose now that $x\in \R^{2,d}$ and $x>0$, and consider
\be
	X'=(e^{-i\f},e^{i\f}x^2,x).
\ee
The condition $\Im(X'^\mu/X'^+)>0$ becomes $x\sin\f>0$, which is satisfied as long as $0<\f<\pi$. As $\f\to \pi$, $\arg X'^+\to -\pi$ and so the second Poincare patch is reached in the limit. This argument shows that for $X=(1,x^2,x)$ and $x>0$ we can write
\be
	e^{i\pi D}\cO(X)|0\>=\cO(R_{AB}(X))|0\>
\ee
and therefore, replacing $\cO\to\cO^\dagger$ and taking Hermitian conjugate,
\be
	\<0|\cO(X)e^{i\pi  D}=\<0|\cO(R_{AB}(X)).
\ee

Similarly, if $X=(-1,-x^2,-x)$ with $x<0$, we can consider
\be
	X'=(e^{i\f-i\pi},-e^{-i\f}x^2,-x)
\ee
The condition $\Im(X'^\mu/X'^+)>0$ becomes again $x\sin{\f}>0$, which is satisfied as long as $0<\f<\pi$. As $\f\to \pi$, $\arg X'^+\to 0$ and so the first Poincare patch is reached in the limit. This implies that for $X=(-1,-x^2,-x)$ and $x<0$
\be
	e^{-i\pi  D}\cO(X)|0\>=\cO(R_{AB}(X))|0\>.
\ee

\section{Redundancy in ansatz~\eqref{eq:fansatz}}
\label{app:redundant}

In this section we determine the redundancy in parameterizing $\tl\psi$ as~\eqref{eq:fansatz}, i.e.\
\be
	\tl\psi(\b_1,\b_2,\b_3,\b_4)=\sum_{i<j}\Psi(\b_i,\b_j).
\ee
In other words, we look for functions $\Psi(p,q)$ which satisfy
\be\label{eq:nullcondition}
	\sum_{i<j}\Psi(\b_i,\b_j)=0,
\ee
as well as the conditions~\eqref{eq:Psibegin}--\eqref{eq:Psiend}. Note that the arguments $\b_i$ are constrained to satisfy
\be\label{eq:appmomentumconservation}
	\b_1+\b_2+\b_3+\b_4=0.
\ee

We first note that differential operators $d_{ij}=\frac{\ptl}{\ptl\b_i}-\frac{\ptl}{\ptl\b_i}$ are tangent to the condition~\eqref{eq:appmomentumconservation}, and so we can freely apply them to~\eqref{eq:nullcondition}. Applying the composition
\be
	d_{13}d_{14}d_{23}d_{24}d_{12}
\ee
we find the condition
\be\label{eq:32diff}
	\p{(\tfrac{\ptl}{\ptl\b_1})^3(\tfrac{\ptl}{\ptl\b_2})^2-(\tfrac{\ptl}{\ptl\b_1})^2(\tfrac{\ptl}{\ptl\b_2})^3}\Psi(\b_1,\b_2)=0.
\ee
Since any values of $\b_1,\b_2$ can be achieved while satisfying~\eqref{eq:appmomentumconservation}, we can treat $\b_1,\b_2$ as unconstrained in this equation.

Assuming even spin parity, for $\b_1\neq 0$ we can write, using~\eqref{eq:Psibegin}--\eqref{eq:Psiend}
\be
	\Psi(\b_1,\b_2)=|\b_1|^J\Psi(1,\b_2/\b_1).
\ee
Plugging into~\eqref{eq:32diff} we find a 5-th order linear differential equation for $\Psi(1,z)$. One can easily check that the general solution of this equation, at least for generic $z$, is
\be
	\Psi(1,z)=c_1 z^J+c_2 z^{J-1}+c_3 (1+z)^J+c_4 z + c_5,
\ee
where $c_i$ are constants. This shows that locally
\be
	\Psi(p,q)=c_1 |q|^J+ c_2 |p| |q|^{J-1}+c_3 |p+q|^J+c_4|p|^{J-1}|q|+c_5 |p|^{J}.
\ee
Since we haven't been careful about the singularities of the differential equation, we allow the constants $c_i$ to take different values in the 6 regions cut out by the curves $p=0$, $q=0$ and $p+q=0$, subject to~\eqref{eq:Psibegin}--\eqref{eq:Psiend}. Analysing the resulting finite-dimensional space of possible solutions, we find that the unique solution for generic $J$ is given by
\be
	\Psi(p,q)=\begin{cases}
	|p|^J(1+\tfrac{3q}{p})+|q|^J(1+\tfrac{3p}{q}), &\text{even spin parity},\\
	(p+q)|p+q|^{J-1}, &\text{odd spin parity}.
	\end{cases}
\ee

\newpage
\providecommand{\href}[2]{#2}\begingroup\raggedright\endgroup

\end{document}